%% file: draftv4.tex
\documentclass[12pt]{article}

\usepackage{jheppub}
\usepackage{hyperref}
\usepackage{footmisc}
 \usepackage{relsize}
\usepackage{comment}
\newcommand{\gammavec}{\frac{\vec \gamma \cdot \vec k}{k}}
\newcommand{\Gammavec}[1]{\frac{\vec \gamma \cdot \vec #1}{#1}}

\include{chandra-macros}
\allowdisplaybreaks %FOR REMOVING UNNECESSARY WHITE SPACES BETWEEN EQUATIONS 

\begin{document}    
\title{Loops, Recursions, and Soft Limits for Fermionic Correlators in (A)dS}       
\author[a]{Chandramouli Chowdhury,}
\author[a, b]{Pratyusha Chowdhury,} 
\author[a]{Radu N. Moga,} 
\author[c, d]{and Kajal Singh}
\affiliation[a]{Mathematical Sciences and STAG Research Centre, University of Southampton, Highfield, Southampton SO17 1BJ, United Kingdom}
\affiliation[b]{School of Physics \& Astronomy, University of Southampton, Highfield, Southampton, SO17 1BJ, United Kingdom}
\affiliation[c]{Department of Mathematical Sciences, University of Liverpool, Liverpool, L69 7ZL, United Kingdom}
\affiliation[d]{Harish-Chandra Research Institute, Chhatnag Road, Jhunsi, Allahabad, India 211019.}
\emailAdd{C.Chowdhury@soton.ac.uk}
\emailAdd{P.Chowdhury@soton.ac.uk}
\emailAdd{r-n.moga@soton.ac.uk}
\emailAdd{Kajal.Singh@liverpool.ac.uk}

\abstract{Study of correlation functions in AdS/CFT and in-in correlators in de Sitter space often requires the computation of Witten diagrams. Due to the complexity of evaluating radial integrals for these correlators, several indirect approaches have been developed to simplify computations. However, in momentum space, these methods have been limited to fields with integer spin. In this paper, we formulate tools for evaluating Witten diagrams with spin$-\frac12$ fields in momentum space and discuss where they differ from the corresponding integer-spin analysis. We formulate our tools explicitly for massless fermions and present how appropriate Weight shifting operators with respect to the external kinematics can be used to obtain the generalization to fermions with integer mass. We apply these tools to loop Witten diagrams and also discuss their use for evaluating in-in correlators in dS. In cases where we can evaluate the loop integrals, we find their transcendentality is lower than the corresponding scalar field results. Further, we classify the nature of IR divergences encountered for interacting massive scalars and fermions. We also prove a novel Weinberg-like soft theorem for gauge fields coupled to matter in AdS and show that the universal terms in the leading soft factor are sensitive to the spin of the matter field. These generalize the recently discovered soft theorems for pure Yang-Mills to Yang-Mills with matter. }

%\date{\today}
\maketitle

%\tableofcontents
\noindent
\flushbottom
\allowdisplaybreaks

%%%%%%%%%%%%%%%%%%%%%%%%%%%
\section{Introduction and Summary}
%%%%%%%%%%%%%%%%%%%%%%%%%%%
In quantum field theory, correlation functions of local fields are one of the simplest objects that provide information about any given theory and are directly related to experimental observables. Even theoretically these are not only useful in flat space but are necessary to make predictions in cosmology and offer consistency checks within the AdS/CFT correspondence \cite{Maldacena:1997re,Gubser:1998bc,Witten:1998qj}. In flat space, one of the important observables is the S-matrix, or the scattering amplitudes, which are also obtained from time-ordered correlators. Over the past few decades, the computation of scattering amplitudes in perturbation theory has garnered considerable attention, leading to the development of various tools designed to simplify these calculations (see \cite{Badger:2023eqz} for a recent textbook discussion). These often require a departure from the evaluation of Feynman diagrams and instead utilize properties of on-shell scattering amplitudes \cite{Britto:2005fq}, thereby reducing the tedious computation of summing numerous Feynman diagrams to a handful of simple terms. Such techniques have been developed for a while in flat space, there is a dearth of powerful computational tools for correlators in Anti-de Sitter (AdS) space and cosmological correlators. The main difficulty in computing correlation functions in such curved spacetimes is the lack of translational invariance along the direction orthogonal to the boundary, which prevents one from using a Fourier transform along that direction. Similar challenges also exist while computing equal, but finite, time correlators in flat space and more generally for any system with a boundary. Hence, one often has to resort to traditional perturbative techniques and compute a version of Feynman diagrams known as Witten diagrams to evaluate such correlators. For computing correlators in flat space, these are sometimes equivalent to solving the Lippmann-Schwinger equation \cite{Lippmann:1950zz} and reduce to performing ``old-fashioned perturbation theory''. Although it is possible to evaluate the standard S-matrix using these methods, it is not recommended due to their cumbersome nature. 

In a remarkable paper \cite{Raju:2010by}, it was shown how a BCFW-like recursion relation for tree-level correlators in AdS could be used to compute boundary four-point current and stress tensor correlators by starting with the 3-pt functions \cite{Raju:2012zs}. This also resulted in one of the first computations of the 4-graviton correlator in AdS, providing a fascinating example of the complexity involved in such computations. For instance, there now exist three a priori different results for the 4-graviton correlator in AdS$_4$ \cite{Raju:2012zs, Bonifacio:2022vwa, Armstrong:2023phb} and it is not immediately clear how these results are consistent with each other! There has also been progress in bootstrapping tree-level AdS correlators by demanding they have the right analytic structure and satisfy the Ward Identities \cite{Arkani-Hamed:2018kmz}. These computations in \cite{Arkani-Hamed:2018kmz}, strictly apply to the wave function of the universe in de Sitter (dS), but by analytical continuation in cases without IR divergences, this is equivalent to evaluating correlators in Euclidean AdS (EAdS). This provides a theory-independent way to arrive at the answers for correlators without making any reference to radial integrals in the bulk. While this approach has been successful for the perturbative 4-pt function, one needs better tools to extend it to higher point functions and beyond tree level. The simplex representation for scalar theories developed in \cite{Bzowski:2019kwd, Bzowski:2020kfw} precisely attacks this question at arbitrary multiplicity and provides an integral representation for the solutions of the conformal Ward identities for scalars. This gives a non-perturbative representation of the correlators but requires further analysis to generate perturbative results. 

For these reasons, it is often convenient to directly evaluate the correlators for specific theories when the radial integrals are doable. One such approach utilizes a well-known technique for loop integrals in flat space, namely, integration by parts (IBP). IBP enables the evaluation of radial integrals for a given diagram in bulk by expressing them in terms of lower-point diagrams. Therefore, it can be viewed as a set of diagrammatic recursion relations. This technique has been used to compute correlation functions for conformally coupled scalars \cite{Arkani-Hamed:2017fdk} and low point spinning correlators for integer spins \cite{Albayrak:2019asr}. It also provides a way to express the loop integrands in a convenient form as it is void of spurious poles. Via the AdS/CFT correspondence, these correspond to $\frac1N$ corrections of CFT correlators. Most of the progress in evaluating loop diagrams in curved spacetime has been primarily focused to scalars \cite{Albayrak:2020isk, Chowdhury:2023khl, Chowdhury:2023arc, Bertan:2018khc, Heckelbacher:2022fbx, Banados:2022nhj, Bzowski:2023jwt, Carmi:2019ocp, Carmi:2021dsn} and  for a few examples in spinning theories \cite{Giombi:2017hpr, Ankur:2023lum, Carmi:2024tzj}. In momentum space, we only have a few explicit examples of one-loop \cite{Albayrak:2020isk} and two-loop integrals \cite{Chowdhury:2023khl, Chowdhury:2023arc}, which heavily rely on the axisymmetric nature of the integrands. It would be an interesting mathematical problem to study these integrals further and also classify the transcendentality of the functions that appear.

% \subsubsection*{\small Summary}
In this paper, we depart from the conventional cases by analyzing fields with half-integer spin in AdS and focus on the subtleties and structure of generic correlators involving spin-$\frac12$ fields coupled to scalars and gauge fields.  
The immediate complication in using the IBP technique for evaluating massless fermionic correlators is that the radial integrands do not vanish at the boundary. Therefore it requires one to extend the existing tools, such as the ones in \cite{Arkani-Hamed:2017fdk},  to more general cases.  We review the derivation for the IBP technique \cite{Arkani-Hamed:2017fdk} in section \ref{sec:recursion} and also discuss simple extensions containing scalar fields with non-vanishing radial integrands at the boundary. 
We then develop a novel recursion relation for correlators containing massless spin-$\frac12$ fields which trivializes performing radial integrals. In a later section (in section \ref{sec:massive}), we also demonstrate how these extend to massive fields. Combined with the recursion relations developed in \cite{Albayrak:2019asr} for gauge fields, these allow one to compute any correlator for fields in the Standard Model in AdS. This method provides a simple way to perform the radial integrals but one has to add the contribution from all channels, which does not seem to yield simple results the way it does in flat space \cite{Raju:2012zs}. Despite such efficient methods, the answers become complicated very soon; therefore, finding ways to check if the answer is correct is necessary. For tree-level scattering amplitudes in flat space, this reduces to evaluating the residues of the amplitudes at their poles. A similar principle also applies to correlation functions in AdS and is discussed in section \ref{sec:sing}. As an outcome of this analysis, we find a novel constraint for correlators in QED (or QCD) when the momentum of the external photon (or gluon) becomes close to zero (see sections \ref{sec:soft} and appendix \ref{app:class2}). This is a generalization of the Weinberg soft theorem for the flat space S-matrix \cite{Weinberg:1965nx}, to AdS correlators and is discussed in \cite{Chowdhury:2024wwe} for pure YM and GR. We analyze the soft limit of the correlators by studying individual Witten diagrams. We find that the full AdS correlator in the soft limit does not split into a universal soft factor times a lower point correlator. However, it is always possible to identify the universal part of the soft limit of a correlator. These universal terms arise from specific kind of diagrams where the soft photon is emitted from an external vertex, which are also the kind of diagrams that give rise to the leading soft theorem in flat space. For every correlator, the universal part of the soft limit is a sum of two terms. One term is similar to the standard expression for the leading Weinberg soft theorem in flat space, with the same numerator but the soft pole replaced by a derivative w.r.t the momentum of the hard particle. The other universal term is proportional to the spin angular momentum of the matter field coupled to the gauge field. This is a new feature in AdS as it indicates that the universal part of the leading soft theorem in AdS is already sensitive to the spin of the matter fields! This is reminiscent of the subleading soft theorem for the S-matrix in flat space (see \cite{McLoughlin:2022ljp} for a recent review) and hints that the leading soft theorem for correlators in AdS already encodes information about both, the leading and subleading soft theorems of the flat space S-matrix. We also discuss the soft limits of diagrams where the soft photon is emitted from an interior leg in section \ref{app:class2}. In the soft limit of these classes of diagrams, the bulk-bulk propagator from which the soft photon is emitted is dressed with spin factors. Therefore, in the soft limit of the full correlator at $(n+1)$-pts, we also obtain terms other than the original $n$-pt correlator. It would be interesting to explore the exact nature of the constraints imposed and bootstrap correlators using these constraints in the future. 

As a non-trivial application of the recursion relations developed in section \ref{sec:recursion}, we also compute 1-loop polygons in section \ref{sec:fermionloop}. For the triangle diagram with the fermions flowing in the loop (discussed in section \ref{scalarYukawa}) this reduces the complication of adding 100 terms spread across several pages to a couple of lines, demonstrating the power of the recursion relations! In cases when we can perform the integrals, like the bubble diagram, we find that it has a lower transcendentality than its scalar counterpart \cite{Chowdhury:2023khl}. We then demonstrate how the recursion relations also generalize to massive fermions by developing a set of novel differential operators (Weight shifting operators) in section \ref{sec:massive} (see \cite{Benincasa:2019vqr} for a similar discussion on correlators for light scalars in cosmology). For massive external fermions these operators take a particularly simple form. However, for massive fermions present in the internal legs the differential operators are specified for each independent basis element formed by a set of $\gamma$-matrices. This tremendously reduces the complexity of evaluating the correlators and allows a simple way to generalize these to higher points.  We also discuss some applications of these operators to loop level Witten diagrams with massive internal legs in appendix \ref{app:evaluatingloop}. One complication arising for massive fields is that the radial integrals are generically IR divergent. We analyze the kind of integrals one obtains for any general 3-pt function with external fermions and scalars in the Yukawa theory and classify the order of divergences obtained for general masses (see table \ref{tab:divergence}). To arrive at answers that are IR finite one also needs to find counterterms and systematically renormalize the correlators (see \cite{Bzowski:2018fql} for a discussion of IR divergences for correlators of scalar fields) and we believe that the classification of divergences done here will be helpful for such an analysis in the future. Finally, in section \ref{sec:cosmologicalcorr} and appendix \ref{app:ferm-inin}, we demonstrate how these techniques are also useful for computing in-in correlators in dS space (often known as cosmological correlators) at both tree and loop levels. 

% \subsubsection*{Future Directions}
While several future directions are illustrated above, there are also some concrete problems for which the techniques above can be utilized:
%\subsection*{\normalfont\textit{Double Copy}}
one major motivation for studying fermionic correlators was to understand the role of supersymmetry in double-copy in flat space. There has been huge progress in computing supersymmetric gravitational scattering amplitudes via supersymmetric Yang-Mills amplitudes (see \cite{Adamo:2022dcm} for a review), however, not much work has been done in trying to understand the same for supersymmetric correlators in AdS (see \cite{Taylor:2023ajd, Lipstein:2023pih, Armstrong:2023phb, Armstrong:2020woi, Albayrak:2020fyp, Bittleston:2024rqe} for some progress for specific theories) or for equal time correlators in flat space. This is partly because the tools developed for computing correlation functions were mainly restricted to integer spins. Since we now have tools to simplify the computations of fermionic correlators, we hope these results will also aid in the computations of correlation functions in supersymmetric theories. This also requires us to extend our analysis to spin-$\frac32$ fields and one possible approach might be to develop the analog of spin-raising operators as \cite{Baumann:2019oyu} to obtain correlators with a spin-$\frac32$ exchange. 
%\subsection*{\normalfont\textit{Soft Theorems}}

The soft theorems discussed in section \ref{sec:soft} and appendix \ref{app:class2} are useful constraints for bootstrapping correlation functions but we still lack a first principle understanding. In the spirit of the infrared triangle in flat space \cite{Strominger:2017zoo}, it would be interesting to understand the role of symmetries in the soft limits and possibly connect these to asymptotic symmetries in AdS \cite{Poole:2018koa, Compere:2020lrt} and to also understand these from the perspective of recently discovered geometric soft theorems \cite{Cheung:2021yog}. It would also be interesting to see if they can be used to derive constraints on higher-spin theories in AdS \cite{Maldacena:2011jn, Sleight:2017pcz} as they do in flat space \cite{Strominger:2017zoo}. Although we analyze the soft limits of AdS correlators in momentum space, these also have an implication for correlators in position space. In particular, the momentum space soft limits also constrain the behavior of integrated correlators \cite{Binder:2019jwn} in position space as discussed towards the end of section \ref{sec:soft}. It would be interesting to explore this connection in the future. 
%\subsection*{\normalfont\textit{Cosmological Polytope for Spinning Correlators}}

One major avenue for studying scattering amplitudes in flat space has been the unexpected connection with positive geometry (see \cite{Herrmann:2022nkh} for a recent review). There has been a similar attempt to study scalar correlators in AdS via the cosmological polytope developed in \cite{Arkani-Hamed:2017fdk} and extended further in \cite{Arkani-Hamed:2018bjr, Benincasa:2018ssx, Benincasa:2024leu}. However, this has not yet been extended to spinning fields. Due to the similarity in the recursive structure for obtaining gluon correlators in AdS \cite{Albayrak:2019asr}, it would be interesting to explore if there exists a version of a ``scaffolded cosmological polytope'' \cite{Arkani-Hamed:2023jry} which gives the gluonic correlators in AdS. For correlators of massless fermions, one obtains two kinds of ``building blocks'' (see \eqref{antisymmgraph}) where one of them naturally fits into the existing polytopic picture while the other one requires some new ideas and we hope some progress can be made on this soon. 

The paper is organized as follows. We first review the diagrammatic rules for scalars, fermions, and gauge fields in section \ref{sec:setup}. While the details are presented for fields that are conformally coupled (hence massless fermions), we also discuss the non-conformal cases in a later section  \ref{sec:massive} and provide an appendix \ref{app:fermionquant} reviewing the derivation of the propagators in detail. In section \ref{sec:recursion} we develop the recursion relations for massless fermions and use them to discuss the singularities of tree-level correlators in section \ref{sec:sing}. We also discuss the universal terms of the soft limit of QED in section \ref{sec:soft} and the remaining terms are discussed in appendix \ref{app:class2}. We then use the techniques of the previous sections to evaluate one-loop diagrams with massless fermions in section \ref{sec:fermionloop}. In section \ref{sec:massive} we demonstrate how the recursion relations also extend to massive fermions and compute the Weight shifting operators that allow us to get massive fermions by acting with simple differential operators on conformally coupled scalars at tree level. We also provide an appendix \ref{app:evaluatingloop} discussing examples of massive fields in loops and the use of Weight-shifting operators for computing these.  Finally in section \ref{sec:cosmologicalcorr} we explain how these results can be used to compute in-in correlators in dS at both tree and loop level. We also provide an appendix \ref{app:ferm-inin} which contains the details of evaluating the path integral and more examples.

%%%%%%%%%%%%%%%%%%%%%%%%%%%
\section{Setup and Diagrammatic Rules}\label{sec:setup}
%%%%%%%%%%%%%%%%%%%%%%%%%%%
We shall work in Euclidean AdS$_4$ (EAdS) and evaluate Witten diagrams corresponding to boundary CFT correlators. By the standard relation between the dS wave function and correlators in EAdS \cite{Maldacena:2002vr}, these are equivalent to evaluating wave function coefficients in dS and we shall return to this in section \ref{sec:cosmologicalcorr}. 

We work in Poincare coordinates with the metric given as (we set $l_{AdS} = 1$)
\begin{eqn}
ds^2 = \frac{dz^2 + dt^2 + dx^2 + dy^2}{z^2}~.
\end{eqn}
Following the conventions of \cite{Kawano:1999au} the vielbein is $e^\nu_\mu = \frac{1}{z} \delta^\nu_\mu$ with $\mu, \nu \in (z, t, x, y).$ Several computations of Witten diagrams exist for bosonic theories (see \cite{Benincasa:2022gtd} for a review). These include self-interacting scalars, scalar QED, gluons coupled to scalars, and gravitons coupled to scalars \cite{Baumann:2020dch}. In this paper, we develop the tools to extend these analyses to theories with spin$-\frac12$ Dirac fermions. We evaluate AdS correlators in momentum space for four kinds of interacting theories, namely, Yukawa, QED, QCD, and 4-Fermi interaction. We assume them to have the standard Lagrangian 
\begin{eqn}\label{eq:Lag}
\mathcal L = \bar \psi (\slashed{\mathcal D}  - m_f)\psi + \frac14 F_{\mu\nu} F^{\mu\nu}  + \frac{1}{4} G^{a}_{\mu\nu} G^{a \mu\nu} +\frac12 \phi(-\Box + m_s^2)\phi +  \bar\psi \psi \phi + \frac12 (\bar\psi \psi)^2,
\end{eqn}
where $\mathcal D$ contains the spin-gauge-covariant derivative, $F_{\mu\nu} = \p_\mu A_\mu - \p_\nu A_\mu$ denotes the field strength tensor for the photon and $G^{\a}_{\mu\nu} = \p_\mu B^{\a}_\nu - \p_\nu B^{\a}_\mu + g_{YM} f^{\a \b \gamma} [B^\beta_\mu, B^\gamma_\nu]$ denotes the field strength tensor of the gluon\footnote{The gluon color indices are denoted by the beginning Greek letters $(\a, \b, \gamma, \cdots)$,  the 3+1-dimensional spacetime indices are denoted by middle Greek letters ($\mu, \nu, \cdots$) and the 3-dimensional boundary indices are denoted by the Latin letters $a, b, i, j, \cdots$. In cases when the spinor indices are displayed explicitly, they shall be denoted by $s, r_1, r_2$. The coupling constants shall mostly be set to 1 unless indicated otherwise.}. For the correlators we analyze, the gauge field only propagates at tree level and therefore we do not add a ghost term to the Lagrangian \eqref{eq:Lag}. We shall also work in the Axial gauge $A_z = 0$ and $B^\a_z =0$. We begin by analyzing conformally coupled fields and also discuss the non-conformal cases in section \ref{sec:massive} and appendix \ref{app:evaluatingloop}.

%%%%%%%%%%%%%%%%%%%%%%%%%%%
\subsection{Propagators}\label{sec:prop}
The bulk Dirac Fermion is represented by $\psi_{AdS}$. The massless free fermion solves the Dirac equation in AdS (we review the derivation in detail in appendix \ref{app:fermionquant})
\begin{eqn}
\slashed D  \psi_{AdS} = 0~,
\end{eqn}
where the action of $\slashed D$ on $\psi_{AdS}$ is given as 
\begin{eqn}
\slashed D \psi_{AdS} = \Big(z \gamma^\mu \p_\mu - \frac{3}{2}\gamma^z \Big) \psi_{AdS},
\end{eqn}
where $\mu \in (z, t, x, y)$ and $\gamma^\mu$ satisfy the Clifford algebra $\{\gamma^\mu, \gamma^\nu \} = 2 \delta^{\mu\nu}$. The equation of motion is accompanied by the following boundary condition 
\begin{eqn}
\lim_{z \to 0} P_- \psi_{AdS}(z, \vec x) = \lim_{z \to0} z^{3/2} \chi_{AdS}(\vec x)~,
\end{eqn}
where $P_{\pm} =\frac12 (1 \pm \gamma^z)$ and $\chi_{AdS}$ is the boundary limit of the bulk-spinor and can be related to a lower dimensional boundary-spinor via a rectangular matrix (see equation (A.8) of \cite{Loganayagam:2020eue}). The bulk-spinor can be expressed in terms of $\chi_{AdS}$ as
\begin{eqn}\label{bulkbound1}
\psi_{AdS}(z, \vec x) = \int \frac{d^3 k}{(2\pi)^3} z^{3/2} e^{- k z} e^{ i \vec k \cdot \vec x} \Big(1 + i \gammavec\Big) \chi_{AdS}(\vec x)~,
\end{eqn}
with $\chi_{AdS}$ satisfying $\gamma^z \chi_{AdS} = - \chi_{AdS}$, which gives the bulk-boundary propagator. Similarly, for $\bar \psi_{AdS}$ we have
\begin{eqn}\label{barbulkbound1}
\bar\psi_{AdS}(z, \vec x) = \int \frac{d^3 k}{(2\pi)^3} z^{3/2} e^{- k z} e^{- i \vec k \cdot \vec x} \bar \chi_{AdS}(\vec x) \Big(1 + i \gammavec \Big) ~,
\end{eqn}
with $\bar \chi_{AdS}\gamma^z =  \bar \chi_{AdS}$ and $\bar \psi_{AdS}$ and $\psi_{AdS}$ are related by
\begin{eqn}
\bar\psi_{AdS}(z, \vec x) = \psi_{AdS}^\dagg(z, \vec x) \gamma^t~.
\end{eqn}
This choice is not unique and any linear combination of $\gamma^t, \gamma^x, \gamma^y$ would be equally valid as we are working in Euclidean AdS. The bulk-bulk propagator satisfies the differential equation 
\begin{eqn}
\slashed D_{z_1} S_{AdS}(z_1, z_2, \vec k) = (z_1 \gamma^z \p_z + i z_1 \vec \gamma \cdot \vec k - \frac32 \gamma^z) S_{AdS}(z_1, z_2, \vec k) = \delta(z_1- z_2)~,
\end{eqn}
along with the boundary condition
\begin{eqn}\label{bndycond1}
\lim_{z_1 \to 0} P_- S_{AdS}(z_1, z_2, \vec k) \sim z_1^{3/2} (\cdots)~,
\end{eqn}
where $\cdots$ denotes terms regular in $z_2$. This completely fixes the bulk-bulk propagator 
\begin{align}\label{fermionprop0}
&S_{AdS}(z_1, z_2, \vec k) \\
&= \frac{(z_1 z_2)^{3/2}}{2}  \Bigg[ \left( \gamma^z - i \gammavec \right)  \Theta(z_1 - z_2) e^{- k (z_1 - z_2)} - \left( \gamma^z + i \gammavec \right)  \Theta(z_2 - z_1) e^{- k (z_2 - z_1)}\nno \\
&\qquad\qquad\qquad  - \left( 1 + i \gamma^z \gammavec \right) e^{- k (z_1 + z_2)}    \Bigg] \nno.
\end{align}
Our result agrees with the bulk-bulk propagator derived in  \cite{Kawano:1999au}. We refer the reader to appendix \ref{app:fermionquant} for details of these derivations.

The massless Dirac fermion in AdS$_4$ can be mapped to a massless Dirac fermion in 4-dimensional Euclidean flat spacetime with a boundary at $z = 0$ via a Weyl transformation. For any conformally flat metric of the form (for AdS $a(z) = \frac1z)$
\begin{eqn}
ds^2 = a^2(z) \big( dz^2 + dt^2 + dx^2 + dy^2\big)~,
\end{eqn}
the conformally coupled fields are transformed using the following Weyl transformations to map them to Euclidean flat space, 
\begin{eqn}\label{weyl}
\phi &\to  a(z)^{1} \phi, \qquad \psi \to  a(z)^{3/2} \psi, \qquad \bar\psi \to  a(z)^{3/2}\bar \psi, \\
g_{\mu\nu} &\to a(z)^{-2} g_{\mu\nu},\qquad
\sqrt{g} \to a(z)^{-4} \sqrt{g}, \qquad
A_\mu \to a(z)^{1} A_\mu. 
\end{eqn}
By performing this for the action in equation \eqref{eq:Lag}, the effective Yukawa, QED, and the 4-Fermi interactions terms pick up the following $z$-dependent factors
\begin{eqn}
V_{\bar\psi\psi\phi} &= a(z)^{0} \bar\psi\psi\phi, \quad
V_{\bar\psi\psi A} = a(z)^{0} \bar\psi\slashed A\psi, \quad
V_{(\bar\psi\psi)^2} = a(z)^{-1} (\bar\psi\psi)^2~.
\end{eqn}
This shows that the Yukawa interaction and QED are conformal invariant. Therefore the effective propagators in Euclidean flat space are given as\footnote{We shall drop the $AdS$ subscripts to distinguish these from equations \eqref{bulkbound1}, \eqref{barbulkbound1} and \eqref{fermionprop0}.} 
\begin{align}\label{fermionprop1}
\psi(z, \vec x) &= \int \frac{d^3 k}{(2\pi)^3} e^{- k z} e^{ i \vec k \cdot \vec x} \left(1 + i\gammavec \right) \chi , \\
\bar\psi(z, \vec x) &= \int \frac{d^3 k}{(2\pi)^3}  e^{- k z} e^{-i \vec k \cdot \vec x} \bar \chi \left(1 + i \gammavec\right) , \\
S(z_1, z_2, \vec k) &= \frac{1}{2}  \Bigg[ \left( \gamma^z - i \gammavec \right)  \Theta(z_1 - z_2) e^{- k (z_1 - z_2)} - \left( \gamma^z + i\gammavec \right)  \Theta(z_2 - z_1) e^{- k (z_2 - z_1)}\nno \\
&\qquad - \left( 1 + i \gamma^z \gammavec \right) e^{- k (z_1 + z_2)}    \Bigg]~.
\label{flat-fermion-bulkbulk}\end{align}
The first line of the bulk-bulk propagator in \eqref{flat-fermion-bulkbulk} resembles the Feynman propagator in Euclidean flat space and can be obtained by a 1-dimensional Fourier transform of the standard expression in momentum space. The last term is the homogeneous solution of the differential equation which ensures that the propagator satisfies the right boundary condition, $\lim_{z_1 \to 0} P_- S(z_1, z_2, \vec k) = 0$. This is the flat space version of \eqref{bndycond1}.

The propagators for a conformally mapped scalar field  with $\Delta = 2$ are given below  
\begin{eqn}\label{scalarprop}
\phi(z, \vec k) &= e^{-k z}, \\
G_D(z_1, z_2, \vec k) &= \frac1\pi \intinf \frac{d\omega}{\omega^2 + k^2} \sin(\omega z_1) \sin(\omega z_2) \\
&= \frac{1}{2k} \Bigg[ \Theta(z_1 - z_2) e^{- k (z_1 - z_2)}  + \Theta(z_2 - z_1) e^{- k (z_2 - z_1)} - e^{- k (z_1 + z_2)}\Bigg]~.
\end{eqn}
We review the derivation for propagators of scalar fields of general mass in appendix \ref{app:fermionquant}. It is useful to contrast these with the field that satisfies Neumann boundary conditions,
\begin{eqn}\label{scalarshadowprop}
\bar\phi(z, \vec k) &= -e^{-k z}, \\
G_N(z_1, z_2, \vec k) &= \frac1\pi \intinf \frac{d\omega}{\omega^2 + k^2} \cos(\omega z_1) \cos(\omega z_2) \\
&= \frac{1}{2k} \Bigg[ \Theta(z_1 - z_2) e^{- k (z_1 - z_2)}  + \Theta(z_2 - z_1) e^{- k (z_2 - z_1)}+e^{- k (z_1 + z_2)}\Bigg]~.
\end{eqn}
These can be viewed as the propagators for fields with scaling dimension $\Delta = 1$ in AdS$_4$ when conformally mapped to flat space.  The fermionic bulk-bulk propagator \eqref{flat-fermion-bulkbulk} can be expressed as a combination of the two scalar propagators \eqref{scalarprop} and \eqref{scalarshadowprop} via,
 \begin{eqn}\label{fermionDscalar}
S(z_1, z_2, \vec k) = - (\gamma^z \p_{z_1} + i \vec\gamma\cdot \vec k)\big[ P_+ G_D(z_1, z_2, \vec k) +  P_- G_N(z_1, z_2, \vec k) \big]~.
\end{eqn}

For the gauge fields we work in Axial gauge $A_z = 0$, $B_z = 0$ and the propagators are given as \cite{Liu:1998ty, Raju:2011mp} 
\begin{align}\label{gaugeprop}
A_i(z, \vec k) &= \e_i(\vec k) e^{- k z}, \nno \\
G^{(A)}_{ij}(z_1, z_2, \vec k) &= \frac1\pi \intinf \frac{d\omega}{\omega^2 + k^2} \sin(\omega z_1) \sin(\omega z_2) \left( \delta_{ij} + \frac{k_i k_j}{\omega^2} \right) \\
&= \frac{\Pi^T_{ij}}{2k} \Bigg[ \Theta(z_1 - z_2) e^{- k (z_1 - z_2)}  + \Theta(z_2 - z_1) e^{- k (z_2 - z_1)} - e^{- k (z_1 + z_2)}\Bigg] \nno \\
& - \frac{k_i k_j}{2k^3} \Bigg[  k (z_1 - z_2) \Theta(z_1 - z_2) +  k (z_2 - z_1)  \Theta(z_2 - z_1)  - k (z_1 + z_2)  \Bigg]~,\nno
\end{align}
where $\Pi^T_{ij} =\delta_{ij} - \frac{k_i k_j}{k^2} $ and $\e_i(\vec k)$ denotes the polarization vector for a photon in axial gauge.
The propagators for the gluon field $B^\a_\mu$ have the same structure up to color factors.

%%%%%%%%%%%%%%%%%%%%%%%%%%%
%%%%%%%%%%%%%%%%%%%%%%%%%%%
\subsection{Diagrammatic Rules}\label{sec:diagramRule}
We state the diagrammatic rules required for evaluating any (conformally mapped) Witten diagram for the Lagrangian in equation \eqref{eq:Lag}. 
%%%%%%%%%%%%%%%%%%%%%%%%%%%
\subsubsection{Fermions}
The bulk-boundary propagators are represented by (momentum is along the lines of the arrow)
\begin{eqn}\label{fermion-bulkBnd}
\begin{tikzpicture}[baseline]
\draw[very thick] (0, 0) circle (1.5);
\draw[fermionbar] ({1.5*cos(140)},{(1.5*sin(140)})  -- (0,0);
\node at ({1.8*cos(140)},{(1.8*sin(140)}) {$\psi$};
\node at (0, -0.25) {$z$};
\end{tikzpicture} 
= u(\vec k) e^{- k z},  \qquad 
\begin{tikzpicture}[baseline]
\draw[very thick] (0, 0) circle (1.5);
\draw[fermion] ({1.5*cos(140)},{(1.5*sin(140)})  -- (0,0);
\node at ({1.8*cos(140)},{(1.8*sin(140)}) {$\bar\psi$};
%\node at (0, 0) {\textbullet};
\node at (0, -0.25) {$z$};
\end{tikzpicture} 
= \bar u(\vec k) e^{- k z}, 
\end{eqn}
where the spinors $u(\vec k)$, $\bar u(\vec k)$ are given in equation \eqref{fermionprop1} and are explicitly written as
\begin{eqn}
u(\vec k) \equiv  \left(1 + i \gammavec\right) \chi',  \qquad 
\bar u(\vec k) \equiv  \bar \chi' \left(1 + i \gammavec\right)
\end{eqn}
with $\chi'$ and $\bar\chi'$ being any $k$-independent spinors which satisfy $\gamma^z \chi' = - \chi'$ and $\bar\chi' \gamma^z = \bar\chi'$. The bulk-bulk propagator is represented as
\begin{eqn}
\begin{tikzpicture}[baseline]
\draw[very thick] (0, 0) circle (1.5);
\draw[fermion] (-0.75, 0) -- (0.75, 0);
\node at (-0.75, -0.25) {$z_1$};
\node at (0.75, -0.25) {$z_2$};
\node at (0, 0.25) {$k$};
\end{tikzpicture} \equiv S(z_1, z_2, \vec k).
\end{eqn}

%%%%%%%%%%%%%%%%%%%%%%%%%%%
\subsubsection{Scalars}
The propagators for the scalar field satisfying Dirichlet boundary conditions \eqref{scalarprop}  are represented as 
\begin{eqn}\label{phibulk}
\begin{tikzpicture}[baseline]
\draw[very thick] (0, 0) circle (1.5);
\draw ({1.5*cos(140)},{(1.5*sin(140)})  -- (0,0);
\node at ({1.8*cos(140)},{(1.8*sin(140)}) {$\phi$};
\node at (0, -0.25) {$z$};
\end{tikzpicture} = e^{-k z}, 
\qquad 
\begin{tikzpicture}[baseline]
\draw[very thick] (0, 0) circle (1.5);
\draw (-0.75, 0) -- (0.75, 0);
\node at (-0.75, -0.25) {$z_1$};
\node at (0.75, -0.25) {$z_2$};
\node at (0, 0.25) {$k$};
\end{tikzpicture} = G_D(z_1, z_2, \vec k).
\end{eqn} 

Propagators of the scalar field satisfying Neumann boundary conditions \eqref{scalarshadowprop} are represtend by
\begin{eqn}
\begin{tikzpicture}[baseline]
\draw[very thick] (0, 0) circle (1.5);
\draw[dashed] ({1.5*cos(140)},{(1.5*sin(140)})  -- (0,0);
\node at ({1.8*cos(140)},{(1.8*sin(140)}) {$\bar\phi$};
\node at (0, -0.25) {$z$};
\end{tikzpicture} &= -e^{-k z}, 
\qquad 
\begin{tikzpicture}[baseline]
\draw[very thick] (0, 0) circle (1.5);
\draw[dashed] (-0.75, 0) -- (0.75, 0);
\node at (-0.75, -0.25) {$z_1$};
\node at (0.75, -0.25) {$z_2$};
\node at (0, 0.25) {$k$};
\end{tikzpicture} = G_N(z_1, z_2,\vec k).
\end{eqn}

%%%%%%%%%%%%%%%%%%%%%%%%%%%
\subsubsection{Gauge Field}
The propagators for the gauge field \eqref{gaugeprop} are represented as 
\begin{eqn}
\begin{tikzpicture}[baseline]
\draw[very thick] (0, 0) circle (1.5);
\draw[photon] ({1.5*cos(140)},{(1.5*sin(140)})  -- (0,0);
\node at ({1.8*cos(140)},{(1.8*sin(140)}) {$A_i$};
\node at (0, -0.25) {$z$};
\end{tikzpicture} = \e_i(\vec k) e^{-k z}, \qquad 
\begin{tikzpicture}[baseline]
\draw[very thick] (0, 0) circle (1.5);
\draw[photon] (-0.75, 0) -- (0.75, 0);
\node at (-0.75, 0.25) {$z_1$};
\node at (0.75, 0.25) {$z_2$};
\node at (-0.75, -0.25) {$i$};
\node at (0.75, -0.25) {$j$};
\node at (0, 0.25) {$k$};
\end{tikzpicture} = G^{(A)}_{ij}(z_1, z_2, \vec k).
\end{eqn}
 We shall use \begin{tikzpicture}
\draw[photon] (-1, 0) -- (1, 0);
\end{tikzpicture} to denote a photon and \begin{tikzpicture}
\draw[gluon] (-1, 0) -- (1, 0);
\end{tikzpicture} to denote a gluon. The expressions for the propagators remain the same for both up to color factors. 

%%%%%%%%%%%%%%%%%%%%%%%%%%%
\subsubsection{Interaction Terms}\label{sec:feynmanrules}
The interaction vertices that appear in Yukawa theory, QED, and QCD are given as 
\begin{eqn}
\scalebox{0.75}{\begin{tikzpicture}[baseline]
\draw[very thick] (0,0) circle (2);
\draw[fermion] (-1, -1) -- (0, 0);
\draw[fermionbar] (-1, 1) -- (0, 0);
\draw (1, 0) -- (0, 0);
\end{tikzpicture}} = 1, \qquad 
\scalebox{0.75}{\begin{tikzpicture}[baseline]
\draw[very thick] (0,0) circle (2);
\draw[fermion] (-1, 0) -- (0, 0);
\draw[fermion] (0, 0) -- (1, 1);
\draw[photon] (0, 0) -- (1, -1);
\node at (0, -0.25) {$i$};
\end{tikzpicture} }
&= \gamma_i , \qquad 
\scalebox{0.75}{\begin{tikzpicture}[baseline]
\draw[very thick] (0,0) circle (2);
\draw[fermion] (-1, 0) -- (0, 0);
\draw[fermion] (0, 0) -- (1, 1);
\draw[gluon] (0, 0) -- (1, -1);
\node at (0, -0.25) {$i$};
\end{tikzpicture} }
=  \gamma_i ,
\end{eqn}
where $i \in (t, x, y)$ as, in Axial gauge, the $z$-component of the gauge field is set to zero. The self-interaction vertices for the gluon are (all momenta are ingoing) \cite{Albayrak:2019asr}
\begin{eqn}
\scalebox{0.75}{\begin{tikzpicture}[baseline]
\draw[very thick] (0,0) circle (2);
\draw[gluon] (-1, 0) -- (0, 0);
\draw[gluon] (0, 0) -- (1, 1);
\draw[gluon] (0, 0) -- (1, -1);
\node at (-0.25,-0.25) {$l$};
\node at (0.75,-0.25) {$m$};
\node at (0,0.5) {$n$};
\node at (-1.25, 0) {$\vec{k_1}$};
\node at (1.25, -1.25) {$\vec k_2$};
\node at (1.25, 1.25) {$\vec k_3$};
\end{tikzpicture} }
&= \frac{i}{\sqrt{2}} \Big[ \delta_{lm} (\vec k_1 - \vec k_2)_n + \delta_{mn} (\vec k_2 - \vec k_3)_l + \delta_{nl} (\vec k_3 - \vec k_1)_m \Big], \\
\scalebox{0.75}{\begin{tikzpicture}[baseline]
\draw[very thick] (0,0) circle (2);
\draw[gluon] (-1, -1) -- (1, 1);
\draw[gluon] (1, -1) -- (-1, 1);
\node at (-1.2, 1.2) {$i$};
\node at (-1.2, -1.2) {$j$};
\node at (1.2, -1.2) {$k$};
\node at (1.2, 1.2) {$l$};
\end{tikzpicture} }
&= i  \Bigg[ \delta^{ik} \delta^{jl} - \frac{1}{2}\big( \delta^{ij} \delta^{kl} + \delta^{il} \delta^{jk}  \big) \Bigg]~.
\end{eqn}
We ignore the color indices as they will not be relevant for the correlators we compute in this paper.

%%%%%%%%%%%%%%%%%%%%%%%%%%%
\section{Recursion Relations for Massless Fermions}\label{sec:recursion}
%%%%%%%%%%%%%%%%%%%%%%%%%%%
In this section, we discuss a set of recursion relations that simplify the evaluation of Witten diagrams and also reveal their singularity structure. The latter shall be discussed in more detail in the following sections. 

%%%%%%%%%%%%%%%%%%%%%%%%%%%
\subsection{Review of Recursions for Scalars}\label{sec:rec-scalar}
A Witten diagram with a single scalar exchange satisfying Dirichlet boundary condition has the following expression
\begin{eqn}\label{scalarcorr1}
\scalebox{0.75}{\begin{tikzpicture}[baseline]
\draw[very thick] (0,0) circle (2);
\draw[dotted] ({2*cos(150)},{(2*sin(150)}) -- (-1, 0) ;
\draw[dotted] ({2*cos(210)},{(2*sin(210)}) -- (-1, 0) ;
\draw[dotted] ({2*cos(30)},{(2*sin(30)}) -- (1, 0) ;
\draw[dotted] ({2*cos(-30)},{(2*sin(-30)}) -- (1, 0) ;
\draw (-1, 0) -- (1, 0);
\node at (1.4, 0) {$x_2$};
\node at (-1.4, 0) {$x_1$};
\end{tikzpicture}}
=
F_L F_R \intsinf dz_1 dz_2 e^{- x_1 z_1} G_D(z_1, z_2, \vec k) e^{- x_2 z_2} 
\end{eqn}
where $F_L$ and $F_R$ can be either scalars, fermions or gauge fields that would appear in the bulk-boundary propagators and are $z$-independent and are denoted by the dotted lines above. $x_1$ and $x_2$ denote the collective energies entering the graph from the left or right respectively. The recursion relations for these scalars are obtained by evaluating the integral \eqref{scalarcorr1} using integration by parts (IBP) \cite{Arkani-Hamed:2017fdk, Benincasa:2022gtd}.  To see this explicitly we insert the $z$-translation operator inside the integral and obtain 
\begin{eqn}\label{scalarcorr2}
F_L F_R \intsinf \left(\pd{}{z_1} + \pd{}{z_2} \right) dz_1 dz_2 e^{- x_1 z_1} G_D(z_1, z_2, \vec k) e^{- x_2 z_2}  = 0
\end{eqn}
where the equality above follows from $G_D(0, z, \vec k) = G_D(z, 0, \vec k) =G_D(\infty, z, \vec k) = G_D(z, \infty, \vec k)   =0$ as evident from the propagator in \eqref{scalarprop}. The number of terms appearing in the expressions upon using IBP crucially depends on the boundary conditions of the Green function. Since the scalar Green function $G_D(z_1, z_2, \vec k)$, given in \eqref{phibulk}, satisfies Dirichlet boundary conditions, and goes to zero as both $z \to 0$ and $z\to \infty$, it is simplest case. By explicitly evaluating the derivative acting on $G_D(z_1, z_2, \vec k) $ given in \eqref{scalarprop} we obtain
\begin{eqn}\label{scalarcorr3}
\left( \pd{}{z_1} +  \pd{}{z_2} \right) G_D(z_1, z_2, \vec k)  = e^{- k z_1} e^{- k z_2}
\end{eqn}
This equation demonstrates how the insertion of the $z$-translation operator in the integrand enables us to the ``snip'' the bulk-bulk Green function into simpler pieces, which in this case simply reduces to a product of bulk-boundary correlators. Therefore the equation \eqref{scalarcorr2} becomes
\begin{eqn}
(x_1 + x_2)F_L F_R \intsinf dz_1 dz_2 e^{- x_1 z_1} G_D(z_1, z_2, \vec k) e^{- x_2 z_2}  =
F_L F_R  \intsinf dz_1 dz_2 e^{- (x_1 + k) z_1- (x_2 + k)z_2} .
\end{eqn}
The LHS contains the expression for the diagram we wanted to evaluate, and by transferring the factor of $(x_1 + x_2)$ to the RHS we obtain 
\begin{eqn}\label{delta2scalar}
\scalebox{0.75}{\begin{tikzpicture}[baseline]
\draw[very thick] (0,0) circle (2);
\draw[dotted] ({2*cos(150)},{(2*sin(150)}) -- (-1, 0) ;
\draw[dotted] ({2*cos(210)},{(2*sin(210)}) -- (-1, 0) ;
\draw[dotted] ({2*cos(30)},{(2*sin(30)}) -- (1, 0) ;
\draw[dotted] ({2*cos(-30)},{(2*sin(-30)}) -- (1, 0) ;
\draw (-1, 0) -- (1, 0);
\node at (1.4, 0) {$x_2$};
\node at (-1.4, 0) {$x_1$};
\end{tikzpicture}}
&= \frac{F_L F_R}{(x_1 + x_2) (x_1 + k)(x_2 + k)}~.
\end{eqn}
One can recognize the RHS as a product of the total energy\footnote{We shall refer to $|\vec k_i|$ as the energy of the $i^{th}$ field.} pole $\frac{1}{x_1 + x_2}$ times a product of two lower point functions with the external energies given by $x_1 + k$ and $x_2 + k$. This shows how IBP reduces the complexity in evaluating the integrals and also introducing certain recursion relations for Witten diagrams. This method also avoids summing over terms with spurious poles as argued in \cite{Arkani-Hamed:2017fdk}. We sometimes use the shorthand notation to denote the equation above,
\begin{eqn}\label{4ptrec}
\begin{tikzpicture}[baseline]
\draw (-0.5, 0) -- (0.5, 0);
\node at (-0.5, 0) {\textbullet};
\node at (0.5, 0) {\textbullet};

\node at (-0.7, -0.25) {$x_1$};
\node at (0.7, -0.25) {$x_2$};
\node at (-0.8, 0.3) {$F_L$};
\node at (0.8, 0.3) {$F_R$};
\node at (0, -0.25) {$k$};
\end{tikzpicture}
= \frac{1}{x_1 + x_2} 
F_L \begin{tikzpicture}[baseline]
\node at (0, 0) {\textbullet};
\node at (0, -0.25) {$x_1 + k$};
\end{tikzpicture}
\begin{tikzpicture}[baseline]
\node at (0, 0) {\textbullet};
\node at (0, -0.25) {$x_2 + k$};
\end{tikzpicture}
F_R~,
\end{eqn}
where the dot in the RHS arises from an integral over a product of bulk-boundary propagators, evaluating to $\begin{tikzpicture}[baseline] 
\node at (0,0) {\textbullet};
\node at (0,-0.25) {$x$}; \end{tikzpicture} \equiv \frac1x$ ~.

A similar trick can be used to express a diagram with two propagators in terms of this diagram. For example, 
\begin{eqn}
\begin{tikzpicture}[baseline]
\node at (-1, -0.25) {$x_1$};
\node at (0, -0.25) {$x_2$};
\node at (1, -0.25) {$x_3$};
\node at (-1, 0) {\textbullet};
\node at (0, 0) {\textbullet};
\node at (1, 0) {\textbullet};
\node at (-0.5, 0.25) {$y_1$};
\node at (0.5, 0.25) {$y_2$};
\draw (-1, 0) -- (1, 0);
\end{tikzpicture}
= \intsinf dz_1 dz_2 dz_3 e^{- x_1 z_1}e^{- x_2 z_2}e^{- x_3 z_3} G_D(z_1, z_2, \vec y_1) G_D(z_2, z_3, \vec y_2).
\end{eqn}
In this case, we would insert the operator $\pd{}{z_1} + \pd{}{z_2} + \pd{}{z_3}$ inside the integrand and repeat the same procedure as before to obtain the following the following recursion relation 
\begin{eqn}
\begin{tikzpicture}[baseline]
\node at (-1, -0.25) {$x_1$};
\node at (0, -0.25) {$x_2$};
\node at (1, -0.25) {$x_3$};
\node at (-1, 0) {\textbullet};
\node at (0, 0) {\textbullet};
\node at (1, 0) {\textbullet};
\node at (-0.5, 0.25) {$y_1$};
\node at (0.5, 0.25) {$y_2$};
\draw (-1, 0) -- (1, 0);
\end{tikzpicture}
= \frac{1}{x_1 + x_2 + x_3} \Big[
\begin{tikzpicture}[baseline]
\draw (-0.5, 0) -- (0.5, 0);
\node at (-0.5, 0) {\textbullet};
\node at (0.5, 0) {\textbullet};

\node at (-0.7, -0.25) {$x_1$};
\node at (0.5, -0.25) {$x_2 + y_2$};
\node at (0, +0.25) {$y_1$};
\end{tikzpicture}
\begin{tikzpicture}[baseline] 
\node at (0,0) {\textbullet};
\node at (0,-0.25) {$x_3 + y_2$};
 \end{tikzpicture}
 + 
 \begin{tikzpicture}[baseline] 
\node at (0,0) {\textbullet};
\node at (0,-0.25) {$x_1 + y_1$};
 \end{tikzpicture}
 \begin{tikzpicture}[baseline]
\draw (-0.5, 0) -- (0.5, 0);
\node at (-0.5, 0) {\textbullet};
\node at (0.5, 0) {\textbullet};

\node at (-0.5, -0.25) {$x_2 + y_1$};
\node at (0.7, -0.25) {$x_3$};
\node at (0, +0.25) {$y_2$};
\end{tikzpicture}
 \Big],
\end{eqn}
where the result is expressed in terms of the 4-pt graph \eqref{4ptrec} and hence is a recursion relation. This procedure can be repeated for an arbitrary $n$-point function in order to obtain recursion relations for scalars satisfying Dirichlet boundary conditions. However, we can also consider diagrams with propagators that satisfy Neumann boundary conditions \eqref{scalarshadowprop}. Consider the example with a 4-pt function in $\bar\phi^3$ theory,
\begin{eqn}\label{scalarNcorr1}
\scalebox{0.75}{\begin{tikzpicture}[baseline]
\draw[very thick] (0,0) circle (2);
\draw[dashed] ({2*cos(150)},{(2*sin(150)}) -- (-1, 0) ;
\draw[dashed] ({2*cos(210)},{(2*sin(210)}) -- (-1, 0) ;
\draw[dashed] ({2*cos(30)},{(2*sin(30)}) -- (1, 0) ;
\draw[dashed] ({2*cos(-30)},{(2*sin(-30)}) -- (1, 0) ;
\draw[dashed] (-1, 0) -- (1, 0);
\node at (1.4, 0) {$x_2$};
\node at (-1.4, 0) {$x_1$};
\end{tikzpicture}}
\equiv 
\begin{tikzpicture}[baseline]
\draw[dashed] (-0.5, 0) -- (0.5, 0);
\node at (-0.5, 0) {\textbullet};
\node at (0.5, 0) {\textbullet};

\node at (-0.5, -0.25) {$x_1$};
\node at (0.5, -0.25) {$x_2$};
\node at (0, +0.25) {$k$};
\end{tikzpicture}
=  \intsinf dz_1 dz_2 e^{- x_1 z_1} G_N(z_1, z_2, \vec k) e^{- x_2 z_2} ~.
\end{eqn}
We follow the procedure described before and insert the  translation operator $\pd{}{z_1} + \pd{}{z_2}$ in the integrand,
\begin{eqn}\label{scalarNcorr2}
 \intsinf dz_1 dz_2\left( \pd{}{z_1} + \pd{}{z_2} \right) e^{- x_1 z_1} G_N(z_1, z_2, \vec k) e^{- x_2 z_2} ~.
\end{eqn}
Unlike the previous case, the Green function $G_N(z_1, z_2, \vec k)$ does not go to zero as $z_i \to 0$, 
\begin{eqn}\label{boundaryN}
G_N(z, 0, \vec k) = G_N(0, z, \vec k) = \frac{1}{k}e^{- k z}~.
\end{eqn}
This results in a contribution from the boundary terms $z_1, z_2 \to 0$ in the IBP. However these boundary contributions also follow a recursive structure \cite{Chowdhury:2023arc}
\begin{eqn}
&  \intsinf dz_2 G_N(0, z_2, \vec k) e^{- x_2 z_2} + \intsinf dz_1 e^{- x_1 z_1} G_N(z_1, 0, \vec k)
=  \frac{1}{k} \Bigg[ \frac{1}{x_2 + k} + \frac{1}{x_1 + k}\Bigg] .
\end{eqn}
The contribution from the bulk of the integral in \eqref{scalarNcorr2} is exactly the same as \eqref{scalarcorr3}, up to a minus sign. This is obvious due to the structural similarity between the Green functions $G_N(z_1, z_2, \vec k)$ and $G_D(z_1, z_2, \vec k)$. Thus, the two contributions to the integral \eqref{scalarNcorr2} are given as,
\begin{eqn}\label{delta1scalar}
&\intsinf dz_1 dz_2\left( \pd{}{z_1} + \pd{}{z_2} \right) e^{- x_1 z_1} G_N(z_1, z_2, \vec k) e^{- x_2 z_2} \\
&\overset{\p}{=}- \frac{1}{k} \Bigg[ \frac{1}{x_2 + k} + \frac{1}{x_1 + k}\Bigg]  \\
&\overset{\mbox{\tiny{bulk}}}{=} - (x_1 + x_2)
\begin{tikzpicture}[baseline]
\draw[dashed] (-0.5, 0) -- (0.5, 0);
\node at (-0.5, 0) {\textbullet};
\node at (0.5, 0) {\textbullet};

\node at (-0.5, -0.25) {$x_1$};
\node at (0.5, -0.25) {$x_2$};
\node at (0, +0.25) {$k$};
\end{tikzpicture}
- \intsinf dz_1  e^{- (x_1 + k) z_1} \intsinf dz_2 e^{-(x_2 + k) z_2} ~.
\end{eqn}
Combining the two equalities, we obtain the value of the diagram
\begin{eqn}
\begin{tikzpicture}[baseline]
\draw[dashed] (-0.5, 0) -- (0.5, 0);
\node at (-0.5, 0) {\textbullet};
\node at (0.5, 0) {\textbullet};

\node at (-0.5, -0.25) {$x_1$};
\node at (0.5, -0.25) {$x_2$};
\node at (0, +0.25) {$k$};
\end{tikzpicture}
  &= \frac{k + x_1 + x_2}{k (x_1 + x_2) (k + x_1)(k + x_2)}~.
\end{eqn}
We can similarly write down recursion relations for other diagrams at higher point by keeping track of the boundary contributions from the IBP. For diagrams where $G_D$ and $G_N$ emanate from the same leg, that point does not give a boundary contribution because $G_D$ goes to zero. 

%%%%%%%%%%%%%%%%%%%%%%%%%%%
%%%%%%%%%%%%%%%%%%%%%%%%%%%
\subsection{Review of Recursions for Gauge Theories }\label{sec:rec-gauge}
These recursion relations were first discussed in \cite{Albayrak:2019asr}. They follow from the structure of the gauge propagator \eqref{gaugeprop} by rewriting it as
\begin{eqn}
&G_{ij}^{(A)}(z_1, z_2, \vec k) = \intinf \frac{d\omega}{\omega^2 + k^2} \sin(\omega z_1) \sin(\omega z_2) \left( \delta_{ij} + \frac{k_i k_j}{\omega^2} \right) \\
&\equiv \Pi_{ij}^T \intinf \frac{d\omega}{\omega^2 + k^2} \sin(\omega z_1) \sin(\omega z_2)  + \Pi_{ij}^L \intinf \frac{d\omega}{\omega^2 } \sin(\omega z_1) \sin(\omega z_2)  ~,
\end{eqn}
where we have introduced the transverse and the longitudinal propagators $\Pi_{ij}^T = \delta_{ij} - \frac{k_i k_j}{k^2}$ and $\Pi_{ij}^L = \frac{k_i k_j}{k^2}$. The integral in the second term can be understood as the $k\to0$ limit of the first integral which gives the following expression of the propagator in the axial gauge
\begin{eqn}\label{gaugeprop2}
G_{ij}^{(A)}(z_1, z_2, \vec k) = \big( \Pi_{ij}^T + \Pi^L_{ij} \lim_{k \to 0} \big) G_D(z_1, z_2, \vec k)~,
\end{eqn}
where $ G_D(z_1, z_2, \vec k)$ is the propagator for the conformally coupled scalar satisfying Dirichlet boundary conditions as given in equation \eqref{scalarprop}. Since the limit in equation \eqref{gaugeprop2} is w.r.t the momenta $\vec k$ it can be taken in the final step, reducing the computation of Witten diagrams with gauge propagators to only computing diagrams with scalar propagators. It is not surprising that the bulk-bulk propagator of the gauge field can be expressed linearly in terms of the propagator of the conformally coupled scalar as the former is also conformal invariant in 4-dimensions. Due to this property, the same set of recursion relations derived in section \ref{sec:rec-scalar} applies to the gauge fields. In section \ref{qcdexamples} we show how this is used to compute correlators in QCD.

%%%%%%%%%%%%%%%%%%%%%%%%%%%
%%%%%%%%%%%%%%%%%%%%%%%%%%%
\subsection{Recursions for Fermions}\label{sec:rec-fermion}
The bulk-bulk propagator for the massless fermion is given as \eqref{flat-fermion-bulkbulk} 
\begin{eqn*}
S(z_1, z_2, \vec k) &\equiv \frac{1}{2} \Bigg[  \left(\gamma^z - i \gammavec  \right) \Theta(z_1 - z_2) e^{- k (z_1 - z_2)} - \left(\gamma^z  + i \gammavec \right) \Theta(z_2 - z_1) e^{- k (z_2 - z_1)} \\
&\qquad -  \left(1 + i \gamma^z  \gammavec  \right)  e^{- k (z_1 + z_2)} \Bigg]~.
\end{eqn*}

Because of the exponential dampings the propagator naturally vanishes as $z_i \to \infty$. However, it does not completely vanish as $z_i \to 0$. Its value as $z_1, z_2 = 0$ becomes
\begin{eqn}\label{boundaryS}
S(0, z_2, \vec k) &= - P_+ (1 + i \gammavec) e^{- k z_2}, \\
S(z_1, 0, \vec k) &=  -(1 + i \gammavec) P_-  e^{- k z_1} ~,
\end{eqn}
with $P_\pm = \frac12 (1 \pm \gamma^z)$. Note that this representation automatically ensures that $S(z_1, z_2, \vec k)$ satisfies: $P_- S(0, z_2, \vec k) = 0, \ S(z_1, 0, \vec k) P_+ = 0 $~, which are the boundary conditions in flat space.

Consider a generic 4-pt diagram with a fermion exchange\footnote{In our conventions the diagrams are read from left to right, which for the standard S-matrix corresponds to evaluating the hermitian conjugate of $S$.},
\begin{eqn}
\begin{tikzpicture}[baseline]
\draw[dotted] ({1.8*cos(160)}, {1.8*sin(160)}) -- (-1, 0);
\draw[dotted] ({1.8*cos(200)}, {1.8*sin(200)}) -- (-1, 0);

\draw[dotted] ({1.8*cos(20)}, {1.8*sin(20)}) -- (1, 0);
\draw[dotted] ({1.8*cos(-20)}, {1.8*sin(-20)}) -- (1, 0);

\node at (-.75, -.25) {$F_L$};
\node at (.75, -.25) {$F_R$};

\draw[fermion] (-1, 0) -- (1, 0);

\draw[very thick] (0,0) circle (1.8);

\node at (-1.3, 0) {$x_1$};
\node at (1.3, 0) {$x_2$};

\end{tikzpicture}
\equiv 
\begin{tikzpicture}[baseline]
\draw[fermion] (-0.5, 0) -- (0.5, 0);
\node at (-0.5, 0) {\textbullet};
\node at (0.5, 0) {\textbullet};

\node at (-0.7, -0.25) {$x_1$};
\node at (0.7, -0.25) {$x_2$};
\node at (-0.7, 0.3) {$F_L$};
\node at (0.7, 0.3) {$F_R$};
\node at (0, -0.25) {$k$};
\end{tikzpicture}
= F_L  \intsinf dz_1 dz_2e^{- x_1 z_1 - x_2 z_2}  S(z_1, z_2, \vec y) F_R,
\end{eqn}
where $F_L, F_R$ are independent of $z_i$ and denote the matrices that arise from the left and right part of the diagram and depend on which theory one considers.  

Proceeding in the same manner as before we derive a set of recursion relations by using IBP and inserting the $z$-translational operator inside the integral. The action of the $z$-translation operator on the bulk-bulk Green function is given as,
\begin{eqn}
\bigg( \pd{}{z_1} + \pd{}{z_2} \bigg)  S(z_1, z_2, \vec k) &= k \Big( 1 + i \gamma^z  \gammavec \Big) e^{- k (z_1 + z_2)} ~.
\end{eqn}
Similar to the Dirichlet and Neumann case, this gives us the product of bulk-boundary propagators with an additional factor depending on the product of $\gamma-$matrices. Proceeding similarly as the $\Delta = 1$ scalar, the contribution to this diagram from the bulk and the boundary values of the $z$-integrals are,
\begin{eqn}\label{recursion1}
&F_L  \intsinf dz_1 dz_2 \left(\pd{}{z_1} + \pd{}{z_2} \right) e^{- x_1 z_1} e^{- x_2 z_2} S(z_1, z_2, \vec k) F_R \\
&\overset{bulk}{=\joinrel=} - (x_1 + x_2) I  +  \frac{k}{x_1 + k} \frac{1}{x_2 + k}  F_L \Big(1 + i \gamma^z \gammavec \Big) F_R \\
&\overset{\p}{=\joinrel=} - F_L \intsinf dz_2 e^{- x_2 z_2} S(0, z_2, \vec k) F_R -  F_L \intsinf dz_1 e^{- x_1z_1} S(z_1, 0, \vec k) F_R \\
&= F_L  \Bigg[  \frac{1}{x_2 + k}  P_+(1 + i \gammavec)  + \frac{1}{x_1 + k}  (1 + i \gammavec) P_-\Bigg] F_R~.
\end{eqn}

Equating the two contributions from above, i.e, the {\it bulk} and the boundary $\p$, the 4-pt diagram evaluates to the following,
\begin{eqn}\label{4ptrec1}
&(x_1 + x_2) \begin{tikzpicture}[baseline]
\draw[fermion] (-0.5, 0) -- (0.5, 0);
\node at (-0.5, 0) {\textbullet};
\node at (0.5, 0) {\textbullet};

\node at (-0.5, -0.25) {$x_1$};
\node at (0.5, -0.25) {$x_2$};
\node at (-0.8, 0.3) {$F_L$};
\node at (0.8, 0.3) {$F_R$};
\node at (0, -0.25) {$k$};
\end{tikzpicture} \\
&= F_L\Bigg\{ \frac{1}{(x_1 + k)(x_2 + k)} \Pi_1(\vec k) 
- \Big[  \frac{1}{x_2 + k}  P_+ \Pi_2(\vec k)  + \frac{1}{x_1 + k}  \Pi_2(\vec k) P_- \Big]  \Bigg\}F_R ~,
\end{eqn}
where $\Pi_1 = k + i \gamma^z \vec \gamma \cdot \vec k$ and $\Pi_2 = 1 + i \gammavec$\footnote{These satisfy the following useful relations,
\begin{eqn}\label{Pirels}
\Pi_1(\vec k) \gamma^z &= k (\gamma^z - i \gammavec), \qquad \gamma^z \Pi_1(\vec k) = k(\gamma^z + i\gammavec), \\
\Pi_2(\vec k) P_- &= - (\gamma^z - i \gammavec) P_- , \qquad 
P_+ \Pi_2(\vec k) = P_+ (\gamma^z + i \gammavec)~.
\end{eqn}

}. The terms in the square bracket are the contributions arising from the boundary value of the propagator. These are absent when a scalar of dimension $\Delta = 2$ (see equation \eqref{delta2scalar}) is exchanged but are present for scalars with $\Delta = 1$ (see equation \eqref{delta1scalar}).

For a general diagram with a fermion exchange, we can diagrammatically express the recursion relation as follows 
\begin{eqn}\label{fermionrec-diagram}
(x_1 + x_2) \begin{tikzpicture}[baseline]
\draw[fermion] (-0.5, 0) -- (0.5, 0);
\node at (-0.5, 0) {\textbullet};
\node at (0.5, 0) {\textbullet};

\node at (-0.7, -0.25) {$x_1$};
\node at (0.7, -0.25) {$x_2$};
\node at (-0.5, 0.3) {$F_L$};
\node at (0.5, 0.3) {$F_R$};
\node at (0, -0.25) {$k$};
\end{tikzpicture} 
&= 
F_L 
\Bigg\{
 \begin{tikzpicture}[baseline]
\node at (0, 0) {\textbullet};
\node at (0, -0.25) {$x_1 + k$};
\end{tikzpicture}
\Pi_1(\vec k)
\begin{tikzpicture}[baseline]
\node at (0, 0) {\textbullet};
\node at (0, -0.25) {$x_2 + k$};
\end{tikzpicture}\\
&\qquad -\Big[
 \begin{tikzpicture}[baseline]
\node at (0, 0) {\textbullet};
\node at (0, -0.25) {$x_1 + k$};
\end{tikzpicture}
\Pi_2(\vec k) P_-
+ 
 \begin{tikzpicture}[baseline]
\node at (0, 0) {\textbullet};
\node at (0, -0.25) {$x_2 + k$};
\end{tikzpicture}
P_+ \Pi_2(\vec k) 
\Big] \Bigg\}
F_R.
\end{eqn}

Since the diagrammatic expression above is slightly abstract, we summarize the algorithm below and provide examples in the following section
\begin{enumerate}
\item Strip off the bulk-bulk propagators carrying momentum $\vec y$ between two vertices and add an extra energy of $|\vec y|$ to the vertices joining it along with a factor of $\Pi_1(\vec y)$ in between the two vertices. 
\begin{enumerate}
\item Hence if the energy entering that vertex before stripping this propagator was $|\vec x|$, after stripping it becomes $|\vec x| + |\vec y|$. For internal vertices, the ones which are not connected to any bulk-boundary propagator, $|\vec x| = 0$. This is equivalent to the second step of \eqref{recursion1}.
%\item The expression obtained in step (a) is the second step of \eqref{recursion1}
\item Add all such terms to obtain the ``bulk part of the answer''. 
\end{enumerate}

\item The rest of the answer (the ``boundary contribution'') is also obtained recursively and corresponds to the third step of \eqref{recursion1}. In this step, we obtain as many terms as the number of internal vertices. The recursive way to go about that is as follows. 
\begin{enumerate}
\item For a vertex $\mathcal V$, remove the bulk-bulk propagator(s) emanating from it.
\item Add the energy of the amputated bulk-bulk propagator $|\vec y|$ to the adjoining vertices.
\item If the vertex $\mathcal V$ is to the left [right] of the bulk-bulk propagator, attach a factor of $P_- \Pi_2(\vec y) $ $\big[ \Pi_2(\vec y) P_+ \big]$ in place of the propagator. This should be dictated by the sign of the arrow in the bulk-bulk propagator. As shown in \eqref{fermionrec-diagram}, Left/Right is defined w.r.t to the direction of arrow on the propagator.
\item Note that if a bulk-bulk propagator emanating from a vertex $\mathcal V$ is that of a scalar field with $\Delta \geq 2$ (or is of a massless gauge field) then the contribution from that vertex is zero. This is because the bulk-bulk propagators for such fields satisfy Dirichlet boundary condition and falls off sufficiently fast as $z \to 0$. 
\item For a vertex $\mathcal V$ with multiple legs emanating from it, all the momenta of the surrounding bulk-bulk propagators gets added to the vertex $\mathcal V$ after amputation. 
\end{enumerate} 
\item Adding the answers obtained from step 1 and step 2 gives the full answer. 
\end{enumerate}

%%%%%%%%%%%%%%%%%%%%%%%%%%%
\subsection{Examples of Tree Level Witten Diagrams}\label{sec:N-witten}
%%%%%%%%%%%%%%%%%%%%%%%%%%%
In this section, we provide examples of Witten diagrams at tree level and also discuss their singularities in the following section \ref{sec:sing}. We also discuss explicit examples at 1-loop in section \ref{sec:fermionloop}. 

%%%%%%%%%%%%%%%%%%%%%%%%%%%
\subsubsection{Yukawa theory}
\begin{comment}
Following the Feynman rules given in section \ref{sec:feynmanrules} the 3-pt function is given as 
\begin{eqn}
\begin{tikzpicture}[baseline]
\draw[fermion] ({1.5*cos(120)}, {1.5*sin(120)}) -- (0, 0);
\draw[fermionbar] ({1.5*cos(-120)}, {1.5*sin(-120)}) -- (0, 0);
\draw ({1.5*cos(0)}, {1.5*sin(0)}) -- (0, 0);
\draw[very thick] (0,0) circle (1.5);

\node at  ({1.75*cos(120)}, {1.75*sin(120)}) {$1$};
\node at  ({1.75*cos(0)}, {1.75*sin(0)}) {$2$};
\node at  ({1.75*cos(-120)}, {1.75*sin(-120)}) {$3$};
\end{tikzpicture} 
= \frac{1}{k_1 + k_2 + k_3} (1 + i \Gammavec{k_1}) (1 + i \Gammavec{k_3})~,
\end{eqn}
where we have suppressed the dependence on the boundary limit of the spinors $\chi$.
\end{comment}
Consider the 4-pt function $\braket{\bar\psi \psi \bar \psi \psi}$ in Yukawa theory. The s-channel contribution is given as,
\begin{eqn}\label{Yukawa-4pt}
\scalebox{0.75}{\begin{tikzpicture}[baseline]
\draw[very thick] (0,0) circle (2);
\draw[fermion] ({2*cos(150)},{(2*sin(150)}) -- (-1, 0) ;
\draw[fermionbar] ({2*cos(210)},{(2*sin(210)}) -- (-1, 0) ;
\draw[fermionbar] ({2*cos(30)},{(2*sin(30)}) -- (1, 0) ;
\draw[fermion] ({2*cos(-30)},{(2*sin(-30)}) -- (1, 0) ;
\draw (-1, 0) -- (1, 0);
\node at (1.4, 0) {$k_{34}$};
\node at (-1.4, 0) {$k_{12}$};
\node at (0, 0.25) {$k$};

\node at ({2.2*cos(150)},{(2.2*sin(150)}) {$2$};
\node at ({2.2*cos(210)},{(2.2*sin(210)}) {$1$};
\node at ({2.2*cos(30)},{(2.2*sin(30)}) {$3$};
\node at ({2.2*cos(-30)},{(2.2*sin(-30)}) {$4$};

\end{tikzpicture}} 
&=  \bar u_2 u_1 \bar u_4  u_3  \intsinf dz_1 dz_2 e^{- {k_{12} z_1}} e^{- {k_{34} z_2}} G_D(z_1, z_2, \vec k)\\
&=\frac{\bar u_2 u_1 \bar u_4  u_3 }{(k_{12} + k_{34})(k_{12} + k)(k_{34} + k)}    
\end{eqn}
where we use the notation,
\begin{eqn*}
k_{a_1 a_2 \cdots a_n} \equiv k_{a_1} + k_{a_2} + \cdots + k_{a_n}~.
\end{eqn*}
 Since this diagram only contains a scalar exchange we get the usual singularity structure \cite{Arkani-Hamed:2017fdk}. The only poles that appear are when the sum of energies $|\vec k_i|$ entering a vertex go to zero. Also note that since we are in Euclidean space, there exists no singularity in the physical configuration of momenta. We shall return to the interpretation of the residue at the poles in the next section \ref{sec:sing}. For a diagram with a fermion exchange, we can consider the s-channel contribution of the correlator $\braket{\bar\psi \phi \bar \psi \phi}$,
\begin{eqn}\label{yukawa-4pt-fermion-1}
\scalebox{0.75}{\begin{tikzpicture}[baseline]
\draw[very thick] (0,0) circle (2);
\draw ({2*cos(150)},{(2*sin(150)}) -- (-1, 0) ;
\draw[fermionbar] ({2*cos(210)},{(2*sin(210)}) -- (-1, 0) ;
\draw ({2*cos(30)},{(2*sin(30)}) -- (1, 0) ;
\draw[fermion] ({2*cos(-30)},{(2*sin(-30)}) -- (1, 0) ;
\draw[fermionbar] (-1, 0) -- (1, 0);
\node at (1.4, 0) {$k_{34}$};
\node at (-1.4, 0) {$k_{12}$};
\node at (0, 0.25) {$k$};

\node at ({2.2*cos(150)},{(2.2*sin(150)}) {$2$};
\node at ({2.2*cos(210)},{(2.2*sin(210)}) {$1$};
\node at ({2.2*cos(30)},{(2.2*sin(30)}) {$3$};
\node at ({2.2*cos(-30)},{(2.2*sin(-30)}) {$4$};

\end{tikzpicture}} 
&=\frac{\bar u_{4}}{E}  \Bigg[ \frac{\Pi_1(\vec k)}{E_L E_R}- \Big( \frac{P_+ \Pi_2(\vec k) }{E_L}+ \frac{ \Pi_2 (\vec k) P_- }{E_R} \Big)  \Bigg] u_{1},
\end{eqn}
where $\Pi_1 = k - i \gamma^z \vec \gamma \cdot \vec k, \ \Pi_2 = 1 + i \gammavec$ and $E_L = k_{12} + k$, $E_R = k_{34} + k$, $E = k_{12} + k_{34}$. A similar computation in position space was carried out in \cite{Kawano:1999au}. Note that the poles appearing in this are the same as the ones appearing with a scalar exchange \eqref{Yukawa-4pt}.

%%%%%%%%%%%%%%%%%%%%%%%%%%%
\subsubsection*{5-pt}
Via the recursion relations we can evaluate higher point exchange graphs using the result derived before. For example, at 5-pts we have the following diagram, 
\begin{eqn}\label{5ptFermion}
\scalebox{0.75}{\begin{tikzpicture}[baseline]
\draw[very thick] (0,0) circle (2);
\draw ({2*cos(150)},{(2*sin(150)}) -- (-1, 0) ;
\draw[fermion] ({2*cos(210)},{(2*sin(210)}) -- (-1, 0) ;
\draw ({2*cos(30)},{(2*sin(30)}) -- (1, 0) ;
\draw[fermionbar] ({2*cos(-30)},{(2*sin(-30)}) -- (1, 0) ;
\draw[fermion] (-1, 0) -- (1, 0);
%\node at (1.4, 0) {$k_{34}$};
%\node at (-1.4, 0) {$k_{12}$};
\node at (-0.5, -0.25) {$y_1$};
\node at (0.5, -0.25) {$y_2$};

\draw (0, 0) -- (0, 2);

\node at ({2.2*cos(150)},{(2.2*sin(150)}) {$2$};
\node at ({2.2*cos(210)},{(2.2*sin(210)}) {$1$};
\node at ({2.2*cos(90)},{(2.2*sin(90)}) {$3$};
\node at ({2.2*cos(30)},{(2.2*sin(30)}) {$4$};
\node at ({2.2*cos(-30)},{(2.2*sin(-30)}) {$5$};

\end{tikzpicture}} 
=  \intsinf dz_1 dz_2 dz_3 \bar u_{1} e^{-k_{12} z_1} S(z_1, z_2, y_1) e^{- k_3 z_2} S(z_2, z_3, y_2) e^{- k_{45} z_3} u_{5}.
\end{eqn}

This is trivially written in terms of the lower point graphs using the recursion relations given in the previous section \ref{sec:rec-fermion},
\begin{eqn}\label{5ptrec}
k_{12345} \scalebox{0.75}{\begin{tikzpicture}[baseline]
\draw[very thick] (0,0) circle (2);
\draw ({2*cos(150)},{(2*sin(150)}) -- (-1, 0) ;
\draw[fermion] ({2*cos(210)},{(2*sin(210)}) -- (-1, 0) ;
\draw ({2*cos(30)},{(2*sin(30)}) -- (1, 0) ;
\draw[fermionbar] ({2*cos(-30)},{(2*sin(-30)}) -- (1, 0) ;
\draw[fermion] (-1, 0) -- (1, 0);
%\node at (1.4, 0) {$k_{34}$};
%\node at (-1.4, 0) {$k_{12}$};
\node at (-0.5, -0.25) {$y_1$};
\node at (0.5, -0.25) {$y_2$};

\draw (0, 0) -- (0, 2);

\node at ({2.2*cos(150)},{(2.2*sin(150)}) {$2$};
\node at ({2.2*cos(210)},{(2.2*sin(210)}) {$1$};
\node at ({2.2*cos(90)},{(2.2*sin(90)}) {$3$};
\node at ({2.2*cos(30)},{(2.2*sin(30)}) {$4$};
\node at ({2.2*cos(-30)},{(2.2*sin(-30)}) {$5$};

\end{tikzpicture}}
&= \bar u_1 \Bigg\{
\begin{tikzpicture}[baseline]
\draw[fermion] (-1, 0) -- (0, 0);
\node at (-1, 0) {\textbullet};
\node at (-1, -0.25) {$x_1$};
\node at (0, 0) {\textbullet};
\node at (0, -0.25) {$x_2 + y_2$};

\end{tikzpicture}
\Pi_1(\vec y_2)
\begin{tikzpicture}[baseline]
\node at (0, 0) {\textbullet};
\node at (0, -0.25) {$x_3 + y_2$};
\end{tikzpicture}
+ 
\begin{tikzpicture}[baseline]
\node at (0, 0) {\textbullet};    
\node at (0, -0.25) {$x_1 + y_1$};  
\end{tikzpicture}
\Pi_1(\vec y_1)
\begin{tikzpicture}[baseline]
\draw[fermion] (-1, 0) -- (0, 0);
\node at (-1, 0) {\textbullet};
\node at (-1, -0.25) {$x_2 + y_2$};
\node at (0, 0) {\textbullet};
\node at (0, -0.25) {$x_3$};

\end{tikzpicture}
\\
&\qquad -
\begin{tikzpicture}[baseline]
\draw[fermion] (-1, 0) -- (0, 0);
\node at (-1, 0) {\textbullet};
\node at (-1, -0.25) {$x_1$};
\node at (0, 0) {\textbullet};
\node at (0, -0.25) {$x_2 + y_2$};

\end{tikzpicture} \Pi_2(\vec y_2) P_+
- P_- \Pi_2(\vec y_1)
\begin{tikzpicture}[baseline]
\draw[fermion] (0, 0) -- (1, 0);
\node at (0, 0) {\textbullet};
\node at (1, 0) {\textbullet};

\node at (0, -0.25) {$x_2 + y_1$};
\node at (1, -0.25) {$x_3$};
\end{tikzpicture} 
\Bigg\} u_5
\end{eqn}
where $x_1 = k_{12}$, $x_2 = k_3, x_{3} = k_{45}$ and we use the shorthand notation from the previous section.

%%%%%%%%%%%%%%%%%%%%%%%%%%%
%%%%%%%%%%%%%%%%%%%%%%%%%%%
\subsubsection{QED}
We can similarly evaluate correlators in QED and QCD. 
%%%%%%%%%%%%%%%%%%%%%%%%%%%
\subsection*{3-pt Function}
The 3-pt function $\braket{\psi A \bar\psi }$ in Axial gauge is 
\begin{eqn}
\begin{tikzpicture}[baseline]
\draw[fermion] ({1.5*cos(120)}, {1.5*sin(120)}) -- (0, 0);
\draw[fermionbar] ({1.5*cos(-120)}, {1.5*sin(-120)}) -- (0, 0);
\draw[photon] ({1.5*cos(0)}, {1.5*sin(0)}) -- (0, 0);
\draw[very thick] (0,0) circle (1.5);

\node at  ({1.75*cos(120)}, {1.75*sin(120)}) {$1$};
\node at  ({1.75*cos(0)}, {1.75*sin(0)}) {$2$};
\node at  ({1.75*cos(-120)}, {1.75*sin(-120)}) {$3$};
\end{tikzpicture} 
= \frac{\vec \e_i(k_2)}{k_1 + k_2 + k_3} \bar u_1\gamma^i u_3~,
\end{eqn}
 This was also computed in position space in \cite{Mueck:1998iz} \footnote{See equation 69 of \cite{Mueck:1998iz} for the 3-pt function of QED in position space. Note that equation 71 of \cite{Mueck:1998iz} also verifies the Ward identity as a consistency check. It is not easy to verify the same in our case from the final result since we are working in a specific gauge and for the consistency of the Ward identity, we also require the $A_z \bar\psi \psi$ vertex. However, as shown in \cite{Baumann:2020dch}, it is possible to completely fix the 3-pt vertex and obtain the standard result of charge conservation at a vertex by studying the general solution of the Ward identity in momentum space.  It would be interesting to use the Ward identity to fix the general 3-pt function for all half-integer spins coupled to a spin-1 current of any mass. See \cite{Jain:2021wyn, Jain:2021vrv} for similar discussions. 

} and it can be easily verified that our results are in agreement by restricting to axial gauge in the computation of \cite{Mueck:1998iz}.

%%%%%%%%%%%%%%%%%%%%%%%%%%%
\subsection*{4-pt Function}
We consider the s-channel of correlator $\braket{\bar\psi(\vec k_1)A(\vec k_2) \psi(\vec k_3) A(\vec k_4) }$. The recursion relation trivializes this computation and the result follows similarly to the Yukawa case, 
\begin{eqn}\label{qed4pt}
\scalebox{0.75}{\begin{tikzpicture}[baseline]
\draw[very thick] (0,0) circle (2);
\draw[photon] ({2*cos(150)},{(2*sin(150)}) -- (-1, 0) ;
\draw[fermionbar] ({2*cos(210)},{(2*sin(210)}) -- (-1, 0) ;
\draw[photon] ({2*cos(30)},{(2*sin(30)}) -- (1, 0) ;
\draw[fermion] ({2*cos(-30)},{(2*sin(-30)}) -- (1, 0) ;
\draw[fermionbar] (-1, 0) -- (1, 0);
%\node at (1.4, 0) {$k_{34}$};
%\node at (-1.4, 0) {$k_{12}$};
\node at (0, 0.25) {$k$};

\node at ({2.2*cos(150)},{(2.2*sin(150)}) {$2$};
\node at ({2.2*cos(210)},{(2.2*sin(210)}) {$1$};
\node at ({2.2*cos(30)},{(2.2*sin(30)}) {$3$};
\node at ({2.2*cos(-30)},{(2.2*sin(-30)}) {$4$};

\end{tikzpicture}} 
&=\frac{\bar u_{4} \e_{2i} \gamma^i }{k_{12} + k_{34}}   \Bigg[ \frac{\Pi_1(\vec k)}{(k_{12} + k) (k_{34} + k)}- \Big( \frac{P_+ \Pi_2(\vec k) }{k_{12} + k}+ \frac{ \Pi_2(\vec k) P_- }{k_{34} + k} \Big)  \Bigg] \gamma^j\e_{3j} u_{1}.
\end{eqn}

%%%%%%%%%%%%%%%%%%%%%%%%%%%
%%%%%%%%%%%%%%%%%%%%%%%%%%%
\subsubsection{QCD}\label{qcdexamples}
The 4-pt function $\braket{\bar\psi(\vec k_1) B(\vec k_2) B(\vec k_3) \psi(\vec k_4) }$ in QCD is similar to that of QED but receives an additional contribution from the self-interacting gluon vertices as shown below
\begin{eqn}
\scalebox{0.75}{\begin{tikzpicture}[baseline]
\draw[very thick] (0,0) circle (2);
\draw[fermion] ({2*cos(150)},{(2*sin(150)}) -- (-1, 0) ;
\draw[fermionbar] ({2*cos(210)},{(2*sin(210)}) -- (-1, 0) ;
\draw[gluon] ({2*cos(30)},{(2*sin(30)}) -- (1, 0) ;
\draw[gluon] ({2*cos(-30)},{(2*sin(-30)}) -- (1, 0) ;
\draw[gluon] (-1, 0) -- (1, 0);
%\node at (1.4, 0) {$k_{34}$};
%\node at (-1.4, 0) {$k_{12}$};
\node at (0, 0.25) {$k$};

\node at ({2.2*cos(150)},{(2.2*sin(150)}) {$4$};
\node at ({2.2*cos(210)},{(2.2*sin(210)}) {$1$};
\node at ({2.2*cos(30)},{(2.2*sin(30)}) {$3$};
\node at ({2.2*cos(-30)},{(2.2*sin(-30)}) {$2$};

\end{tikzpicture}}  
&= \bar u_{4}  \gamma^m u_{1} V^{ijn}(\vec k_2, \vec k_3, \vec k) \e_{2i} \e_{3j}  \intsinf dz_1 dz_2 e^{- k_{14} z_1}  G^{(A)}_{mn}(z_1, z_2, \vec k) e^{-k_{23} z_2} . 
\end{eqn}
The contraction with the vertex factor is given as 
\begin{eqn}
V^{ijn}(\vec k_2, \vec k_3, \vec k) \e_{2i} \e_{3j}  = \frac{i}{\sqrt{2}} \Big[ \e_2 \cdot \e_3 (\vec k_2 - \vec k_3)^n + 2 (\vec k_3 \cdot \e_2) \e_3^n - 2 (\vec k_2 \cdot \e_3) \e_2^n \Big]~.
\end{eqn}
By performing the radial integral using the recursion relation for the gauge field \eqref{gaugeprop2} we obtain 
\begin{eqn}
\scalebox{0.75}{\begin{tikzpicture}[baseline]
\draw[very thick] (0,0) circle (2);
\draw[fermion] ({2*cos(150)},{(2*sin(150)}) -- (-1, 0) ;
\draw[fermionbar] ({2*cos(210)},{(2*sin(210)}) -- (-1, 0) ;
\draw[gluon] ({2*cos(30)},{(2*sin(30)}) -- (1, 0) ;
\draw[gluon] ({2*cos(-30)},{(2*sin(-30)}) -- (1, 0) ;
\draw[gluon] (-1, 0) -- (1, 0);
%\node at (1.4, 0) {$k_{34}$};
%\node at (-1.4, 0) {$k_{12}$};
\node at (0, 0.25) {$k$};

\node at ({2.2*cos(150)},{(2.2*sin(150)}) {$4$};
\node at ({2.2*cos(210)},{(2.2*sin(210)}) {$1$};
\node at ({2.2*cos(30)},{(2.2*sin(30)}) {$3$};
\node at ({2.2*cos(-30)},{(2.2*sin(-30)}) {$2$};

\end{tikzpicture}}  
= \frac{\bar u_4  \gamma_m u_1 V^{ijk}(\vec k_2, \vec k_3, \vec k) \e_{2i} \e_{3j}}{k_{1234}}   \Big[  \frac{\Pi_T^{mk} }{(k_{14} + k) (k_{23} + k)}  + \frac{\Pi_L ^{mk} }{ k_{14}k_{23} } \Big]~,
\end{eqn}
where the projection tensors are given as $\Pi_T^{ij} = \delta^{ij} - \frac{k^i k^j}{k^2}$ and $\Pi_L^{ij} = \frac{k^i k^j}{k^2}$ and $\vec k = -( \vec k_2 + \vec k_3)$.

%%%%%%%%%%%%%%%%%%%%%%%%%%%
\section{Singularities of Correlators and Soft Limits}\label{sec:sing}
%%%%%%%%%%%%%%%%%%%%%%%%%%%
The S-matrix for tree-level amplitudes admits several key properties, such as factorization on poles, soft limits, etc. These are directly linked to physical properties such as unitarity \& locality, and form a major ingredient for bootstrapping the S-matrix at higher points. However, unlike the S-matrix, correlators in EAdS do not admit any poles for the physical configuration of the momenta. This is made explicit from the computations in the previous section and we see that they admit singularities only when the momenta are analytically continued to complex values. In this section, we explore the physical significance of these poles.

%%%%%%%%%%%%%%%%%%%%%%%%%%%
%%%%%%%%%%%%%%%%%%%%%%%%%%%
\subsection{Flat Space Limit }\label{sec:flatlimit}
One stark difference between the flat space S-matrix and correlators in EAdS is the absence of an energy-conserving delta function in the latter. Correlation functions in AdS do not admit momentum conservation along the radial direction (which is referred to as energy conservation) due to the lack of translational invariance in $z$. This is not surprising, as the standard radial integrals that give rise to energy-conserving delta functions for the S-matrix arise from $\intinf dz e^{- E z} \sim \delta(E)$, whereas for correlation functions these integrals are cut off as $z \to 0$ and hence one obtains poles in energy from $\intsinf dz e^{- E z} \sim \frac1E $ rather than delta functions. In particular, the pole in the sum over all external energies is known as the {\it total energy singularity} as the residue at this pole recovers the S-matrix in flat space \cite{Raju:2012zr}. Although the original proof for the flat space limit \cite{Raju:2012zr} is given for current and stress-tensor correlators in AdS, a similar reasoning also goes through for fermionic correlators. For simplicity, we demonstrate this with an explicit example at 4-pts for massless correlators, but the same set of arguments is applicable for any fermionic graph. Consider the following 4-pt graph for Yukawa theory,
\begin{eqn}
\scalebox{0.75}{\begin{tikzpicture}[baseline]
\draw[very thick] (0,0) circle (2);
\draw ({2*cos(150)},{(2*sin(150)}) -- (-1, 0) ;
\draw[fermion] ({2*cos(210)},{(2*sin(210)}) -- (-1, 0) ;
\draw ({2*cos(30)},{(2*sin(30)}) -- (1, 0) ;
\draw[fermionbar] ({2*cos(-30)},{(2*sin(-30)}) -- (1, 0) ;
\draw[fermion] (-1, 0) -- (1, 0);
%\node at (1.4, 0) {$k_{34}$};
%\node at (-1.4, 0) {$k_{12}$};
\node at (0, 0.25) {$k$};

\node at ({2.2*cos(150)},{(2.2*sin(150)}) {$2$};
\node at ({2.2*cos(210)},{(2.2*sin(210)}) {$1$};
\node at ({2.2*cos(30)},{(2.2*sin(30)}) {$3$};
\node at ({2.2*cos(-30)},{(2.2*sin(-30)}) {$4$};

\end{tikzpicture}} 
&= \frac{\bar u_{1}}{2}  \intsinf dz_1 dz_2 e^{- k_{12} z_1} e^{- k_{34} z_2} \Bigg[ (\gamma^z - i \gammavec) \Theta(z_1 - z_2)e^{- k (z_1 - z_2)} \\
&\quad- (\gamma^z + i \gammavec) \Theta(z_2 - z_1)e^{- k (z_2 - z_1)}   - (1 + i \gamma^z \gammavec ) e^{- k(z_1 + z_2)} \Bigg] u_4,
\end{eqn}
where we have explicitly shown the fermionic bulk-bulk propagator. In order to obtain the S-matrix for the corresponding process in flat space, we simply extend the limits of the $z-$integral from $(-\infty, \infty)$ (which amounts to ignoring the boundary at $z = 0$). Since the first two terms of the bulk-bulk propagator arise from the Feynman propagator in flat space, they end up contributing as the expected answer for the flat space S-matrix
\begin{eqn}
&\intinf dz_1 dz_2 e^{- k_{12} z_1} e^{- k_{34} z_2} \Bigg[ (\gamma^z - i \gammavec) \Theta(z_1 - z_2)e^{- k (z_1 - z_2)}- (\gamma^z + i \gammavec) \Theta(z_2 - z_1)e^{- k (z_2 - z_1)} \Bigg] \\
&  \sim \frac{\gamma^z (k_1 + k_2) + i \gamma \cdot (\vec k_1 + \vec k_2) }{(k_1 + k_2)^2 - (\vec k_1 + \vec k_2)^2 } \delta(k_{12} + k_{34})~.
\end{eqn}

However, from the last term in the bulk-bulk propagator (which was obtained from the homogeneous solution) we get
\begin{eqn}
\intinf dz_1 dz_2 e^{- k_{12} z_1} e^{- k_{34} z_2} e^{-k (z_1 + z_2)} \sim \delta(k_{12} + k) \delta(k_{34}+ k)~,
\end{eqn}
which is zero for generic configurations of momenta. This example illustrates that the homogeneous piece of the bulk-bulk propagators does not contribute to the flat space limit.  By a simple contour integral manipulation, the process corresponding to extending the limits of the $z$-integral to obtain the energy-conserving delta function can be converted to evaluating a residue at the total energy pole. Thus for any graph, the only terms containing the total energy pole for any correlator arise from the part of the bulk-bulk propagator containing the $\Theta$ functions and hence give rise to the expected flat-space limit. Therefore, correlation functions in AdS already contain the flat space S-matrix and this often aids in several consistency checks and bootstrapping spinning correlators \cite{Baumann:2021fxj, Mei:2024abu}. Following the same procedure one recovers the expected flat space 4-pt amplitude in QED \eqref{qed4pt}. As an example, we also demonstrate this for the 5-pt function given in \eqref{5ptFermion}. By computing the residue at the total energy pole, we obtain
\begin{eqn}
&\Res{k_{12345} = 0}\scalebox{0.75}{\begin{tikzpicture}[baseline]
\draw[very thick] (0,0) circle (2);
\draw ({2*cos(150)},{(2*sin(150)}) -- (-1, 0) ;
\draw[fermion] ({2*cos(210)},{(2*sin(210)}) -- (-1, 0) ;
\draw ({2*cos(30)},{(2*sin(30)}) -- (1, 0) ;
\draw[fermionbar] ({2*cos(-30)},{(2*sin(-30)}) -- (1, 0) ;
\draw[fermion] (-1, 0) -- (1, 0);
%\node at (1.4, 0) {$k_{34}$};
%\node at (-1.4, 0) {$k_{12}$};
\node at (-0.5, -0.25) {$y_1$};
\node at (0.5, -0.25) {$y_2$};

\draw (0, 0) -- (0, 2);

\node at ({2.2*cos(150)},{(2.2*sin(150)}) {$2$};
\node at ({2.2*cos(210)},{(2.2*sin(210)}) {$1$};
\node at ({2.2*cos(90)},{(2.2*sin(90)}) {$3$};
\node at ({2.2*cos(30)},{(2.2*sin(30)}) {$4$};
\node at ({2.2*cos(-30)},{(2.2*sin(-30)}) {$5$};

\end{tikzpicture}} = \bar u_{1} \frac{\slashed{y}_1^{flat}}{{y^{flat}_1}^2}   \frac{\slashed{y}_2^{flat}}{{y^{flat}_2}^2} u_{5}~,
\end{eqn}
where the 4-vector $y_{2}^{flat} = (i x_3, \vec y_2)$ and $y^{flat}_1 = (ix_1, - \vec y_1)$ and $\slashed y^{flat}_1 =  \gamma^z x_1 - i \vec \gamma \cdot \vec y_1 $. Thus we recover the expected flat space S-matrix in this limit. In general, this prescription gives the high energy limit of the flat space S-matrix and therefore is not sensitive to the mass of the field being exchanged. While it is possible to recover the S-matrix for massive fields by taking a flat space limit at the level of the $ z$-integrand \cite{vanRees:2022zmr, Marotta:2024sce} (this can be viewed as obtaining the S-matrix from a path integral \cite{Jain:2023fxc}), it would be interesting to explore if this can be used to constrain the nature of $z$-integrals.

%%%%%%%%%%%%%%%%%%%%%%%%%%%
\subsection{Partial Energy Singularities}\label{sec:partialsing}
Along with the total energy singularity, correlators also contain other poles. The structure of these poles are also constrained and always arises when the sum of energies entering any given vertex goes to zero. Therefore these are referred to as {\it partial energy singularities}. The residue of the correlator at these poles is constrained by lower point correlators and amplitudes \cite{Arkani-Hamed:2017fdk, Baumann:2020dch, Jazayeri:2021fvk}. These can be viewed as a corollary for the recursion in section \ref{sec:recursion} relations and are useful for bootstrapping correlation functions. We review the properties of correlators with scalar exchanges and then discuss the fermionic case. 

%%%%%%%%%%%%%%%%%%%%%%%%%%%
\subsection*{Scalar Exchange}
In AdS$_4$, there are two possible values of $\Delta$ for conformally coupled scalars. One whose propagators satisfies the Dirichlet boundary conditions ($\Delta = 2$), and other which satisfies Neumann boundary conditions ($\Delta = 1$). 

%%%%%%%%%%%%%%%%%%%%%%%%%%%
\subsubsection*{$\Delta = 2$ exchange}
This case is already studied in \cite{Arkani-Hamed:2017fdk} and we review the argument with a proof for the simplest case. Consider the two site graph below
\begin{eqn}
\begin{tikzpicture}[baseline]
\node at (-0.5, 0) {\textbullet};
\node at (0.5, 0) {\textbullet};
\draw (-0.5, 0)-- (0.5, 0);
\node at (-0.5, -0.25) {$x_1$};
\node at (0.5, -0.25) {$x_2$};
\node at (0, +0.25) {$y$};
\end{tikzpicture}
=\intsinf dz_1 dz_2 G_D(z_1, z_2, \vec y) e^{-x_1 z_1} e^{- x_2 z_2}~.
\end{eqn}
 Evaluating the residue at a partial energy pole, say $x_2 + y = 0$ tantamounts to extending the limit of the $z_2$ integral from $(-\infty, \infty)$. This roughly corresponds to ``stretching a part of the diagram such that it stops feeling the effect of the boundary''. Performing this gives us 
 \begin{eqn}
\intsinf dz_1 \intinf dz_2 G_D(z_1, z_2, \vec y) e^{-x_1 z_1} e^{- x_2 z_2} =\frac{\delta(x_2 + y)}{(x_1 + y)( x_1 - y)} ~.
 \end{eqn}
Hence we obtain a lower point correlator times a lower point amplitude along with a partial energy-conserving delta function. The energy is conserved in the leg between the 1-site graph (the lower point correlator) and the 1-site scattering amplitude (the lower point amplitude). 
By stripping off the delta function in the form of residue, we can diagrammatically represent this as
\begin{eqn}
\Res{x_2 + y = 0} 
\begin{tikzpicture}[baseline]
\node at (-0.5, 0) {\textbullet};
\node at (0.5, 0) {\textbullet};
\draw (-0.5, 0)-- (0.5, 0);
\end{tikzpicture}
&=\frac{1}{x_1 + y} \times \frac{1}{x_1 - y} 
\equiv \begin{tikzpicture}[baseline]
\node at (0,0) {\textbullet};
\node at (0, -0.25) {$x_1 + y$};
\end{tikzpicture}
\tilde\otimes \quad 
\begin{tikzpicture}[baseline]
\draw (-0.5, 0) -- (0,0);
\draw (0.5, 0.5) -- (0,0);
\draw (0.5, -0.5) -- (0,0);
\end{tikzpicture}~,
\end{eqn}
where the rightmost object is a Feynman diagram in flat space. This relation trivially generalizes to higher points,
\begin{align}
\Res{x_2 + y_1 + y_2 = 0} 
\begin{tikzpicture}[baseline]
\node at (-1, 0) {\textbullet};
\node at (0, 0) {\textbullet};
\node at (1, 0) {\textbullet};
\draw (-1, 0) -- (1, 0);
\end{tikzpicture}
&=
\begin{tikzpicture}[baseline]
\node at (0, 0.25) {\textbullet};
\node at (0, -0.25) {$x_1 + y_1$};
\end{tikzpicture} 
\tilde{\otimes}\quad 
\begin{tikzpicture}[baseline]
\draw (-0.5, 0) -- (0,0);
\draw (0.5, 0.5) -- (0,0);
\draw (0.5, -0.5) -- (0,0);
\end{tikzpicture}\quad 
\tilde{\otimes}
\begin{tikzpicture}[baseline]
\node at (0, 0.25) {\textbullet};
\node at (0, -0.25) {$x_3 + y_2$};
\end{tikzpicture} \\
&= \frac{1}{x_1 + y_1}  \times \frac{1}{(x_1 - y_1)(x_3 - y_2)} \times \frac{1}{x_3 + y_2}, \nno \\
\Res{x_2 + x_3 + y_1 = 0} \begin{tikzpicture}[baseline]
\node at (-1, 0) {\textbullet};
\node at (0, 0) {\textbullet};
\node at (1, 0) {\textbullet};
\draw(-1, 0) -- (0, 0);
\draw (1, 0) -- (0, 0);
\end{tikzpicture}
&=\begin{tikzpicture}[baseline]
 \node at (0,0) {\textbullet};
  \node at (0,-0.25) {$x_1 + y_1$};
 \end{tikzpicture} \quad \tilde \otimes \quad 
  \begin{tikzpicture}[baseline]
 \draw (-1, 0.5) -- (-0.5, 0);
  \draw (-1, -0.5) -- (-0.5, 0);
   \draw (1, 0.5) -- (0.5, 0);
  \draw (1, -0.5) -- (0.5, 0);
  \draw (-0.5, 0) -- (0.5, 0);
 \end{tikzpicture}\\
 &=  \frac{1}{x_1 + y_1} \times
 \frac{1}{x_1 - y_1 } \times
 \begin{tikzpicture}[baseline]
 \draw (-1, 0.5) -- (-0.5, 0);
  \draw (-1, -0.5) -- (-0.5, 0);
   \draw (1, 0.5) -- (0.5, 0);
  \draw (1, -0.5) -- (0.5, 0);
  \draw (-0.5, 0) -- (0.5, 0);
 \end{tikzpicture} ~.\nno
\end{align}

%%%%%%%%%%%%%%%%%%%%%%%%%%%
\subsubsection*{$\Delta = 1$ exchange}
There exists a similar rule for graphs with scalar exchanges of $\Delta = 1$. The only difference is that due to the difference in the boundary value of the bulk-bulk propagator, the partial energy-conserving delta function gets modified by a small factor. We demonstrate this with several examples below. Consider the 2-site graph with the $\Delta = 1$ exchange. By an explicit evaluation, we get
\begin{eqn}
\begin{tikzpicture}[baseline]
\node at (-1, -0.25) {$x_1$};
\node at (1, -0.25) {$x_2$};
\node at (0, -0.25) {$y$};
\draw[dashed] (-1, 0) -- (1, 0);
\node at (-1, 0) {\textbullet};
\node at (1, 0) {\textbullet};
\end{tikzpicture}
&= \intsinf dz_1 dz_2 e^{- x_1 z_1} e^{- x_2 z_2} G_N(z_1, z_2, \vec y)
= \frac{1}{(x_1 + y)(x_2 + y)} \left( \frac{1}{x_1 + x_2} + \frac{1}{y} \right)~.
\end{eqn}
The residue at the total energy pole is
 \begin{eqn}
\Res{E = 0}
\begin{tikzpicture}[baseline]
\node at (-1, -0.25) {$x_1$};
\node at (1, -0.25) {$x_2$};
\node at (0, -0.25) {$y$};
\draw[dashed] (-1, 0) -- (1, 0);
\node at (-1, 0) {\textbullet};
\node at (1, 0) {\textbullet};
\end{tikzpicture} 
= \frac{1}{x_1^2 - y^2} ~,
\end{eqn}
which is the same as the one with $\Delta = 2$ exchange. This is not surprising as the residue at the total energy pole is insensitive to the presence of the homogeneous term in the bulk-bulk propagator (see section \ref{sec:flatlimit}). The residue at the partial energy pole $x_1 + y = 0$ is given as
\begin{eqn}\label{partialN1}
\Res{x_1 + y = 0} \begin{tikzpicture}[baseline]
\node at (-1, -0.25) {$x_1$};
\node at (1, -0.25) {$x_2$};
\node at (0, -0.25) {$y$};
\draw[dashed] (-1, 0) -- (1, 0);
\node at (-1, 0) {\textbullet};
\node at (1, 0) {\textbullet};
\end{tikzpicture}  
&= 1 \times \frac{1}{x_2 - y} \Big(\frac{x_2}{y} \Big) \times \frac{1}{x_2 + y}\\
&= \begin{tikzpicture}[baseline]
\draw (-0.5, 0) -- (0,0);
\draw (0.5, 0.5) -- (0,0);
\draw (0.5, -0.5) -- (0,0);
\end{tikzpicture} \quad \tilde\otimes \quad 
\begin{tikzpicture}[baseline]
\node at (0,0) {\textbullet};
\node at (0, -0.25) {$x_2 + y$};
\end{tikzpicture}
\end{eqn}
where the $\tilde\otimes$ is now defined with an extra factor of $\frac{x_2}{y}$. This rule easily generalizes to higher point graphs including ones with both $\Delta = 1$ and $\Delta = 2$ exchanges,
\begin{eqn}
\Res{x_2 + x_3 + y_1 = 0 } \begin{tikzpicture}[baseline]
\node at (-1, 0) {\textbullet};
\node at (0, 0) {\textbullet};
\node at (1, 0) {\textbullet};
\draw[dashed] (-1, 0) -- (0, 0);
\draw (1, 0) -- (0, 0);
\end{tikzpicture}
 &= \begin{tikzpicture}[baseline]
 \node at (0,0) {\textbullet};
  \node at (0,-0.25) {$x_1 + y_1$};
 \end{tikzpicture} \tilde\otimes
 \begin{tikzpicture}[baseline]
 \draw (-1, 0.5) -- (-0.5, 0);
  \draw (-1, -0.5) -- (-0.5, 0);
   \draw (1, 0.5) -- (0.5, 0);
  \draw (1, -0.5) -- (0.5, 0);
  \draw (-0.5, 0) -- (0.5, 0);
 \end{tikzpicture}\\
 &=\begin{tikzpicture}[baseline]
 \node at (0,0) {\textbullet};
  \node at (0,-0.25) {$x_1 + y_1$};
 \end{tikzpicture} \times
 \frac{1}{x_1 - y_1 } \Big( \frac{x_1}{y_1} \Big) \times
 \begin{tikzpicture}[baseline]
 \draw (-1, 0.5) -- (-0.5, 0);
  \draw (-1, -0.5) -- (-0.5, 0);
   \draw (1, 0.5) -- (0.5, 0);
  \draw (1, -0.5) -- (0.5, 0);
  \draw (-0.5, 0) -- (0.5, 0);
 \end{tikzpicture}~.
\end{eqn}

%%%%%%%%%%%%%%%%%%%%%%%%%%%
\subsection*{Massless Fermion Exchange}
We can similarly write down the residue at the partial energy singularities for the massless Fermion exchange. Consider a 2-site graph in Yukawa theory that was evaluated before in equation \eqref{4ptrec1},
\begin{eqn*}
\begin{tikzpicture}[baseline]
\node at (-0.5, 0) {\textbullet};
\node at (0.5, 0) {\textbullet};
\draw[fermion] (-0.5, 0) -- (0.5, 0);
\node at (-0.5, -0.25) {$x_1$};
\node at (0.5, -0.25) {$x_2$};
\node at (0, 0.25) {$y$};
\end{tikzpicture}
= \frac{\bar u_1}{x_1 + x_2} \Bigg\{ \frac{ \Pi_1(\vec y)}{(x_1 + y)(x_2 + y)}  - \frac{P_+ \Pi_2(\vec y)}{(x_2 + y)}  - \frac{\Pi_2(\vec y)P_-}{(x_1 + y)}  \Bigg\} u_4~. 
\end{eqn*}
Due to the asymmetry of the bulk-boundary correlator, certain signs in the residue are sensitive to the particular pole one is interested in. For example, evaluating the residue at $x_2 + y = 0$ and using equation \eqref{Pirels} we get
\begin{eqn}
\Res{x_2 + y = 0} \begin{tikzpicture}[baseline]
\node at (-0.5, 0) {\textbullet};
\node at (0.5, 0) {\textbullet};
\draw[fermion] (-0.5, 0) -- (0.5, 0);
\node at (-0.5, -0.25) {$x_1$};
\node at (0.5, -0.25) {$x_2$};
\node at (0, 0.25) {$y$};
\end{tikzpicture}
= -   \frac{\bar u_1}{x_1 + y} \times   \left(\frac{x_1}{y} P_+ +  P_- \right) \frac{y \gamma^z + i \vec \gamma \cdot \vec y}{x_1 - y} \times u_4~,
\end{eqn}
which reduces to a product of a lower point wave function dressed with the partial energy-conserving piece and an additional term in the numerator $y \gamma^z + i \vec\gamma \cdot \vec y$ which resembles the inverse propagator in flat space and arises from a completeness relation of polarization of spinning particles. A similar relation holds for the residue at the left energy pole
\begin{eqn}
\Res{x_1 + y = 0}\begin{tikzpicture}[baseline]
\node at (-0.5, 0) {\textbullet};
\node at (0.5, 0) {\textbullet};
\draw[fermion] (-0.5, 0) -- (0.5, 0);
\node at (-0.5, -0.25) {$x_1$};
\node at (0.5, -0.25) {$x_2$};
\node at (0, 0.25) {$y$};
\end{tikzpicture}
= \bar u_1 \times \frac{y \gamma^z - i \vec \gamma \cdot \vec k }{x_2 - y}  \left(\frac{x_2}{y} P_- +  P_+ \right)  \times \frac{u_4}{x_2 + y} 
\end{eqn}
where the difference in signs arises because of the asymmetry of the propagator. This shows how the relations in the previous section \eqref{partialN1} generalize to fermions. For more general theories, have gamma matrices entering the interaction vertices and therefore, the residues at the partial energy poles will also depend on the kind of interaction vertices. This shows how the pattern of residues at the partial singularities, generalizes to Witten diagrams with fermionic exchanges.

%%%%%%%%%%%%%%%%%%%%%%%%%%%
%%%%%%%%%%%%%%%%%%%%%%%%%%%
\subsection{Soft Limits}\label{sec:soft}
In this section, we consider another limit of correlators in fermionic QED. Motivated by the soft theorems in flat space, we study correlators in QED with soft photons and derive universal constraints on correlation functions. Soft theorems for S-matrices in flat spacetime \cite{Low:1958sn, Weinberg:1965nx} are associated with conservation laws arising from asymptotic symmetries at the boundary of flat space \cite{Strominger:2017zoo}. The behavior of correlation functions in the soft limits and its connection to symmetries are not well understood in AdS, although there has been some progress in understanding both, the symmetries \cite{Creminelli:2012ed, Pimentel:2013gza, Hinterbichler:2013dpa, Compere:2020lrt} and the diagrammatic soft limits in YM and GR \cite{Chowdhury:2024wwe}. In this section, we show how the soft limits discovered in \cite{Chowdhury:2024wwe} in YM also hold in QED with photons coupled to matter fields and lead to novel properties of the correlators. This trivially generalizes to QCD.

Correlation functions in QED with fermionic matter receive contributions from several Witten diagrams. However, as argued below, the universal part in the soft limit arises from diagrams where the soft photon is attached to an external vertex. The same class of diagrams gives rise to leading soft theorems for the S-matrix in flat space. We also discuss the case of a photon coupled to an internal vertex in appendix \ref{app:class2} and combined with the analysis here these are all classes of diagrams that contribute to the soft limit of correlators in fermionic QED. Consider the following diagram with a soft photon attached to an external vertex,
\begin{eqn}\label{class1}
\mathcal F_{n+1} \equiv \scalebox{0.8}{\begin{tikzpicture}[baseline]
\draw[very thick] (0, 0) circle (2);
\node at (-1.75, 0) {\vdots};
\draw[fill = lightgray] (-1, 0) circle (0.5);
\draw[fermion] ({2*cos(150)},{(2*sin(150)}) -- (-1.3, 0.4);
\draw[fermionbar] ({2*cos(210)},{(2*sin(210)}) -- (-1.3, -0.4);
\draw[fermion] (-.5, 0) -- (1, 0);

\draw[photon] ({2*cos(30)},{(2*sin(30)}) -- (1, 0);
\draw[fermionbar] ({2*cos(-30)},{(2*sin(-30)}) -- (1, 0);

\node at ({2.25*cos(30)},{(2.25*sin(30)}) {$\vec k_s$};
\node at ({2.25*cos(-30)},{(2.25*sin(-30)}) {$\vec k_h$};

\node at (0, -0.25) {$\vec k$};
\node at (-1, 0) {$F$};
\end{tikzpicture} }
= \e_i(k_s) \intsinf dz_1 dz_2 F(z_1)  S(z_1, z_2, \vec k) e^{- (k_s + k_h) z_2}  \gamma^i u(\vec k_h),
\end{eqn}
where $F(z_1)$ denotes the blob in gray.  In the soft limit $\vec k_s \to 0$, via momentum conservation at the vertex, we have $\vec k = \vec k_h$. Performing the integral over $z_2$ in \eqref{class1} we get,
\begin{align}
&\intsinf dz_2 e^{- k_h z_2} S(z_1, z_2, \vec k) \gamma^i u(\vec k) \\
&=\frac12 e^{- k z_1} \Bigg[ z_1 (\gamma^z - i \gammavec) - \frac{1}{2k} (\gamma^z + i \gammavec)  - \frac{1}{2k} (1 + i \gamma^z \gammavec)  \Bigg] \gamma^i  u(\vec k)~. \nno
\end{align}
Consider the term proportional to $z_1$. After some algebra we find
\begin{eqn}\label{class1-2}
z_1 e^{- k z_1} \Bigg\{ \gamma^i (- \gamma^z + i \gammavec) - 2 i \frac{k_i}{k} \Bigg\}u(\vec k)&= - 2 i \frac{z_1 k_i}{k}  e^{- k z_1} u(\vec k)~,
\end{eqn}
where we have used the equation of motion for $u(\vec k)$. This is similar to how the equation of motion for the external fermion is required to get the soft factor for photons coupled to fermions in the flat space S-matrix \cite{Low:1958sn, Weinberg:1965nx}. By expressing the factor $z_1$ in terms of a derivative w.r.t $\p_k$, its contribution to the soft limit is given as
\begin{eqn}
2 i \frac{\vec k \cdot \vec \e_s}{k} \p_k \intsinf dz_1 F(z_1) e^{- k z_1} u(\vec k) ~.
\end{eqn}
From the other two terms in \eqref{class1-2} we obtain\footnote{Relations \eqref{class1-2} and \eqref{class1-3} are true for any choice of $\chi_{AdS}$ as long as $A(\vec k)$ is \eqref{Adefn1}. This shows that the relations below are insensitive to the boundary conditions on $u(\vec k)$. } 
\begin{eqn}\label{class1-3}
e^{- k z_1} \Bigg\{ - \frac{1}{2k} (\gamma^z + i \gammavec)  \Bigg\} \gamma^i  u(\vec k)
&= - e^{- k z_1}  \frac{2i}{k^2} k_j \Sigma^{ji} u(\vec k), \\
e^{- k z_1} \Bigg\{ - \frac{1}{2k} (1 + i \gamma^z \gammavec)  \Bigg\} \gamma^i  u(\vec k)
&= -e^{- k z_1} \frac{2i}{k^2} k_j \Sigma^{ji}\gamma^z u(\vec k) ~,
\end{eqn}
where $\Sigma^{ji} = \frac14 [\gamma^j, \gamma^i]$. Hence the net contribution to diagrams of the kind \eqref{class1} in the soft limit is given as 
\begin{align}\label{fermionsofttheorem}
&\lim_{k_s \to 0} \mathcal F_{n+1} 
= \frac{i}{k_h} \Big\{ (\vec k_h \cdot \vec \e_s) \delta_{r_1 r_2} \p_{k_h}  + \frac{2}{k_h} (\vec k_h)_j (\vec \e_s)_i (\Sigma^{ji} P_+)_{r_1 r_2} \Big\} \intsinf dz F^{r_1}(z) e^{- k_h z} u^{r_2}( \vec k_h)  \nno \\
&\equiv \hat S_{r_1 r_2}
\intsinf dz F^{r_1}(z) e^{- k_h z} u^{r_2}( \vec k_h) ~.
\end{align}
where the spinor indices ($r_1, r_2$) are shown explicitly. We also summarize this diagrammatically below,
\begin{eqn}
\lim_{k_s \to 0} \scalebox{0.8}{\begin{tikzpicture}[baseline]
\draw[very thick] (0, 0) circle (2);
\node at (-1.75, 0) {\vdots};
\draw[fill = lightgray] (-1, 0) circle (0.5);
\draw[fermion] ({2*cos(150)},{(2*sin(150)}) -- (-1.3, 0.4);
\draw[fermionbar] ({2*cos(210)},{(2*sin(210)}) -- (-1.3, -0.4);
\draw[fermion] (-.5, 0) -- (1, 0);

\draw[photon] ({2*cos(30)},{(2*sin(30)}) -- (1, 0);
\draw[fermionbar] ({2*cos(-30)},{(2*sin(-30)}) -- (1, 0);

\node at ({2.25*cos(30)},{(2.25*sin(30)}) {$\vec k_s$};
\node at ({2.25*cos(-30)},{(2.25*sin(-30)}) {$\vec k_h$};

\node at (0, -0.25) {$\vec k$};
\node at (-1, 0) {$F$};
\end{tikzpicture} }
=  \hat S_{r_1 r_2}
\scalebox{0.8}{\begin{tikzpicture}[baseline]
\draw[very thick] (0, 0) circle (2);
\node at (-1.75, 0) {\vdots};
\draw[fill = lightgray] (-1, 0) circle (0.5);
\draw[fermion] ({2*cos(150)},{(2*sin(150)}) -- (-1.3, 0.4);
\draw[fermionbar] ({2*cos(210)},{(2*sin(210)}) -- (-1.3, -0.4);
\draw[fermion] (-.5, 0) -- (2, 0);

%\draw[photon] ({2*cos(30)},{(2*sin(30)}) -- (1, 0);
%\draw[fermionbar] ({2*cos(-30)},{(2*sin(-30)}) -- (1, 0);

%\node at ({2.25*cos(30)},{(2.25*sin(30)}) {$\vec k_s$};
%\node at ({2.25*cos(-30)},{(2.25*sin(-30)}) {$\vec k_h$};

\node at (2.75, 0) {$u^{r_2}(\vec k_h)$};
\node at (-1, 0) {$F^{r_1}$};
\end{tikzpicture} }
\end{eqn}
Equation \eqref{class1} shows that the soft factor $\hat S_{r_1 r_2}$ is a sum of two terms. The first one, containing a derivative, has the same numerator as the standard leading soft theorem in flat space and hence is the Weinberg-like term. This term is present for all gauge theories and is independent of the matter content\footnote{See equations 4.24 and 4.35 of \cite{Chowdhury:2024wwe} for the soft theorems in pure YM and GR.}. The second term in the soft factor is spin-dependent, and, in this case, it effectively replaces the bulk-boundary propagator $u(\vec k_h)$ with $P_+ u(\vec k_h)$, which can be viewed as a ``shifted bulk-boundary propagator''\footnote{For the case when the soft photon emits from $\bar \psi$ the propagator gets modified to $\bar u(\vec k_h) P_- $ which is consistent with the shift in the propagator $u(\vec k_h)$.}. 

To obtain the soft limit of the full correlator, one has to sum over all external hard legs and also consider diagrams where the soft photon is attached to an internal leg (these contribute to the subleading soft theorem for the scattering amplitude in flat space \cite{Sen:2017xjn}). These diagrams are analyzed in detail in appendix \ref{app:class2}. However, as argued below, the two terms in soft operator  $\hat S_{r_1 r_2}$ \eqref{fermionsofttheorem} can be uniquely identified in the full correlator and are thus called the universal terms. This is consistent with the explicit analysis of the other class of diagrams in appendix \ref{app:class2}.

The term with the derivative in $k_h$ can be distinguished from the rest as it always has a double pole in total energy $E_{tot}$ because of the derivative w.r.t $k_h$. Due to the double pole in $E_{tot}$ this is also the term that dominates in the flat space limit and its coefficient resembles the coefficient of the leading Weinberg soft pole for the flat space S-matrix. Note that this also shows that the soft limit and the flat space limit do not commute since the standard Weinberg soft pole, $\frac{1}{2 k_{\rm soft}^\mu k_{{\rm hard}  \mu}} $, does not appear upon taking the soft limit before the flat space limit. This non-commutativity of the order of limits can be explained by the following intuitive picture. The flat space limit probes the short-wavelength regime of correlators and therefore taking the flat space limit before the soft limit would not be sensitive to the AdS curvature. This is in contrast with taking the soft limit first, as that probes the long-wavelength regime and hence it is sensitive to the presence of the curvature. Therefore, depending on which limit is taken first, the results are expected to probe different regimes in the momentum configuration, and therefore, the limits are not expected to commute.

The other term in equation \eqref{fermionsofttheorem}, which is dependent on the spin of the field that the photon couples to (proportional to $\Sigma^{ij}$), can be separately identified as it has the highest pole in $k_h$. This is evident from the singularity structure of general diagrams discussed in the previous section \ref{sec:partialsing} as all singularities for any given correlator only appear at the sum of energies entering a vertex. The spin-dependent term is absent for scalars coupled to photons\footnote{While this is true in general, there is a subtlety for conformally coupled scalars with $\Delta =1$ coupled to a gauge field as we get a term with coefficients similar to the spin-dependent terms here. Hence the second term generally is sensitive to a combination of the spin and $\Delta$ whose significance would be interesting to explore in the future.} but is present for pure YM and GR where it is convenient to express it in terms of a derivative w.r.t the polarization tensor of the hard fields \cite{Chowdhury:2024wwe}\footnote{Although it is not obvious on how to combine both terms in \eqref{fermionsofttheorem} into a term with a momentum derivative w.r.t $\vec k_i$, it would be interesting if the same pattern observed in \cite{Chowdhury:2024wwe} also generalized to fermions. We thank Arthur Lipstein for this suggestion. }. This is reminiscent of the subleading soft theorem for scattering amplitudes in flat space which depends on the angular momentum operator and hence, \eqref{fermionsofttheorem} suggests that the leading soft limit in AdS already encodes some information about both, the leading and the subleading soft theorems in flat space. This establishes the universality of the two terms appearing in the soft operator \eqref{fermionsofttheorem} in QED and is one of the central results of this section.

 Note that there is an important difference between the soft limit of scattering amplitudes in flat space and of correlators in AdS. In flat space, scattering amplitudes factorize into a universal factor times a lower point amplitude in the soft limit. Unfortunately, this is not true in AdS as for the diagrams where the soft photon is attached to an internal leg one does not observe a factorization of the correlator into a soft factor times the original lower point correlator in the soft limit (see appendix \ref{app:class2} for details). Therefore, the soft limits constrain certain terms in the correlators, as discussed above, but still leave some freedom\footnote{ It is instructive to compare the soft limits for AdS correlators with the soft limits of an  ``off-shell S-matrix'' where we define this by setting at least one of the hard particles off-shell in the standard S-matrix. In the soft limit of this object, we no longer have the Weinberg pole and instead, end up with expressions of order $k_{soft}^0$ like the ones we obtain in \eqref{fermionsofttheorem}. For diagrams where the soft particle emanates from a vertex of 4 or higher points, we would no longer see a factorization into a lower point off-shell S-matrix, which is similar to the behavior of AdS correlators.}. There is also another important difference between the implication of the soft theorem for the S-matrix vs the soft limit of the correlators. The soft factor in the flat space S-matrix of QED is directly linked to the conservation of charges as a consequence of gauge invariance \cite{Weinberg:1965nx}. To carry out a similar analysis for the correlation functions, we need to better investigate their behavior under bulk gauge transformations. For the residual gauge transformations $\vec \e_s \to \vec \e_s + \a \vec k_s$, we get lower order terms in $O(\vec k_s)$, which are subleading in the soft limit. It would be interesting to explore the behavior of such terms and if they can be removed by adding boundary counterterms. We hope to address this point in the future. 
 
 It would also be interesting to understand how our soft limit relates to the $l_{AdS}$ corrections to the soft limit of the S-matrix in flat space studied in \cite{Hijano:2020szl, Bhatkar:2022qhz, Banerjee:2022oll, Cheng:2022xyr, Duary:2022pyv} and of the soft limit of CFT correlators in Mellin space \cite{Fitzpatrick:2012cg}\footnote{In section 5 of \cite{Fitzpatrick:2012cg} the authors have studied the behavior of the soft limit of a  CFT correlator in the embedding space formalism.  In this formalism, the variables used to denote momentum are different from our definition of momentum in momentum space and therefore it is not easy to directly compare the two expressions and it would be interesting to make the connection between the two limits more precise. We thank Alok Laddha for pointing this reference to us and for several useful suggestions and discussions on this subject.}. Another possible direction of investigation are the constraints imposed by the momentum space soft limits on position space correlators. Since the two are related by a Fourier transform, the soft limits in momentum space correlators are related to integrated correlators in position space \cite{Binder:2019jwn}. This is because the Fourier transform of the soft limit of an $(n+1)$-pt correlator in momentum space is obtained by an $(n+1)$-pt correlator in position space with the soft field $A(\vec k_s)$ replaced by $\int d^3 x A(\vec x)$. Hence it would also be interesting to study these limits from that perspective.

%%%%%%%%%%%%%%%%%%%%%%%%%%%
\section{Fermion Loops}\label{sec:fermionloop}
%%%%%%%%%%%%%%%%%%%%%%%%%%%
In this section, we utilize the recursion relations developed in section \ref{sec:recursion} to compute one-loop diagrams with fermions propagating in the loop. We also compare the transcendentality of the loop integrals with similar diagrams for scalars. The analysis of higher point loop diagrams also provides a useful application of the recursion relations as it vastly reduces the complication of evaluating the loop integrand. We shall restrict to one-loop diagrams and leave the study of higher loop diagrams to a future project. 

%%%%%%%%%%%%%%%%%%%%%%%%%%%
%%%%%%%%%%%%%%%%%%%%%%%%%%%
\subsection{1-Loop Bubble}\label{sec:1loopbubble}
Consider a general 1-loop bubble diagram with a massless fermion running in the loop
\begin{eqn}\label{bubble-x1x2}
\scalebox{0.7}{\begin{tikzpicture}[baseline]
\draw[very thick] (0, 0) circle (2);
\draw ({2*cos(150)},{2*sin(150)}) -- (-1, 0);
\draw ({2*cos(210)},{2*sin(210)}) -- (-1, 0);
\node at (-1.5, 0) {$\vdots$};

\draw ({2*cos(30)},{2*sin(30)}) -- (1, 0);
\draw ({2*cos(-30)},{2*sin(-30)}) -- (1, 0);
\node at (1.5, 0) {$\vdots$};

\draw[fermion] (0,0) circle (1);
\node at (-2.25, 0) {$x_1$};
\node at (2.25, 0) {$x_2$};

\node at (0, 1.25) {$\vec l_1$};
\node at (0, -1.25) {$\vec l_2$};
\end{tikzpicture} }
= \int d^3 l \intsinf dz_1 dz_2 e^{- x_1 z_1} e^{- x_2 z_2} \tr \big[ S(z_1, z_2, \vec l_1)  S(z_2, z_1, \vec l_2) \big] ~.
\end{eqn}
where $S(z_1, z_2, \vec k)$ is the bulk-bulk propagator for the massless fermion given in \eqref{flat-fermion-bulkbulk}. While it is possible to use the recursion relations to evaluate this, we shall first evaluate it brute force and show the simplicity of using the recursion relation to evaluate such one-loop polygons in the next section. The non-zero traces in the equation above are evaluated using the standard formulas \cite{Weinberg:1995mt}
\begin{eqn}
&\tr \left\{ \Big( \gamma^z - i \Gammavec{l_1} \Big) \Big( \gamma^z + i \Gammavec{l_2} \Big) \right\} = \tr \left\{ \Big( \gamma^z + i \Gammavec{l_1} \Big) \Big( \gamma^z - i \Gammavec{l_2} \Big) \right\}\\
&= \tr \left\{ \Big( 1 + i \gamma^z  \Gammavec{l_1} \Big) \Big( 1 + i \gamma^z \Gammavec{l_2} \Big) \right\}  = 4\Big( 1+ \frac{\vec l_1 \cdot \vec l_2}{l_1 l_2} \Big)~.
\end{eqn}

This gives the bubble loop integrand 
\begin{eqn}\label{bubbleintegrand1}
\int d^3 l  \Bigg\{ 1 + \frac{\vec l_1 \cdot \vec l_2}{l_1 l_2} \Bigg\} &\Bigg[ - \frac{1}{(l_1 + l_2 + x_2)(x_1 + x_2)} - \frac{1}{(l_1 + l_2 + x_1)(x_1 + x_2)} \\
&\quad + \frac{1}{(l_1 + l_2 + x_1)(l_1 + l_2 + x_2)} \Bigg]~.
\end{eqn}
The flat space limit of the loop integrand is obtained by evaluating the residue $\Res{x_1 + x_2 = 0}$ and this reduces to the loop integrand in flat space but with $l_0$ integrated out. We perform the loop integral using hard-cutoff\footnote{In the context of inflationary cosmology the standard hard cut-off prescription needs to be modified \cite{Senatore:2009cf} as it gives a logarithmic growth. } \cite{Albayrak:2020isk, Chowdhury:2023khl, Chowdhury:2023arc} and obtain 
\begin{eqn}\label{bubble-x1x2-ans}
&\scalebox{0.65}{\begin{tikzpicture}[baseline]
\draw[very thick] (0, 0) circle (2);
\draw ({2*cos(150)},{2*sin(150)}) -- (-1, 0);
\draw ({2*cos(210)},{2*sin(210)}) -- (-1, 0);
\node at (-1.5, 0) {$\vdots$};

\draw ({2*cos(30)},{2*sin(30)}) -- (1, 0);
\draw ({2*cos(-30)},{2*sin(-30)}) -- (1, 0);
\node at (1.5, 0) {$\vdots$};

\draw[fermion] (0,0) circle (1);
\node at (-2.25, 0) {$x_1$};
\node at (2.25, 0) {$x_2$};

\node at (0, 1.25) {$\vec l_1$};
\node at (0, -1.25) {$\vec l_2$};
\end{tikzpicture} }
= - \frac{2 \Lambda^2}{x_1 + x_2} + 2 \Lambda \\
&\quad + \frac{1}{3 (x_1 - x_2)(x_1 + x_2)} \Bigg\{ -\left(x_1-x_2\right) \big\{-k^2 (1+\log (8))+3 k \left(x_1+x_2\right)\\
&\hspace{2.5cm}+\left(x_1^2+x_2 x_1+x_2^2\right) \log (8)\big\}
+3 x_1 \left(x_1-k\right) \left(k+x_1\right) \log \left(\frac{k+x_1}{\Lambda }\right)\\
&\hspace{6cm} +3 x_2 \left(k-x_2\right) \left(k+x_2\right) \log \left(\frac{k+x_2}{\Lambda }\right)\Bigg\} ~.
\end{eqn}
Note that the final answer is a Log instead of a $\mbox{Li}_2$ which appears for the scalar bubble in AdS (see equation (3.25) of \cite{Chowdhury:2023khl}). This hints that fermionic loops have lower transcendentality than their corresponding scalar counterparts. For the bubble, this can be explained by comparing the integrands of the fermionic bubble  \eqref{bubbleintegrand1} and the scalar bubble (see equation (3.26) of \cite{Chowdhury:2023khl}). By doing so, we see that the scalar bubble contains an extra pole $(x_1 + x_2 + 2l)^{-1}$ which disappears for the fermionic bubble because the trace of an odd number of $\gamma$-matrices is zero. Instead, for the fermionic bubble we have an extra pole $l^{-1}$. However, due to the integration measure $d^3 l$, the pole at $l = 0$ disappears and therefore the fermionic bubble has a lesser number of poles than the scalar and hence has lower transcendentality. This feature is generically true for all one-loop even-polygons but not for one-loop odd-polygons as demonstrated in the next section \ref{scalarYukawa}.

For $x_1 = x_2 = k$ (which results in the self-energy correction in the standard Yukawa interaction) we get, 
\begin{eqn}
\scalebox{0.65}{\begin{tikzpicture}[baseline]
\draw[very thick] (0, 0) circle (2);
\draw ({2*cos(180)},{2*sin(180)}) -- (-1, 0);
\draw ({2*cos(0)},{2*sin(0)}) -- (1, 0);

\draw[fermion] (0,0) circle (1);
\node at (-2.25, 0) {$k$};

\node at (0, 1.25) {$\vec l_1$};
\node at (0, -1.25) {$\vec l_2$};
\end{tikzpicture} }
= \frac{\Lambda^2}{k} + 2 \Lambda + \frac{k}{6}(-5 + 6 \log \frac{k}{\Lambda})~.
\end{eqn}

Due to the presence of the cut-off $\Lambda$ the answer does not satisfy the conformal ward identity (CWI). This can be remedied by using a alternate regularization scheme, \textit{analytic regularization} introduced in \cite{Chowdhury:2023arc}. However, for this particular example it is possible to convert this answer to something that does satisfy the CWI at 2-pts by removing the polynomial divergence in $\Lambda$ and setting $\Lambda = k$. 
\begin{eqn}\label{selfenergy-ans}
\scalebox{0.6}{\begin{tikzpicture}[baseline]
\draw[very thick] (0, 0) circle (2);
\draw ({2*cos(180)},{2*sin(180)}) -- (-1, 0);
\draw ({2*cos(0)},{2*sin(0)}) -- (1, 0);

\draw[fermion] (0,0) circle (1);
\node at (-2.25, 0) {$k$};

\node at (0, 1.25) {$\vec l_1$};
\node at (0, -1.25) {$\vec l_2$};
\end{tikzpicture} } &=\frac{k}{\kappa} - \frac{5}{6} k~,
\end{eqn}
where $\kappa$ is the regulator that arises by replacing the divergent part of the answer in hard-cutoff, $\log(\frac{k}{\Lambda})$, by $\frac1\kappa$. Note that in this regularization scheme, the flat space limit is zero, which is consistent with the loop integral in flat space when evaluated in analytic regularization \cite{Smirnov:2012gma}.

%%%%%%%%%%%%%%%%%%%%%%%%%%%
%%%%%%%%%%%%%%%%%%%%%%%%%%%
\subsection{Triangle}\label{scalarYukawa}
In this section, we demonstrate the power of using the recursion relations for a more complicated diagram in the Yukawa theory. We consider the simplest one-loop odd-gon, the triangle diagram,
\begin{eqn}
\scalebox{1.5}{$\Delta$}\equiv \scalebox{0.75}{\begin{tikzpicture}[baseline]
\draw[very thick] (0, 0) circle (2);
\draw ({2*cos(-30)},{(2*sin(-30)}) -- (0.75, -0.75);
\draw ({2*cos(210)},{(2*sin(210)}) -- (-0.75, -0.75);
\draw (0,2) -- (0, 0.75);

\draw[fermion] (0.75, -0.75) -- (0, 0.75);
\draw[fermion] (0, 0.75) -- (-0.75, -0.75);
\draw[fermion] (-0.75, -0.75) -- (0.75, -0.75);

\node at (0.25, 1.25) {$x_1$};
\node at (1.25, -1.1) {$x_2$};
\node at (-1.25, -1.1) {$x_3$};

\node at (0.75, 0.35) {$y_{12}$};
\node at (0, -0.5) {$y_{23}$};
\node at (-0.75, 0.35) {$y_{31}$};

\end{tikzpicture} } 
\end{eqn}
which when written explicitly,
\begin{eqn}\label{triangle1}
\scalebox{1.5}{$\Delta$} =  \intsinf dz_1 dz_2 dz_3 e^{- x_1 z_1} e^{- x_2 z_2} e^{- x_3 z_3} \tr \big[ S(z_1, z_2, \vec y_{12}) S(z_2, z_3, \vec y_{23}) S(z_3, z_1, \vec y_{31})\big]~.
\end{eqn}
We use the recursion relations derived in section \ref{sec:recursion} and write the $z$-integral as a sum of two contributions: ``Bulk'' (denoted by  $\scalebox{1.5}{$\Delta$}_{B}$) and ``Boundary'' (denoted by  $\scalebox{1.5}{$\Delta$}_{\p}$),
\begin{eqn}
\scalebox{1.5}{$\Delta$}  = \scalebox{1.5}{$\Delta$}_{\p} + \scalebox{1.5}{$\Delta$}_{B}~.
\end{eqn}
Due to the cyclic structure of the integral \eqref{triangle1}, there is a permutation symmetry in both the terms. The boundary term is given as 
\begin{eqn}
\scalebox{1.5}{$\Delta$}_{\p} &= \intsinf dz_2 dz_3  e^{- x_2 z_2} e^{- x_3 z_3} \tr \big[ S(0, z_2, \vec y_{12}) S(z_2, z_3, \vec y_{23}) S(z_3,0, \vec y_{31})\big] +\mbox{perms}~,
\end{eqn}
where perms denote the permutation cycle $(x_2, x_3, x_1)(
\vec y_{12}, \vec y_{23}, \vec y_{31})$. Using \eqref{boundaryS} the trace evaluates to zero 
\begin{eqn}
 &\tr \big[ S(0, z_2, \vec y_{12}) S(z_2, z_3, \vec y_{23}) S(z_3,0, \vec y_{31})\big] = \tr  \big[ P_+ \Pi_2(\vec y_{12}) S(z_2, z_3, \vec y_{23}) \Pi_2(\vec y_{31}) P_- \big] \\
  &= \tr  \big[ P_- P_+ \Pi_2(\vec y_{12}) S(z_2, z_3, \vec y_{23}) \Pi_2(\vec y_{31})  \big] = 0~,
\end{eqn}
resulting in 
\begin{eqn}
\scalebox{1.5}{$\Delta$}_{\p}  = 0 ~.
\end{eqn}
This simplification occurs for any one-loop polygon with  Yukawa interaction and more generally for any interaction vertex that is proportional to the identity matrix or $\gamma^z$. The bulk term $\scalebox{1.5}{$\Delta$}_{B} $ does not apriori receive any such simplification and simplifies to
\begin{eqn}\label{triangle2}
\scalebox{1.5}{$\Delta$}_{B} &= \intsinf dz_1 dz_2 dz_3 e^{- (x_1 + y_{12}) z_1} e^{- (x_2 + y_{23}) z_2} e^{- x_3 z_3} \\
&\qquad \qquad\times \tr \Big[ (y_{12} + i \gamma^z \vec \gamma \cdot \vec y_{12}) S(z_2, z_3, \vec y_{23})  S(z_3, z_1, \vec y_{31}) \Big]  + \mbox{perms.}
\end{eqn}
where perms. denotes cycling between $(x_1, x_2, x_3)(\vec y_{12}, \vec y_{23}, \vec y_{31} )$. 
Therefore the $z$-integrals reduce to the tree level 5-pt function \eqref{5ptFermion}. For this particular case, it is simple enough to evaluate the integrals and traces at this stage directly. The two kinds of traces that appear in \eqref{triangle2} are 
\begin{eqn}
\tr\Big[ (\gamma^z + i \vec \gamma \cdot \vec p_1)  (\gamma^z + i  \vec \gamma \cdot \vec p_2) \Big] &= 4 (1 - \vec p_1 \cdot \vec p_2), \\
\tr\Big[ (i \gamma^z \vec \gamma \cdot \vec p_1) (\gamma^z + i \vec \gamma \cdot \vec p_2)  (\gamma^z + i \vec \gamma \cdot \vec p_3) \Big] &= 4 \vec p_1 \cdot (\vec p_3 - \vec p_2). 
\end{eqn}
This leaves us with four $z$-integrals in \eqref{triangle2} which are given as,
\begin{eqn}
\mathcal I_1 &= \frac{1}{\left(x_1'+x_2'+x_3\right) \left(y_{3 1}+x_2'+x_3\right) \left(y_{2 3}+x_2'\right)} , \\
\mathcal I_2 &= \frac{1}{\left(x_1'+x_2'+x_3\right) \left(y_{3 1}+x_1'\right) \left(y_{2 3}+x_2'\right)},\\
\mathcal I_3 &= \frac{y_{3 1}+y_{2 3}+x_1' +x_2'+2 x_3}{\left(x_1' +x_2'+x_3\right) \left(y_{3 1}+x_2'+x_3\right) \left(y_{2 3}+x_1' +x_3\right) \left(y_{3 1}+y_{2 3}+x_3\right)}, \\
\mathcal I_4 &= \frac{1}{\left(x_1' +x_2'+x_3\right) \left(y_{3 1}+x_1' \right) \left(y_{2 3}+x_1' +x_3\right)}
\end{eqn}
where  $x_1' = x_1 +  y_{12}, \ x_2' = x_2 +  y_{12}$. Summing everything up we obtain the full answer,
\begin{eqn}
\scalebox{1.5}{$\Delta$} &= \frac{4}{x_1 + x_2 + x_3} \Bigg[  y_{12} \Big\{  (\mathcal I_1 + \mathcal I_4) (1 - \frac{\vec y_{23} \cdot \vec y_{31}}{y_{23} y_{31}} ) -  (\mathcal I_2 + \mathcal I_3) (1+ \frac{\vec y_{23} \cdot \vec y_{31}}{y_{23} y_{31}} )\Big\} \\
&\qquad+ \vec y_{12} \cdot \Big\{  - (\mathcal I_1 - \mathcal I_4) (\frac{\vec y_{31}}{y_{31}} - \frac{\vec y_{23}}{y_{23}}  ) +  (\mathcal I_2 - \mathcal I_3) (\frac{\vec y_{31}}{y_{31}} + \frac{\vec y_{23}}{y_{23}})\Big\} \Bigg] + \mbox{perms}~.
\end{eqn}
A brute force computation of the same diagram results in summing over 100 individual terms without any apriori pattern! This example shows how the recursion relations combine the result of radial integrals along with the trace of the $\gamma$-matrices into a simple form. By comparing with the integrand of the scalar triangle graph (see equation 3.29 of \cite{Chowdhury:2023khl}) we see that the poles of the fermionic triangle are the same as the scalar triangle and hence for odd-gons, there is no simplification in the transcendentality of the loop integral. Unfortunately, the tools used in the previous section are not useful for performing this loop integral as it is no longer axi-symmetric. While it is possible to express this integral as an infinite series by expanding around the squeezed limit of two momenta as done for the scalar triangle in section 3.2.2 of \cite{Chowdhury:2023khl}, the summation does not yield a closed form expression. However, given that this is a 3-pt function and should ultimately be constrained by conformal invariance, we expect there should be a closed form answer for this integral and we leave a detailed investigation of this integral to a future project.

Note that the recursion relation can also be used for QED in order to write a compact form for the diagram that contributes to the triangle anomaly in flat space, 
\begin{eqn}
& \scalebox{0.75}{\begin{tikzpicture}[baseline]
\draw[very thick] (0, 0) circle (2);
\draw[photon] ({2*cos(-30)},{(2*sin(-30)}) -- (0.75, -0.75);
\draw[photon] ({2*cos(210)},{(2*sin(210)}) -- (-0.75, -0.75);
\draw[photon] (0,2) -- (0, 0.75);

\draw[fermion] (0.75, -0.75) -- (0, 0.75);
\draw[fermion] (0, 0.75) -- (-0.75, -0.75);
\draw[fermion] (-0.75, -0.75) -- (0.75, -0.75);
\end{tikzpicture} } 
= \e_{i_1}(\vec k_1) \e_{i_2}(\vec k_2) \e_{i_3}(\vec k_3) \int d^3l \ \tr\big[ \begin{tikzpicture}[baseline]
\draw[fermion] (-1, 0) -- (0,0);
\draw[fermion] (0, 0) -- (1,0);
\draw[fermion] (1, 0) -- (2,0);

\node at (-1, 0) {\textbullet};
\node at (0, 0) {\textbullet};
\node at (1, 0) {\textbullet};
\node at (2, 0) {\textbullet};
\node at (-1, 0.25) {$\gamma^{i_1}$};
\node at (0, 0.25) {$\gamma^{i_2}$};
\node at (1, 0.25) {$\gamma^{i_3}$};
\node at (-1, -0.25) {1};
\node at (0, -0.25) {2};
\node at (1, -0.25) {3};
\node at (2, -0.25) {1};

\end{tikzpicture} \big]
\end{eqn}
where
\begin{eqn}
\tr\big[
\begin{tikzpicture}[baseline]
\draw[fermion] (-1, 0) -- (0,0);
\draw[fermion] (0, 0) -- (1,0);
\draw[fermion] (1, 0) -- (2,0);

\node at (-1, 0) {\textbullet};
\node at (0, 0) {\textbullet};
\node at (1, 0) {\textbullet};
\node at (2, 0) {\textbullet};
\node at (-1, 0.25) {$\gamma^{i_1}$};
\node at (0, 0.25) {$\gamma^{i_2}$};
\node at (1, 0.25) {$\gamma^{i_3}$};
\node at (-1, -0.25) {1};
\node at (0, -0.25) {2};
\node at (1, -0.25) {3};
\node at (2, -0.25) {1};

\end{tikzpicture}
\big]
&= \tr\big[
\begin{tikzpicture}[baseline]
\node at (0, 0) {\textbullet};
\node at (0, -0.25) {1};
\node at (0, 0.25) {$\gamma^{i_1}$};
\end{tikzpicture}
\Pi_1(\vec y_1)
\begin{tikzpicture}[baseline]
\draw[fermion] (0, 0) -- (1,0);
\draw[fermion] (1, 0) -- (2,0);

\node at (0, 0) {\textbullet};
\node at (1, 0) {\textbullet};
\node at (2, 0) {\textbullet};

\node at (0, 0.25) {$\gamma^{i_2}$};
\node at (1, 0.25) {$\gamma^{i_3}$};

\node at (0, -0.25) {2};
\node at (1, -0.25) {3};
\node at (2, -0.25) {1};
\end{tikzpicture}
 \big] + \mbox{perms} \\
 &- \tr \big[
 \gamma^{i_1} \Pi_2(\vec y_1) P_+
\begin{tikzpicture}[baseline]
\draw[fermion] (0, 0) -- (1, 0);
\node at (0,0) {\textbullet};
\node at (1,0) {\textbullet};
\node at (0,-0.25) {2};
\node at (1,-0.25) {3};
\node at (1,0.25) {$\gamma^{i_3}$};
\end{tikzpicture}
P_- \Pi_2(\vec y_1) 
\big]  + \mbox{perms}~,
\end{eqn}
where $\Pi_1(\vec y)$ and $\Pi_2(\vec y)$ are defined in \eqref{4ptrec1}. While this is sufficient to evaluate the loop integrand, performing this loop integral is an arduous task and it is not yet known how one can perform the integral explicitly. We hope to address this point in the future.

In the following section, we explain how the 1-loop integrands can be obtained via tree-level amplitudes by an alternate method.

%%%%%%%%%%%%%%%%%%%%%%%%%%%
%%%%%%%%%%%%%%%%%%%%%%%%%%%
\subsection{Loops from Trees}\label{sec:treethm}
It is also possible to obtain the one-loop integrand at $n$-pts by using the tree-level graph at $(n+1)$-pts. This is often known as the {\it tree theorem} \cite{Feynman:1963ax, Bierenbaum:2010xg, Caron-Huot:2010fvq} and was explored for the wave function of the universe in dS in \cite{Benincasa:2018ssx} which are equivalent to the correlators in EAdS. These were also recently extended to cosmological correlators in \cite{Benincasa:2024leu}. In this section we demonstrate that a treatment similar to \cite{Benincasa:2018ssx} also applies to diagrams with fermion exchange. 

Consider any general tree-level Witten diagram with massless fermionic exchanges of the form,
\begin{eqn}
\begin{tikzpicture}[baseline]

\draw[fermion] (-1, 0) -- (0, 0);
\draw[fermion] (0, 0) -- (1, 0.25);
\draw[fermion] (1, 0.25) -- (2, -0.25);
\draw[fermion] (2, -0.25) -- (3, -0.1);

\node at (-1, 0) {\textbullet};
\node at (0, 0) {\textbullet};
\node at (1, 0.25) {\textbullet};
\node at (2, -0.25) {\textbullet};
\node at (3, -0.1) {\textbullet};

\node at (-1.5, 0) {$\cdots$};
\node at (3.5, -0.1) {$\cdots$};

\node at (-1, -0.25) {$x_j$};
\node at (0, -0.25) {$x_{j_1}$};
\node at (1, 0.5) {$x_{j_2}$};
\node at (2, -0.5) {$x_{j_3}$};
\node at (3, -0.35) {$x_{j_4}$};

\end{tikzpicture} 
&= \int dz_1 \cdots dz_n e^{- x_1 z_1} V_1 S(z_1, z_2, \vec y_{12}) V_2 e^{- x_2 z_2} V_2' S(z_2, z_3, \vec y_{23}) V_3 \\
&\qquad\times \cdots e^{-x_{n-1} z_{n-1}} V'_{n-1} S(z_{n-1}, z_n, \vec y_{n-1 n}) V_n e^{- x_n z_n}
\end{eqn}
where $V_i$'s are factors appearing through interaction vertices. By deforming two of the $x$'s via $x_{j_1} \to x_{j_1} + \e$ and $x_{j_3} \to x_{j_3} - \e$ and integrating the resulting function in $\e$ over the contour from $(-i\infty, +i\infty)$ we get,
\begin{align}
& \int dz_1 \cdots dz_n e^{- x_1 z_1} V_1 S(z_1, z_2, \vec y_{12}) V_2 e^{- x_2 z_2} V_2' S(z_2, z_3, \vec y_{23}) V_3\\
&\qquad\times  \cdots \int_{-i\infty}^{i\infty} \frac{d\e}{2\pi i}  e^{- \e (z_{j_1} - z_{j_3})} V_{j_1} S(z_{j_1}, z_{j_2}, \vec y_{j_1 j_2}) V_{j_2} e^{- x_{j_2} z_{j_2}} V_{j_2}' S(z_{j_2}, z_{j_3}, \vec y_{j_2 j_3}) V_{j_3}\cdots\nno \\
&\qquad\times \cdots e^{-x_{n-1} z_{n-1}} V'_{n-1} S(z_{n-1}, z_n, \vec y_{n-1 n}) V_n e^{- x_n z_n}~.\nno
\end{align}
This sets $z_{j_1} = z_{j_3}$ and thus, provides the following one-loop integrand from the tree-level diagram,
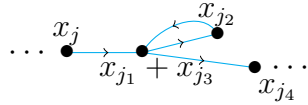
\begin{eqn*}
\begin{tikzpicture}

\draw[fermion] (-1, 0) -- (0, 0);
\draw[fermion] (0, 0) -- (1, 0.25);
\draw[fermion] (1, 0.25) to[out=120,in=60] (0, 0);
\draw[fermion] (0, 0) -- (1.5, -0.2);

\node at (-1, 0) {\textbullet};
\node at (0, 0) {\textbullet};
\node at (1, 0.25) {\textbullet};
\node at (1.5, -0.2) {\textbullet};

\node at (-1.5, -0) {$\cdots$};
\node at (2, -0.2) {$\cdots$};

\node at (-1, 0.25) {$x_j$};
\node at (0.2, -0.25) {$x_{j_1} + x_{j_3}$};
\node at (1, 0.5) {$x_{j_2}$};
\node at (1.8, -0.5) {$x_{j_4}$};

\end{tikzpicture}
\end{eqn*}
By deforming a different set of $x_{j}$'s it is possible to obtain other one-loop diagrams. The nice property of this representation is that it can also be performed after doing the $z$-integrals, which results in summing over residues in the $\e$-plane. Therefore, this version of the tree theorem for Witten diagrams \cite{Benincasa:2018ssx} also applies to half-integer spins. Although this should also work for massive fermions one needs to exercise caution when there are IR divergences as the integrals in $z$ and $\e$ do not always commute. In appendix \ref{app:evaluatingloop} we provide examples of using the tree theorem to evaluate the loop integrand with massive internal legs for cases without IR divergences.

%%%%%%%%%%%%%%%%%%%%%%%%%%%
\section{Massive Fermions and Weight Shifting Operators}\label{sec:massive}
%%%%%%%%%%%%%%%%%%%%%%%%%%%
In this section, we explain how the recursion relations developed in section \ref{sec:recursion} also generalize to massive fermions in AdS. We first consider the case of massive fermions in the external legs. 

%%%%%%%%%%%%%%%%%%%%%%%%%%%
%%%%%%%%%%%%%%%%%%%%%%%%%%%
\subsection{External Legs}
The massive fermion bulk-boundary propagator (for an outgoing leg $\psi$ of mass $m$) is \cite{Mueck:1998iz, Kawano:1999au}
\begin{eqn}\label{bulkbndy-F1}
\mathcal B^{m}(z, \vec k) = z^2 \Big( K_{m+ \frac12}(k z) + i \gammavec K_{m- \frac12}(k z) \Big)~,
\end{eqn}
where we have suppressed the dependence on the boundary spinor $\chi$ for brevity. For $m = 0$ this reduces to \eqref{bulkbound1}. For the case $m = 1$ this becomes 
\begin{eqn}
\mathcal B^{m = 1}(z, \vec k) = z^2 \Big( K_{\frac32}(k z) + i \gammavec K_{\frac12}(k z) \Big)~.
\end{eqn}
 One can now use the following relation for $K_{3/2}(x)$ 
\begin{eqn}
K_{3/2}(kz) &= - \frac{\sqrt{k}}{z} \pd{}{k} \left(\frac{1}{\sqrt{k}} K_{1/2}(kz) \right) = -\sqrt{\frac{\pi}{2}} \frac{\sqrt{k}}{z^{3/2}} \pd{}{k} \left( \frac{e^{- k z}}{k} \right)~.
\end{eqn}

Using this we write $\mathcal B^{m = 1}(z, \vec k)$ as 
\begin{eqn}
\mathcal B^{m = 1}(z, \vec k)  &= z^2 \Bigg( i \gammavec  \sqrt{\frac{\pi}{2}}\frac{e^{- k z}}{\sqrt{kz}} - \sqrt{\frac{\pi}{2}}\frac{\sqrt{k}}{z^{3/2}} \pd{}{k} \left( \frac{e^{- k z}}{k} \right)  \Bigg) \\
&= -\sqrt{\frac{\pi}{2}} \sqrt{\frac{z}{k}} \Bigg\{ i \gammavec + (i \vec \gamma \cdot \vec k + k) \pd{}{k}\Bigg\} \frac{e^{- kz}}{k}~.
\end{eqn}
Hence for any diagram with a fermion of $m = 1$ as an external leg, we can define the following differential operator $\hat D_k$ to map it to a conformally coupled case (it will be convenient to express in terms of a left-handed derivative)
\begin{eqn}\label{diff-ferm-1}
\mathcal B^{m = 1}(z, \vec k)  = \sqrt{z} \hat D_k \frac{e^{-kz}}{k} , \qquad 
\hat D_k = -\sqrt{\frac{\pi}{2}} \frac{1}{\sqrt{k}}  \Bigg\{ i \gammavec + (i \vec \gamma \cdot \vec k + k) \overset{\leftarrow}{\pd{}{k}}\Bigg\}~.
\end{eqn}
The corresponding equation for massive ingoing $\bar\psi$ is
\begin{eqn}
\bar{\mathcal B}^{m = 1}(z, \vec k)  = \sqrt{z}  \frac{e^{-kz}}{k} \hat{\bar D}_k , \qquad 
\hat{\bar D}_k = -\sqrt{\frac{\pi}{2}} \frac{1}{\sqrt{k}}  \Bigg\{ i \gammavec + (i \vec \gamma \cdot \vec k + k) \pd{}{k} \Bigg\}~.
\end{eqn}
These are analogous to {\it Weight shifting operators} for half-integer spin in momentum space \cite{Costa:2018mcg, Sleight:2017fpc, Baumann:2019oyu, Chen:2023xlt, Bzowski:2022rlz}. While these differential operators are trivially used for Witten diagrams at 4-pts, one needs to exercise caution while using the recursion relations along with the differential operators at higher points. This is because additional tensors are obtained from the differential operators above for fermions emanating from internal vertices at higher point graphs. 

To demonstrate its use we consider a 3-pt function in  Yukawa theory of a conformally coupled scalar and fermions of $m = 1$, then we get,
\begin{eqn}
\begin{tikzpicture}[baseline]
\draw (0,0) circle (1);
\draw[fermion, red] ({1*cos(210)}, {1*sin(210)}) -- (0, 0);
\draw[fermionbar, red] ({1*cos(150)}, {1*sin(150)}) -- (0, 0);
\draw ({1*cos(0)}, {1*sin(0)}) -- (0, 0);

\node at ({1.25*cos(210)}, {1.25*sin(210)}) {\scriptsize $1$};
\node at ({1.25*cos(150)}, {1.25*sin(150)}) {\scriptsize $2$};
\node at ({1.25*cos(0)}, {1.25*sin(0)}) {\scriptsize $3$};
\end{tikzpicture} &=  \intsinf \frac{dz}{z^4} \Big( \sqrt{z} \hat{\bar D}_{k_1} \frac{e^{- k_2 z}}{k_2} \Big) \Big( z e^{- k_3 z} \Big) \Big( \sqrt{z} \hat{\bar D}_{k_2} \frac{e^{- k_1 z}}{k_1} \Big)  \\
&= \hat D_{k_1}   \frac{1}{k_1 k_2}\intsinf \frac{dz}{z^2} e^{-k_{123}z} \hat{\bar D}_{k_2} \\
&= \lim_{z_* \to 0} \hat D_{k_1} \Bigg[ \frac{1}{k_1 k_2} \left\{ \frac{1}{z_*} +  k_{123} \log (k_{123} z_* ) + (\gamma_E - 1) k_{123} \right\}\Bigg] \hat{\bar D}_{k_2}~,
\end{eqn}
where we have denoted the IR cutoff of the $z-$integral by $z_*$ and $\gamma_E$ is the Euler–Mascheroni constant. The IR divergences are discussed in more detail in the following section \ref{sec:IRdivergences}. Note that in order to use the differential operators, momentum conservation at the vertex must be imposed after taking the derivative. Diagrams with a scalar exchange and external fermions can be evaluated in a similar manner. For example,
\begin{align}\label{schwinger1}
&\begin{tikzpicture}[baseline]
\draw (0,0) circle (1);
\draw[fermionbar, red] ({1*cos(210)}, {1*sin(210)}) -- (-0.5, 0);
\draw[fermion, red] ({1*cos(150)}, {1*sin(150)}) -- (-0.5, 0);
\draw[fermionbar, red] ({1*cos(30)}, {1*sin(30)}) -- (0.5, 0);
\draw[fermion, red] ({1*cos(30)}, {1*sin(-30)}) -- (0.5, 0);
\draw (-0.5, 0) -- (0.5, 0);
\node at ({1.25*cos(30)}, {1.25*sin(-30)}) {\footnotesize $4$};
\node at ({1.25*cos(30)}, {1.25*sin(30)}) {\footnotesize $3$};
\node at ({1.25*cos(150)}, {1.25*sin(150)}) {\footnotesize $2$};
\node at ({1.25*cos(-150)}, {1.25*sin(-150)}) {\footnotesize $1$};
\end{tikzpicture}\nno \\
 &= \hat{D}_{k_2} \hat{ D}_{k_4}  \frac{1}{k_1 k_2 k_3 k_4} \intsinf \frac{dz_1}{z_1^2} \frac{dz_2}{z_2^2} e^{-k_{12} z_1}G_D (z_1, z_2, \vec k) e^{-k_{34} z_2}  \hat{\bar D}_{k_1}  \hat{\bar D}_{k_3} \\
&=\hat D_{k_2} \hat D_{k_4} \intsinf \frac{s_1 s_2 ds_1 ds_2   }{k_1 k_2 k_3 k_4(k_{12} + k_{34} + s_1 + s_2)(k_{12} + s_1 + k) (k_{34} + s_2 + k)} \hat{\bar D}_{k_1} \hat{\bar D}_{k_3} \nno~,
\end{align}
where we have introduced Schwinger parameters $s_i$ to regulate the IR divergences and express the $z$-integrals in a form that was already evaluated in section \ref{sec:recursion}. The integral can be performed using the hard-cutoff prescription by cutting off the $s-$integral at $s_0 \to \infty$, with $s_0 \sim z_*^{-1}$, 
\begin{eqn}
&\intsinf \frac{s_1 s_2 ds_1 ds_2   }{(E + s_1 + s_2)(E_L + s_1 ) (E_R + k)} \\ 
&= s_0 \log 4 - E - \frac{\pi^2}{12} (E_L - E_R) \\
&+\frac{1}{2 (E -E_L - E_R) }\Bigg[12 E_L(E_L - E) \mbox{Li}_2 \frac{E_L - E}{E_L}  + 12 E_R(E - E_R) \mbox{Li}_2 \frac{E_R}{E - E_R}  \\
&\quad - 2 E_L E_R\log \frac{E_L}{s_0} \log \frac{s_0}{E_R} - 2 E (E_L + E_R + E) \log\frac{E}{s_0} \\
&\quad+ E_L (E - E_L)\log\frac{E_L}{s_0} \log\frac{(E-E_L)}{E_L } + E_R (E - E_R) \frac{E - E_R}{s_0} \log \frac{(E - E_R)s_0}{E_R^2}\Bigg]
\end{eqn}
where $E_L = k_{12} + k$, $E_R = k_{34} + k$, $E = k_{12} + k_{34}$.

Differential operators similar to the ones in equation \eqref{diff-ferm-1} can be defined for other integer $m$ using the following identity\footnote{Equation \eqref{besselid1} is true for odd integers $n$. For other values such a recursion is generically not useful as it would involve an infinite number of derivatives. Alternatively one could use the spectral representation to obtain a correlator in Mellin space \cite{Fitzpatrick:2011ia, Faller:2017hyt, Sleight:2019mgd,  Carmi:2019ocp, Carmi:2021dsn}. However, it is not straightforward to further perform the integral in the Mellin variable to obtain the result in momentum space.},
\begin{eqn}\label{besselid1}
K_{\frac{n}{2}}(kz) = \sqrt{\frac{\pi}{2}} \left( - \frac{1}{z} \right)^{\frac{n-1}{2}} \frac{k^{\frac{n+1}{2}}}{\sqrt{k z}} \left( \frac{1}{k} \pd{}{k} \right)^{\frac{n-1}{2}} \left( \frac{e^{- k z}}{k} \right)~,
\end{eqn}
which gives us 
\begin{eqn}\label{weightshifting1}
&K_{m+\frac12}(k z) + i \gammavec K_{m-\frac12}(kz) \\
&= \sqrt{\frac{\pi}{2}} \frac{(-1)^{m-1}}{z^m} \frac{k^m}{\sqrt{k z}} \left\{ \left( \frac{1}{k}\pd{}{k} \right)^{m} - i z \vec \gamma\cdot \vec k \Big( \frac1k \pd{}{k} \Big)^{m-1} \right\} \frac{e^{- k z}}{k}~.
\end{eqn}

The second term above can be written in the form below in order to obtain the differential operator for any integer $m$~. We provide the first few below,
\begin{align}
z \left( \frac1k \pd{}{k} \right)^1 \frac{e^{- k z}}{k} &= - \left( \frac{\p^2 }{\p k^2} +\frac{1}{k^1} \frac{\p }{\p k} - \frac{1}{k^2}  \right) \frac{e^{- k z}}{k} , \nno \\
z \left( \frac1k \pd{}{k} \right)^2 \frac{e^{- k z}}{k} &=  - \left( \frac{1}{k} \frac{\p^3 }{\p k^3} - \frac{3}{k^3} \frac{\p }{\p k} + \frac{3}{k^4} \right) \frac{e^{- k z}}{k}, \\
z \left( \frac1k \pd{}{k} \right)^3 \frac{e^{- k z}}{k} &=  - \left( \frac{1}{k^2} \frac{\p^4 }{\p k^4} - \frac{2}{k^3} \frac{\p^3 }{\p k^3} - \frac{3}{k^4} \frac{\p^2 }{\p k^2} +\frac{15}{k^5} \frac{\p }{\p k} - \frac{15}{k^6} \right) \frac{e^{- k z}}{k} , \nno \\
z \left( \frac1k \pd{}{k} \right)^4 \frac{e^{- k z}}{k} &=  - \left( \frac{1}{k^3} \frac{\p^5 }{\p k^5} - \frac{5}{k^4} \frac{\p^4 }{\p k^4} + \frac{5}{k^5} \frac{\p^3 }{\p k^3} + \frac{30}{k^6} \frac{\p^2 }{\p k^2} -\frac{105}{k^7} \frac{\p }{\p k} + \frac{105}{k^8} \right) \frac{e^{- k z}}{k} \nno~.
\end{align}

%%%%%%%%%%%%%%%%%%%%%%%%%%%
%%%%%%%%%%%%%%%%%%%%%%%%%%%
\subsection{Internal Legs}
It is also possible to obtain a set of recursion relations for the exchange of massive fermions. The massive Bulk-Bulk propagator is given as \cite{Kawano:1999au}\footnote{We follow the conventions of $m \geq 0$ and therefore this propagator satisfies the same boundary conditions as the massless case, as given in appendix \ref{app:fermionquant}. } 
\begin{eqn}
\mathcal S^{(m)}(z_1, z_2, \vec k) &= k \Bigg[ \Theta(z_1 - z_2) \phi_{(m)}^{(K)}(z_1, \vec k) P_- \bar \phi_{(m)}^{(I)}(z_2, -\vec k) \\
&\qquad - \Theta(z_2 - z_1) \phi_{(m)}^{(I)}(z_1, \vec k) P_- \bar \phi_{(m)}^{(K)}(z_2, -\vec k)  \Bigg]~,
\end{eqn}
where 
\begin{eqn*}
\phi_{(m)}^{(K)}(z, \vec k) &= z^{\frac{d+1}{2}} \Bigg[ K_{m+\frac12}(k z) - i \gammavec K_{m-\frac12}(kz)\Bigg], \quad \bar \phi_{(m)}^{(K)}(z, \vec k) = i \gammavec \phi^{(K)}(z, - \vec k),\\
\phi_{(m)}^{(I)}(z, \vec k) &= z^{\frac{d+1}{2}} \Bigg[ I_{m+\frac12}(k z) + i \gammavec I_{m-\frac12}(kz)\Bigg], \quad
\bar \phi_{(m)}^{(I)}(z, \vec k) = -  i \gammavec \phi^{(I)}(z, - \vec k).
\end{eqn*}
For the massless case, the propagator $\mathcal S^{(m = 0)}$  reduces to equation \eqref{fermionprop0}. Let us consider the case of $m = 1$ for which we get the following bulk-bulk propagator
\begin{align}\label{bulk-bulk-m1}
&\mathcal S^{(m = 1)}(z_1, z_2, \vec k) \\
&= k (z_1 z_2)^2 \Bigg[\Theta(z_1 - z_2) \Big\{ K_{3/2}(k z_1) - i \gammavec K_{1/2}(k z_1) \Big\} P_- \Big\{ i \gammavec I_{3/2}( k z_2) -  I_{1/2}( k z_2) \Big\} \nno\\
&\qquad\quad + \Theta(z_2 - z_1) \Big\{ I_{3/2}(k z_1) + i \gammavec I_{1/2}( k z_1) \Big\} P_- \Big\{ i \gammavec K_{3/2}(k z_2) + K_{1/2}(k z_2) \Big\} \Bigg]\nno~.
\end{align}

To derive a recursion relation for the massive case it is convenient to define the following quantities 
\begin{eqn}
F_{a, b}(z_1, z_2, k) &= (z_1 z_2)^{3/2} K_{a}(k z_1) I_{b}(k z_2), \\
H(z_1, z_2, k) &= (z_1 z_2)^{1/2} K_{\frac12}(k z_1) I_{\frac12}(k z_2).
\end{eqn}
It is easily checked that these satisfy the following relations\footnote{A similar set of relations hold for $m \geq1$,
\begin{eqn}\label{massiverecursion2}
\hat \Delta  \mathcal  F^{m-1}_{m+\frac12, m+\frac12} &=  ((2m-1)(z_1 + z_2) - z_1 z_2 \hat \Delta)\mathcal F^{m-2}_{m-\frac12, m-\frac12}  , \\
\hat \Delta   \mathcal F^{m-1}_{m+\frac12, m-\frac12}  &=((2m-1)(z_1 + z_2) - z_1 z_2 \hat \Delta) \mathcal F^{m-2}_{m-\frac12, m-\frac32} + \frac 1k (2m-1 - z_1 \hat \Delta) \mathcal F^{m-2}_{m-\frac12, m-\frac12} , \\
\hat \Delta  \mathcal F^{m-1}_{m-\frac12, m+\frac12} &=  ((2m-1)(z_1 + z_2) - z_1 z_2 \hat \Delta) \mathcal F^{m-2}_{m-\frac32, m-\frac12} - \frac 1k ( 2m-1 - z_2 \hat \Delta) \mathcal F^{m-2}_{m-\frac12, m-\frac12} ~,
\end{eqn}
where $\mathcal F^{n}_{a, b} = (z_1 z_2)^{n + \frac32} K_a(k z_1) I_b(k z_2)$. These relations hold for $\forall \ m $ but are only useful to us for integer $m$. For $m = 1$ they reduce to \eqref{massiverecursion}.
} 
\begin{eqn}\label{massiverecursion}
\hat \Delta H (z_1, z_2, k)&= e^{- kz_1} e^{- k z_2}, \\
\hat \Delta F_{\frac32, \frac32}(z_1, z_2, k) &=  (z_1 + z_2) H - (z_1 z_2) \hat \Delta H , \\
\hat \Delta F_{\frac32, \frac12}(z_1, z_2, k) &= \Big(z_1 + z_2 + \frac1k \Big) H + \frac{z_2}{k} (1 + kz_1) \hat \Delta H, \\
\hat \Delta F_{\frac12, \frac32}(z_1, z_2, k) &= \Big(z_1 + z_2 - \frac1k \Big) H + \frac{z_2}{k} (1 - k z_1) \hat \Delta H~.
\end{eqn}
where $\hat \Delta = \p_{z_1} + \p_{z_2}$. We also define the radial-ordered combination of $F$'s that enter the propagator,
\begin{eqn}
    \bar F^{(\pm)}_{a,b}(z_1,z_2,k)=\Theta(z_1-z_2)F_{a,b}(z_1,z_2,k)\pm\Theta(z_2-z_1)F_{a,b}(z_2,z_1,k)~.
\end{eqn}

In terms of these, the propagator \eqref{bulk-bulk-m1} becomes \footnote{This equation can be generalized to all $m$ via,
\begin{align}\label{massiverecursion3}
\mathcal S^{(m)}(z_1, z_2, \vec k)&= k (z_1 z_2)^{1/2} \Bigg[  \bar F_{m+\frac{1}{2}, m+\frac{1}{2}}^{(+)}P_- \left( i \gammavec \right) 
+ \bar F_{m-\frac{1}{2}, m-\frac{1}{2}}^{(+)} \left( i \gammavec \right) P_- \nno \\
&\qquad + \frac12 \bar F_{m-\frac12, m+\frac12}^{(+)} + \frac12 \bar F_{m-\frac12, m+\frac12}^{(-)} \gamma^z 
- \frac12 \bar F_{m+\frac12, m-\frac12}^{(+)} + \frac12 \bar F_{m+\frac12, m-\frac12}^{(-)} \gamma^z \Bigg] \\
 &= k (z_1 z_2)^{1/2} \Bigg[ \Big\{  \bar F_{m+\frac{1}{2}, m+\frac{1}{2}}^{(+)} + \bar F_{m-\frac{1}{2}, m-\frac{1}{2}}^{(+)}\Big\} \frac{i}{2} \gammavec + 
\Big\{ \bar F_{m-\frac{1}{2}, m-\frac{1}{2}}^{(+)}-  \bar F_{m+\frac{1}{2}, m+\frac{1}{2}}^{(+)} \Big\} \frac{i}{2} \gamma^z  \gammavec \nno
\\
&\qquad + \Big\{ \frac12 \bar F_{m-\frac12, m+\frac12}^{(+)} - \frac12 \bar F_{m+\frac12, m-\frac12}^{(+)} \Big\} +\Big\{  \frac12 \bar F_{m-\frac12, m+\frac12}^{(-)} 
+ \frac12 \bar F_{m+\frac12, m-\frac12}^{(-)} \Big\} \gamma^z \Bigg]~. \nno
\end{align}
For $m = 1$ this reduces to \eqref{massiveFexchange1}.
}
\begin{eqn}
\mathcal S^{(m = 1)}(z_1, z_2, \vec k) = k (z_1 z_2)^{1/2} &\Bigg[  \bar F_{\frac32, \frac32}^{(+)}P_- \left( i \gammavec \right) 
+ \bar F_{\frac12, \frac12}^{(+)} \left( i \gammavec \right) P_- \\
&\quad + \frac12 \bar F_{\frac12, \frac32}^{(+)} + \frac12 \bar F_{\frac12, \frac32}^{(-)} \gamma^z 
- \frac12 \bar F_{\frac32, \frac12}^{(+)} + \frac12 \bar F_{\frac32, \frac12}^{(-)} \gamma^z \Bigg]~,
\end{eqn}
where we suppress the $(z_1, z_2, k)$ dependence of $\bar F$'s for brevity. Using this we can evaluate the Witten diagram with a massive fermion exchange. As a concrete example, we consider the diagram for the s-channel contribution to the correlator $\braket{\bar\psi \phi \psi \phi}$ in $z^3 \bar \psi \psi \phi$ theory\footnote{For a different power of $z$ in the interaction term, it is possible to use the Schwinger parameters similar to \eqref{schwinger1} in order to convert it to an integral over \eqref{massiveFexchange1}. For generic $z$-powers in the interaction, this would be IR divergent, and the behavior of these are discussed in the upcoming section \ref{sec:IRdivergences}.} where the field $\phi$ is conformally coupled and the mass of $\bar\psi, \psi$ is $m_f=1$,
\begin{eqn}\label{massiveint1}
\scalebox{0.75}{\begin{tikzpicture}[baseline]
\draw[very thick] (0,0) circle (2);
\draw ({2*cos(150)},{(2*sin(150)}) -- (-1, 0) ;
\draw[fermion, red] ({2*cos(210)},{(2*sin(210)}) -- (-1, 0) ;
\draw ({2*cos(30)},{(2*sin(30)}) -- (1, 0) ;
\draw[fermionbar, red] ({2*cos(-30)},{(2*sin(-30)}) -- (1, 0) ;
\draw[fermion, red] (-1, 0) -- (1, 0);
%\node at (1.4, 0) {$k_{34}$};
%\node at (-1.4, 0) {$k_{12}$};
\node at (0, 0.25) {$k$};

\node at ({2.2*cos(150)},{(2.2*sin(150)}) {$2$};
\node at ({2.2*cos(210)},{(2.2*sin(210)}) {$1$};
\node at ({2.2*cos(30)},{(2.2*sin(30)}) {$3$};
\node at ({2.2*cos(-30)},{(2.2*sin(-30)}) {$4$};

\end{tikzpicture}} 
&= \hat{\bar D}_{k_1} \mathcal I  \hat{ D}_{k_4} ~,
\end{eqn}
where we used the differential operators introduced in the previous section to write the bulk integral as
\begin{eqn}\label{massiveFexchange1}
\mathcal I =\frac12 \intsinf dz_1 dz_2 e^{- x_1 z_1} e^{- x_2 z_2}&\Bigg[  \Big\{ F_{\frac12, \frac12}^{(+)} - F_{\frac32, \frac32}^{(+)}  \Big\} i  \gamma^z \gammavec +  \Big\{  F_{\frac12, \frac12}^{(+)} + F_{\frac32, \frac32}^{(+)}  \Big\}  i \gammavec \\
& +  \Big\{ \bar F_{\frac12, \frac32}^{(+)} - \bar F_{\frac32, \frac12}^{(+)} \Big\} +  \Big\{  \bar  F_{\frac12, \frac32}^{(-)} + \bar F_{\frac32, \frac12}^{(-)} \Big\}   \gamma^z  \Bigg]~.
\end{eqn}
This shows there exists a natural decomposition of the diagram \eqref{massiveint1} in terms of independent tensor structures and this pattern trivially generalizes to all masses via \eqref{massiverecursion3}. By performing the integrals via IBP and using the relations \eqref{massiverecursion} we express this in terms of differential operators w.r.t external kinematics,
\begin{eqn}\label{massivefinal1}
\mathcal I &= - 2 i \gamma^z \gammavec \left\{ \frac{1}{x_1 + x_2} ( \p_{x_1} + \p_{x_2} ) \right\}
 \begin{tikzpicture}[baseline]
\draw(-0.5, 0) -- (0.5, 0);
\node at (-0.5, 0) {\textbullet};
\node at (+0.5, 0) {\textbullet};
\end{tikzpicture} \\
&\quad  
+2  i \gammavec  \left\{
 \frac{1}{x_1 + x_2}( \p_{x_1} + \p_{x_2} )
+ \p_{x_1} \p_{x_2} 
 \right\}   \begin{tikzpicture}[baseline]
\draw(-0.5, 0) -- (0.5, 0);
\node at (-0.5, 0) {\textbullet};
\node at (+0.5, 0) {\textbullet};
\end{tikzpicture} 
\\
 &\quad
 + \frac{2}{x_1 + x_2} \Bigg\{ \frac{1}{k}
- (x_1 + x_2) \p_{x_1} \p_{x_2} - (\p_{x_1} + \p_{x_2})
\Bigg\}\begin{tikzpicture}[baseline]
\draw(-0.5, 0) -- (0.5, 0);
\node at (-0.5, 0) {\textbullet};
\node at (+0.5, 0) {\textbullet};
\end{tikzpicture}
  \\
&\quad + 
\frac{2\gamma^z}{x_1 + x_2} \Bigg\{ \left( \p_{x_1} + \p_{x_2}\right) \frac{1}{x_1 + x_2}  - \frac{1}{k(x_1 + x_2 + 2k)}  \Bigg\}  \begin{tikzpicture}[baseline]
\node at (-0.5, 0) {\textbullet};
\draw (-0.5, 0) -- (0.5, 0);
\node at (0.5, 0) {\textbullet};
\node at (0, 0) {$\triangleright$};
\node at (-0.5, -0.25) {\footnotesize$x_1 + k$};
\node at (0.5, 0.25) {\footnotesize$x_2 + k$};
\end{tikzpicture}   ~,
\end{eqn}
where we have used the previously introduced shorthand notation 
\begin{eqn*}
\begin{tikzpicture}[baseline]
\node at (-0.5, 0) {\textbullet};
\draw (-0.5, 0) -- (0.5, 0);
\node at (0.5, 0) {\textbullet};
\end{tikzpicture} &= \frac{1}{(x_1 + x_2)(x_1 + k)(x_2 + k)}
\end{eqn*}
and also introduced a new notation,
\begin{eqn}\label{antisymmgraph}
\begin{tikzpicture}[baseline]
\node at (-0.5, 0) {\textbullet};
\draw (-0.5, 0) -- (0.5, 0);
\node at (0.5, 0) {\textbullet};
\node at (0, 0) {$\triangleright$};
\node at (-0.5, -0.25) {\footnotesize$y_1$};
\node at (0.5, 0.25) {\footnotesize$y_2$};
\end{tikzpicture} = \frac{1}{y_1 + y_2} \left(  \frac{1}{y_1 } - \frac{1}{y_2} \right)~,
\end{eqn}
which naturally arises because of the antisymmetric combination $\bar  F_{\frac12, \frac32}^{(-)} + \bar F_{\frac32, \frac12}^{(-)} $\footnote{
The intermediate step which leads to this combination is
\begin{eqn*}
&\intsinf dz_1 dz_2 e^{- x_1 z_1} e^{- x_2 z_2} \Big[ \Theta(z_1 - z_2) H(z_1, z_2, k) - \Theta(z_2 - z_1) H(z_2, z_1, k)  \Big] \\
&= \frac{1}{(x_1 + x_2)(x_1 + x_2 + 2k)} \Big[ \frac{1}{k + x_2} - \frac{1}{k + x_1} \Big] \\
&= \frac{1}{x_1 + x_2} 
\begin{tikzpicture}[baseline]
\node at (-0.5, 0) {\textbullet};
\draw (-0.5, 0) -- (0.5, 0);
\node at (0.5, 0) {\textbullet};
\node at (0, 0) {$\triangleright$};
\node at (-0.5, -0.25) {\footnotesize$x_2 + k$};
\node at (0.5, 0.25) {\footnotesize$x_1 + k$};
\end{tikzpicture}
\end{eqn*}
}. 
 For $m > 1$, we would obtain a similar expression by using equations \eqref{massiverecursion2} and \eqref{massiverecursion3} with the same differential operators now acting on graphs with exchanged scalars of dimension $\Delta = m +1$. This shows how the massive fermions can be derived from the scalar graphs we saw in section \ref{sec:recursion} and a new antisymmetric graph \eqref{antisymmgraph}, by the action of simple differential operators in a basis of independent gamma matrices. The differential operators can be thought of as the Weight shifting operators for massive fermions at the tree level. The recursion relations can also be generalized to higher point functions by following the same procedure as the massless case given in section \ref{sec:recursion}. In appendix \ref{app:evaluatingloop} we describe how these recursion relations enable us to compute diagrams with massive internal legs at 1-loop.

%%%%%%%%%%%%%%%%%%%%%%%%%%%
%%%%%%%%%%%%%%%%%%%%%%%%%%%
\subsection{IR Divergences} \label{sec:IRdivergences}
In the previous section, we encountered divergent integrals for massive external fermions. This is a feature of Witten diagrams which contain fields and interactions that are not conformally coupled. In these cases, upon performing the $z-$integral (or the integral over the Schwinger parameters $s$ in equation \eqref{schwinger1}) we end up with divergences from the boundary $z = 0$ and these are known as IR divergences. These are studied very systematically for scalar fields in \cite{Bzowski:2013sza, Bzowski:2018fql}. We summarize the basic idea behind the computations from the bulk side in a slightly different language by analyzing the divergences of fermionic correlators. For fields with any generic mass, the maximal IR divergence in the 3-pt function $\braket{\bar\psi \psi\phi}$ in Yukawa theory receives a contribution from the following bulk-integral, 
\begin{eqn}\label{KKK-fermion}
\intsinf dz z^{3/2} K_{m_f + \frac12}(k_1 z) K_{m_f + \frac12}(k_2 z) K_{\nu_s}(k_3 z)~,
\end{eqn}
where $\nu_s = \sqrt{\frac{9}{4} + m_s^2}$ where $m_f, m_s$ are the masses of the fermion and the scalar respectively. This integral is IR divergent for specific values of $m_f, \nu_s$ and there are several prescriptions for regulating such divergences, including the hard-cutoff \cite{Maldacena:2011nz, Anninos:2014lwa, Konstantinidis:2016nio} and also using dimensional regularization \cite{Bzowski:2013sza}. In the following discussion, we shall work with dimensional regularization\footnote{In the language of \cite{Bzowski:2018fql} this is equivalent to setting $v = 0$.}, which results in the following integral
\begin{eqn}\label{KKK-fermion1}
\intsinf dz z^{3/2} z^{\e} K_{m_f + \frac12}(k_1 z) K_{m_f + \frac12}(k_2 z) K_{\nu_s}(k_3 z)~,
\end{eqn}
where we have introduced the regulator $\e$. The general result of this integral can be expressed in terms of an Appell $F_4$ function \cite{Coriano:2013jba, Bzowski:2013sza}. However, we find it convenient to analyze the singularities of the integral directly, without resorting to such a representation. The divergences in this integral can be classified into the following cases depending on the masses of the fields:
\begin{enumerate}
\item $m_f + \frac12 \neq Z$ and $\nu_s \neq Z$ \\
For generic values of $m_f, \nu_s$, the behavior of the integral near $z \to 0$ is given as\footnote{These integrals can be written in terms of the Gamma function, 
\begin{eqn}
\intsinf dz \frac{z^\e}{z^n} e^{- z} = \Gamma(1 - n + \e)~.
\end{eqn}
Since the $\Gamma$ function has poles only at negative integers, such integrals are divergent as $\frac1\e$ only for $n \in \mathbb Z^+$. 
}
\begin{eqn}
\intsinf dz \frac{ z^{\e} }{z^{a}} e^{-z} \sim \mbox{finite}~,
\end{eqn}
where $a \neq \mathbb Z$, which gives is a finite contribution. For the special cases when $2m_f + \nu_s -\frac12 = \mathbb Z^+$ (where $\mathbb Z^+$ refers to positive integers greater than 0) we get divergences of the form
\begin{eqn}
\intsinf dz \frac{z^\e}{z^n} f(z) &\sim \frac{1}{\e}~,
\end{eqn}
where $n \in \mathbb Z^+$.

\item  $m_f \in \mathbb Z^+$ and $\nu_s \in \mathbb Z^+ + \frac12$

In this case we only end up with integrals of the kind 
\begin{eqn}
\intsinf dz \frac{z^\e}{z^n} f(z) &\sim \frac{1}{\e}~.
\end{eqn}

\item  $m_f + \frac12 \in \mathbb Z^+$ and $\nu_s \in \mathbb Z^+ + \frac12$ \\
In this case we encounter Bessel functions with integer arguments. Near $z \to 0$ this has $\log$'s in the expansion around $z \to0$, 
\begin{eqn}
K_{n}(z) \sim \frac{1}{z^{n}} + \cdots +  z^{n -1} + z^{n} \log(z) + \cdots 
\end{eqn}
where $\log(z)$ also appear in higher order terms but one never gets $\log^2(z)$ or higher powers. 

For $\nu_s - 2m_f \geq \frac72$ we get integrals of the form\footnote{These integrals can be written in terms of a product of the Gamma function and PolyGamma function. For example,
\begin{eqn}
\intsinf dz \frac{z^\e}{z^n} e^{- z} \log(z) = \Gamma(1 - n + \e) \psi^{(0)}(-n+\epsilon +1)
\end{eqn}
where the PolyGamma function is defined as $\psi^{(n)}(x) = \frac{d^n}{dx^n} \frac{\Gamma'(x)}{\Gamma(x)}$. As these are derivatives of the Gamma function, they also have poles only at negative integers. Integrals with more number of $\log(z)$ insertions give a product such PolyGamma functions.}
\begin{eqn}
 \intsinf dz \frac{z^\e}{z^n} \log(z) \log(z) e^{-z}&\sim \frac{1}{\e^3}~,
\end{eqn}
and for the case $\nu_s - 2m_f < \frac72$ we get
\begin{eqn}
 \intsinf dz \frac{z^\e}{z^n} \log(z) e^{-z}&\sim \frac{1}{\e^2}~.
\end{eqn}

\end{enumerate}

 We summarize the conditions for divergences in the table below
 \begin{table}[H]
 \centering
 \begin{tabular}{|c|c|c|}
 \hline
   $\nu_s$ & $m_f$ & Max(Divergence) \\
    \hline
    \hline
%     $\neq \mathbb Z$ & $\neq \mathbb Z + \frac12$ & $\e^{0} $\\
 %\hline
%$\mathbb Z$ & $ \mathbb Z + \frac12$ & $\e^{0} $\\ \hline
%$\mathbb Z$ & $ \mathbb Z$ & $\e^{0} $\\ \hline
$\neq \mathbb Z$ & $\neq \mathbb Z + \frac12$, \quad $2m_f + \nu_s - \frac12 = 2\mathbb Z^+$ & $\e^{-1} $\\
\hline
$\mathbb Z^+ + \frac12 $ & $ \mathbb Z^+$ & $\e^{-1} $\\ \hline
$\mathbb Z^+ + \frac12 $ & $ \mathbb Z^+ - \frac12$,\quad $m_f > \frac{\nu_s}{2} - \frac74$ & $\e^{-2} $\\ \hline
$\mathbb Z^+ + \frac12 $ & $ \mathbb Z^+ - \frac12$,\quad $m_f \leq \frac{\nu_s}{2} - \frac74$ & $\e^{-3} $\\ \hline
 \end{tabular}
 \caption{Classification of the order of IR divergences from bulk integrals for 3-pt functions.}
  \label{tab:divergence}
 \end{table}
The order of divergence is sensitive to the choice of the regularization scheme, for other schemes (like the cut-off), the powers and the analytic structure of the divergence would be different.  In the regularization scheme chosen above, the degree of divergences encountered for 3-pt functions of fermions coupled to scalars are similar to the divergences for 3-pt functions of self-interacting scalars in AdS$_4$ as studied in \cite{Bzowski:2013sza}\footnote{ Similar to the analysis performed in \cite{Bzowski:2013sza} in position space, the divergences in momentum space can be categorized as ultralocal or sub-local, depending on their dependence on the momenta. For the 3-pt function, the ultralocal divergences arise when one obtains an expression as a function of $\sum_{i = 1}^3 f(k_i)$, whereas sublocal divergences arise from $\sum_{i = 1}^3 f(k_i, k_{i+1})$ \cite{Maldacena:2011nz}. This is because upon taking the Fourier transform w.r.t $k_i$, these give delta functions in position space and hence only click at coincident points.}.  Therefore, the counterterms analogous to the ones proposed in \cite{Bzowski:2013sza} are expected to also renormalize these correlators and it would be interesting to make this precise in the future.  A similar analysis can be performed for higher dimensions resulting in the same set of divergences. 
 
 For higher point exchange graphs the maximal divergence arises from a product of such contact graphs. This is understood by the following intuitive picture: IR divergences in the integrals arise when $z \to 0$. Hence, for any graph, the maximal divergence arises by squishing the entire Witten diagram to the boundary at $z \to 0$, reducing it to a product of lower point contact diagrams. In general, the limit of taking both radial points in the bulk-bulk propagator close to the boundary makes the radial integrals converge better. Hence the analysis for the maximal divergence performed above is sufficient for all graphs with polynomial interactions. For derivative interactions, the degree of divergence could increase depending on the number of derivatives in the interaction term, but the maximal divergence for any exchange graph is still less than equal to the product of contact graphs.

%%%%%%%%%%%%%%%%%%%%%%%%%%%
\section{Cosmological Correlator}\label{sec:cosmologicalcorr}
%%%%%%%%%%%%%%%%%%%%%%%%%%%
Correlation functions in AdS are also useful for computing in-in correlators in dS and flat spacetime. This is usually done by relating the on-shell action in AdS  to the ground state wave function in dS \cite{Hartle:1983ai} via an analytical continuation of the form \cite{Maldacena:2002vr}
\begin{eqn}\label{AdStoWF}
\Psi_{dS}(\phi, \psi, \bar\psi) = e^{-S^{AdS}_{on-shell}(\phi, \psi, \bar\psi)}~,
\end{eqn}
where $S^{AdS}_{on-shell}(\phi, \psi, \bar\psi)$ has the following expansion\footnote{In this section we use $a, b, c, d$ to denote the Spinor indices.}, 
\begin{align}\label{Psi-ferm}
S^{AdS}_{on-shell}(\phi, \psi, \bar\psi) &=  \int \bar\psi_a \psi_b W^{ab}_2[\bar\psi, \psi] +  \int \phi \phi W_2^{(\phi)}[\phi, \phi] + \int \bar\psi_a \psi_b \phi W_3^{ab}[\bar \psi, \psi, \phi]  \\
&+ \int \bar\psi_a \psi_b \phi \phi W_4^{ab}[\bar \psi, \psi, \phi, \phi]  + \int \bar\psi_a \psi_b \bar\psi_c \psi_d W_4^{abcd }[\bar \psi, \psi, \bar\psi, \psi]   + \cdots \nno
\end{align}
where we suppress the momentum dependence in the integrands to avoid a clutter of notation. $W_n$'s are referred to as the {\it wave function coefficients}. Since these are evaluated via an analytic continuation of the on-shell action in EAdS, they are computed by Witten diagrams and therefore can be evaluated by using the tools in the previous sections. For IR divergent integrals the analytical continuation in \eqref{AdStoWF} has to be performed carefully (see \cite{Bzowski:2023nef} for a recent discussion on IR divergences for scalar theories) and for this section, we shall only restrict to massless fermions coupled to conformally coupled scalars where there are no IR divergences\footnote{The issue of renormalizing IR divergences is more subtle in dS than for AdS and it would be interesting to understand the EFT for fermions in dS by following \cite{Gorbenko:2019rza}. }. The Cosmological Correlator (or the In-In correlator) in the state $\Psi_{dS}$ is defined as,
\begin{eqn}\label{cosmo-corr-defn}
\braket{\Psi_{dS}|\mathcal O_1(\vec k_1) \cdots \mathcal O_n(\vec k_n)|\Psi_{dS}} = \frac{1}{\mathcal N} \int D\bar\psi  D\psi D\phi  \big| \Psi_{dS}[\bar\psi, \psi, \phi]  \big|^2 \mathcal O_1(\vec k_1) \cdots \mathcal O_n(\vec k_n),
\end{eqn}
where $\mathcal O_1(\vec k_1), \cdots, \mathcal O_n(\vec k_n)$ can be either fermionic fields or scalar fields at the conformal boundary and $\mathcal N$ is the normalization factor for the path integral
\begin{eqn}
\mathcal N = \int D\bar\psi  D\psi D\phi  \big| \Psi_{dS}[\bar\psi, \psi, \phi]  \big|^2 ~.
\end{eqn}

This path integral can be directly performed using standard methods \cite{Maldacena:2002vr} and can also be performed using the Schwinger-Keldysh (In-In) approach \cite{Weinberg:2005vy}. For our purpose, we shall directly evaluate the path integral above. Using \eqref{AdStoWF} and \eqref{Psi-ferm}, the full ground state wave function $\Psi_{dS}[\bar\psi, \psi, \phi]$ can be expressed in terms of the wave-function coefficients as 
\begin{align}
\Psi_{dS}[\bar\psi, \psi, \phi] &= \exp\Big(- \int \bar\psi_a \psi_b W^{ab}_2[\bar\psi, \psi] -  \int \phi \phi W_2^{(\phi)}[\phi, \phi] \Big) \nno \\
&\times\exp\Big( - \int \bar\psi_a \psi_b \phi W_3^{ab}[\bar \psi, \psi, \phi]  \Big) \label{Psi-ferm2}\\
&\times \exp \Big(- \int \bar\psi_a \psi_b \phi \phi W_4^{ab}[\bar \psi, \psi, \phi, \phi]  - \int \bar\psi_a \psi_b \bar\psi_c \psi_d W_4^{abcd }[\bar \psi, \psi, \bar\psi, \psi]   \Big) \cdots \nno
\end{align}
where we have explicitly displayed the spinor indices. Squaring the equation above we obtain an effective action for the cosmological correlator \eqref{cosmo-corr-defn},
\begin{align}
\big| \Psi_{dS}[\bar\psi, \psi, \phi]  \big|^2  &= \exp\Big(- 2\re\int \bar\psi_a \psi_b W^{ab}_2[\bar\psi, \psi] - 2\re  \int \phi \phi W_2^{(\phi)}[\phi, \phi] \Big) \nno\\
&\times \exp\Big( - 2\re \int \bar\psi_a \psi_b \phi W_3^{ab}[\bar \psi, \psi, \phi]  \Big) \\
&\times \exp \Big(- 2\re \int \bar\psi_a \psi_b \phi \phi W_4^{ab}[\bar \psi, \psi, \phi, \phi]  - 2\re \int \bar\psi_a \psi_b \bar\psi_c \psi_d W_4^{abcd }[\bar \psi, \psi, \bar\psi, \psi]   \Big) \cdots  \nno
\end{align}
This can be viewed as an effective action for the path integral in \eqref{cosmo-corr-defn} and can be used to compute any correlator of interest using perturbation theory. For concreteness, we evaluate the following 4-pt correlator at leading order in the coupling constant in $\frac{g}{z}\phi^2 \bar\psi \psi $ theory,
\begin{eqn}\label{cosmocorr1}
\braket{\Psi_{dS}| \phi(\vec k_1) \phi(\vec k_2) \phi(\vec k_3) \phi(\vec k_4) | \Psi_{dS}} &= \mathcal N \int D\psi D\bar\psi D\phi \phi(\vec k_1) \phi(\vec k_2) \phi(\vec k_3) \phi(\vec k_4) \big| \Psi_{dS}  \big|^2. 
\end{eqn}
We also provide an appendix \ref{app:ferm-inin} illustrating the details of evaluating the path integral to obtain correlators with external fermions $\braket{\Psi_{dS}| \bar\psi_i(\vec k_1) \phi(\vec k_2) \psi_j(\vec k_3) \phi(\vec k_4) | \Psi_{dS}}$. The leading order contribution to  the cosmological correlator \eqref{cosmocorr1} occurs at $O(g^2)$ and receives a contribution from the following wave function coefficients
\begin{eqn}
W_4^{(1) ab}[\bar\psi, \psi, \phi, \phi] &=   
\scalebox{0.75}{\begin{tikzpicture}[baseline]
\draw[very thick] (0, 0) circle (2);
\draw[fermion] ({2*cos(-30)},{(2*sin(-30)}) -- (0, 0);
\draw[fermionbar] ({2*cos(210)},{(2*sin(210)}) -- (0, 0);
\draw ({2*cos(30)},{(2*sin(30)}) -- (0, 0);
\draw ({2*cos(150)},{(2*sin(150)}) -- (0, 0);

\node at ({2.25*cos(-30)},{(2.25*sin(-30)}) {$1$};
\node at ({2.25*cos(210)},{(2.25*sin(210)}) {$2$};
\node at ({2.25*cos(150)},{(2.25*sin(150)}) {$3$};
\node at ({2.25*cos(30)},{(2.25*sin(30)}) {$4$};

\end{tikzpicture}}
= g \frac{ \bar u_1^a u_2^b }{k_1 + k_2 + k_3 + k_4}, \\
W_4^{(2)}[\phi, \phi, \phi, \phi] &=   
\scalebox{0.75}{\begin{tikzpicture}[baseline]
\draw[very thick] (0, 0) circle (2);
\draw ({2*cos(150)},{2*sin(150)}) -- (-1, 0);
\draw ({2*cos(210)},{2*sin(210)}) -- (-1, 0);

\draw ({2*cos(30)},{2*sin(30)}) -- (1, 0);
\draw ({2*cos(-30)},{2*sin(-30)}) -- (1, 0);

\draw[fermion] (0,0) circle (1);
\node at ({2.25*cos(-30)},{(2.25*sin(-30)}) {$4$};
\node at ({2.25*cos(210)},{(2.25*sin(210)}) {$1$};
\node at ({2.25*cos(150)},{(2.25*sin(150)}) {$2$};
\node at ({2.25*cos(30)},{(2.25*sin(30)}) {$3$};

%\node at (0, 1.25) {$\vec l_1$};
%\node at (0, -1.25) {$\vec l_2$};
\end{tikzpicture} }
= g^2 \times \mbox{equation \eqref{bubble-x1x2-ans}}. 
\end{eqn}
which results in the following contribution to the cosmological correlator at $O(g^2)$ for the s-channel
\begin{eqn}
&\braket{\Psi_{dS}| \phi(\vec k_1) \phi(\vec k_2) \phi(\vec k_3) \phi(\vec k_4) | \Psi_{dS}} \\
&= \int d^3 l\  \re W_4^{(1) ab}[\bar\psi(l_1), \psi(\vec l_2), \phi(\vec k_1), \phi(\vec k_2)]   \re W_4^{(1) ba}[\bar\psi(l_2), \psi(\vec l_1), \phi(\vec k_3), \phi(\vec k_4)]  \\
&\quad + 2 \re W_4^{(2)}[\phi(\vec k_1), \phi(\vec k_2), \phi(\vec k_3), \phi(\vec k_4)]~.
\end{eqn}
The two terms are explicitly given as\footnote{The real part of the product of $\gamma$-matrices is computed using 
\begin{eqn*}
\re \big[(1 + i\Gammavec{l_1})(1 + i\Gammavec{l_2}) \big] = 1- \frac{\vec l_1 \cdot \vec l_2}{l_1 l_2}
\end{eqn*}
which uses the fact that the gamma matrices in the Euclidean signature are hermitian. 
}
\begin{eqn}\label{1loopcosmo}
 \re W_4^{(1) ab}  \re W_4^{(1) ba} 
&= \frac{2}{( l_1 + l_2 + k_{12})(l_1 + l_2 + k_{34})} \Bigg( 1 + \frac{\vec l_1 \cdot \vec l_2}{l_1 l_2} \Bigg)^2, \\
 \re W_4^{(2)}&=  \Bigg[ - \frac{1}{(l_1 + l_2 + k_{34})(k_{12} + k_{34})} - \frac{1}{(l_1 + l_2 + k_{12})(k_{12} + k_{34})} \\
&\qquad + \frac{1}{(l_1 + l_2 + k_{12})(l_1 + l_2 + k_{34})} \Bigg] \Bigg( 1 + \frac{\vec l_1 \cdot \vec l_2}{l_1 l_2} \Bigg)~,
\end{eqn}
where $l_1 = |\vec l|$ and $l_2 = |\vec l + \vec k_1 + \vec k_2|$. As an immediate consistency check, we see that only the second term, $ \re W_4^{(2)}$, contributes to the flat space limit, recovering the expected integrand for the bubble diagram in flat space. The integral over $\vec l$ can be performed in the same manner as the wave function coefficient in section \ref{sec:1loopbubble}. Upon evaluating the integral\footnote{For the interested readers, the full result is recorded in the following \reference{Mathematica} notebook which can be found in \cite{github}. } we find that the disconnected piece, $\re W_4^{(1) ab}  \re W_4^{(1) ba} $, is a $\mbox{Li}_2$ in the external momenta whereas the other term, $ \re W_4^{(2)}$, gives a $\log$.  This suggests that for spinning exchanges, in-in correlators can have a higher transcendentality than the corresponding wave function coefficient. We also discuss another example of a spinning correlator in appendix \ref{app:ferm-inin}. It would be interesting to compare the pole structures of the in-in correlator vs the wave function coefficients more generally for fermions. A similar analysis was carried out for scalars in \cite{Lee:2023kno} and it would be useful to explore this further using the shadow formalism \cite{Sleight:2020obc, Sleight:2021plv, DiPietro:2021sjt, Schaub:2023scu}. This formalism suggests that in-in correlators of fields with conformal dimension $\Delta$ are directly obtained by evaluating Witten diagrams for an effective action in EAdS$_{d+1}$ constructed out of their shadow fields (which have a conformal dimension $d - \Delta$). For scalars, this provides an easier way to evaluate loop integrals for cosmological correlators as shown in \cite{Heckelbacher:2022fbx, Chowdhury:2023arc} and it would be interesting to see if a similar statement also holds for fermions.

%%%%%%%%%%%%%%%%%%%%%%%%%%%
%%%%%%%%%%%%%%%%%%%%%%%%%%%
%%%%%%%%%%%%%%%%%%%%%%%%%%%
\subsection*{Acknowledgements}
We thank several people for discussions and clarifications at various stages of the development of this project: A. H. Anupam, Paolo Benincasa, Ernesto Bianchi, Adrian Garrido, Fridrik Gautason, Alok Laddha, Arthur Lipstein, Ioannis Matthaiakakis, Jiajie Mei, Yuyu Mo, Silvia Nagy, Shivam Sharma, Arvind Shekar,  Kostas Skenderis and Ida Zadeh. We are grateful to Arthur Lipstein, Ioannis Matthaiakakis and Yuyu Mo for very useful feedback on the draft. CC would like to thank Arthur Lipstein and Durham university for hospitality while this work was in progress and for an opportunity to present a part of this paper. CC is supported by the STFC consolidated grant (ST/X000583/1) “New Frontiers in Particle Physics, Cosmology and Gravity”. PC is supported by ``University of Southampton Presidential Scholarship''. RM is supported by an STFC studentship. KS is supported by the STFC research grant (ST/X000699/1) ``Particles, Fields and Strings at Liverpool''.

%%%%%%%%%%%%%%%%%%%%%%%%%%%
%%%%%%%%%%%%%%%%%%%%%%%%%%%
%%%%%%%%%%%%%%%%%%%%%%%%%%%
\begin{appendix}
%%%%%%%%%%%%%%%%%%%%%%%%%%%
\section{Review of perturbation theory for scalars and fermions} \label{app:fermionquant}
%%%%%%%%%%%%%%%%%%%%%%%%%%%
In this appendix we review the derivation of propagators for scalars and fermions in Euclidean AdS in the Poincare patch. 
The metric is given as 
\begin{eqn}
ds^2 = \frac{dz^2 + dx_1^2 + \cdots + dx_d^2}{z^2}~,
\end{eqn}
where $z$ is the radial coordinate which ranges from $(0, \infty)$ and $x_i$'s are the other spatial directions ranging from $(-\infty, \infty)$. The volume element is given as $\sqrt{g} = \frac{1}{z^{d+1}}$.

%%%%%%%%%%%%%%%%%%%%%%%%%%%
%%%%%%%%%%%%%%%%%%%%%%%%%%%
\subsection{Scalars}
Consider a scalar in AdS with the action given as
\begin{eqn}
S = \int d^{d+1}x \sqrt{g}  \left( \frac12 \p_\mu \varphi \p^\mu \varphi + \frac12 m^2 \varphi^2 \right) ~.
\end{eqn}

The equation of motion of the scalar field is
\begin{eqn}
\frac{1}{\sqrt{g}} \p_\mu (\sqrt{g} \p^\mu \varphi) = z^{d+1}\p_z\left(\frac{1}{z^{d-1}} \p_z\varphi \right) + z^2 \p_i^2\varphi = m^2 \varphi~.
\end{eqn}
The field $\phi$ can be decomposed in terms of its Fourier modes,
\begin{eqn}
\varphi(z, \vec x) = \int \frac{d^d k }{(2\pi)^d} e^{i \vec k \cdot \vec x} \phi(z, \vec k)~.
\end{eqn}
This leads to the following differential equation for $\phi(z, \vec k)$, 
\begin{eqn}
z^2 \p_z^2 \phi - (d-1) z \p_z \phi - (k^2 z^2 + m^2)  \phi = 0~,
\end{eqn}
and its solution is given as,
\begin{eqn}
\phi(z, \vec k) = c_1 z^{d/2} J_{\nu}(\sqrt{-k^2} z) + c_2 z^{d/2} Y_{\nu}(\sqrt{- k^2} z)~,
\end{eqn}
where $\nu = \frac{\sqrt{d^2 + 4 m^2}}{2}$. For the massless case when $\nu = \frac{d}2$, expanding this about $z \to \infty$ we find 
\begin{eqn}
\lim_{z \to \infty}\phi(z, \vec k) =  z e^{k z} \frac{(-1)^{1/4}}{\sqrt{2\pi k}} \left( i c_1 - c_2 \right) + \mbox{regular}~.
\end{eqn}
Thus choosing $c_2 =  i c_1$ ensures that the field is normalizable as $z \to \infty$. A similar conclusion holds for any general mass of the scalar field. Therefore for any general mass, the field becomes 
\begin{eqn}
\phi(z, \vec k) = c_2 z^{d/2} \left\{ i  J_{\nu}(\sqrt{- k^2} z) - Y_{\nu}(\sqrt{-k^2} z) \right\}.
\end{eqn}
Absorbing the extra factors into the normalization and demanding that $k^2 < 0$ for spacelike momenta, we obtain the usual solution in terms of the modified Bessel function of the second kind
\begin{eqn}
\phi(z, \vec k) = c_2 z^{d/2} K_{\nu}(k z)~.
\end{eqn}
The normalization constant can be fixed via the boundary value of the scalar field. This becomes the bulk-boundary Green function. The bulk-bulk Green function can be derived in multiple ways. The simplest way to derive this in Euclidean space is to solve the differential equation 
\begin{eqn}
\big\{ z^2 \p_z^2  - (d-1) z \p_z  - (k^2 z^2 + m^2)  \big\} G(z, z', \vec k) = i z^{d+1} \delta(z - z') ~.
\end{eqn}
This has to satisfy the boundary condition that it is normalizable as $ z \to 0$ and $z \to \infty$ and is continuous across $z = z'$. Along with demanding that the Green function is normalizable as $z \to 0$, for special values of mass, we also require it to fall off as $z^\Delta$ as $z\to 0$, where $\Delta = \nu + \frac{d}{2}$. This completely fixes the Green function for all masses and we arrive at the standard expression for the Green function of a scalar field in AdS satisfying normalizable boundary condition (see section 2.1 of \cite{Bzowski:2023nef} for a very systematic derivation),
\begin{eqn}
G(z, z', \vec k) = (z z')^{d/2} \Big\{ \Theta(z - z') K_{\nu}(k z) I_{\nu}(k z') + \Theta(z' - z) K_{\nu}(k z') I_{\nu}(k z) \Big\}~.
\end{eqn}
For $d =3$ and $\Delta = 1, 2$ we recover the expressions given in section \ref{sec:setup}.

\begin{comment}
It can be checked that the continuity across $z = z'$ satisfies 
\begin{eqn}
\lim_{z \to z'} z^2 \p_z \phi = i z'^{d+1} \implies \p_{z'} \phi = i z'^{d-1}
\end{eqn}
\end{comment}

%%%%%%%%%%%%%%%%%%%%%%%%%%%
%%%%%%%%%%%%%%%%%%%%%%%%%%%
\subsection{Fermions}
We\footnote{We are grateful to Shivam Sharma for several helpful discussions and also pointing out typos in the previous version of the equations.} review the derivation of the propagators for massless fermions in AdS$_4$ for simplicity, the general dimensional massive case is conceptually similar and is discussed in detail in \cite{Kawano:1999au, Mueck:1998iz, Henningson:1998cd}\footnote{It is well known that the on-shell action for the free fermion is zero and it needs to be supplemented by a boundary term of the form $\bar\psi \psi$ \cite{Henneaux:1998ch}. For computing tree-level correlators for interacting theories this choice is equivalent to using the bulk-bulk and bulk-boundary propagators given below (see \cite{Giecold:2009tt, Loganayagam:2020eue, Martin:2024mdm} for a discussion of the propagators used in computing out of time-ordered correlators). It would be interesting to investigate this more carefully for loop-level correlators. }. 

The differential equation for the equation of motion of a minimally coupled massless free Dirac fermion in AdS$_4$ is 
\begin{eqn}
\slashed D \psi_{AdS} =  \big(z \gamma^\mu \p_\mu - \frac{3}{2} \gamma^z \big) \psi_{AdS} = 0 ~.
\end{eqn}
We use the following ansatz to solve it
\begin{eqn}\label{bulkbndyapp1}
\psi_{AdS}(z, \vec x) = \int \frac{d^3 k }{(2\pi)^3} A(\vec k) \chi_{AdS} z^{3/2} e^{-k z} e^{i \vec k \cdot \vec x}~,
\end{eqn}
where $A(\vec k)$ is a $4\times4$ matrix and $\chi_{AdS}$ denotes the boundary limit of the bulk spinor\footnote{This can be related with the lower dimensional boundary spinor via a rectangular matrix \cite{Loganayagam:2020eue}.}.  Using this ansatz we obtain the following for $\slashed D \psi_{AdS}$ 
\begin{eqn}\label{fermion-eom1}
\slashed D \psi_{AdS} &= \int  \frac{d^3 k }{(2\pi)^3} e^{i \vec k \cdot \vec x}  \big(z \gamma^z \p_z + i z \gamma^i k_i - \frac{3}{2} \gamma^z \big) (z^{3/2} A \chi_{AdS} e^{- k z}) \\
&=- i \int  \frac{d^3 k }{(2\pi)^3} z^{5/2} e^{i \vec k \cdot \vec x}   \big( i  \gamma^z \p_z - \vec \gamma \cdot \vec k \big) A \chi_{AdS} e^{-k z}~,
\end{eqn}
The boundary condition we impose is the standard Dirichlet boundary condition \cite{Kawano:1999au}
\begin{eqn}\label{fermion-bc1}
\lim_{z \to 0} P_- \psi_{AdS}(z, \vec x) = \lim_{z\to 0} z^{3/2} \chi_{AdS}~,
\end{eqn}
 where the power of $z$ indicates the scaling dimension and $P_{\pm} = \frac12 (1 \pm \gamma^z)$. This boundary condition sets two out of the four spinor components to zero at the boundary. From the ansatz given above \eqref{bulkbndyapp1}, the most general solution of the equation of motion \eqref{fermion-eom1} is obtained via the following matrix $A(\vec k)$,
\begin{eqn}\label{Adefn1}
A(\vec k) =  1 +  i \gamma^z \gammavec~.
\end{eqn}
By imposing the boundary condition \eqref{fermion-bc1} we get the following constraint on the spinor $\chi_{AdS}$,
\begin{eqn}
\gamma^z \chi_{AdS} = -  \chi_{AdS}~.
\end{eqn}
Hence the bulk-boundary propagator \eqref{bulkbndyapp1} can also be written as
\begin{eqn}
\psi_{AdS}(z, \vec x) = \int \frac{d^3 k }{(2\pi)^3} z^{3/2} e^{- k z} e^{i \vec k \cdot \vec x} \left( 1 + i \gammavec \right) \chi_{AdS}~,
\end{eqn}
for $\gamma^z \chi_{AdS} = - \chi_{AdS}$. Similar relations also hold for $\bar \psi_{AdS}(z, x)$, for which we make the standard choice  
\begin{eqn}\label{barpsi}
\bar \psi_{AdS} (z, \vec x) = \psi_{AdS}^\dagg(z, \vec x) \gamma^t ~,
\end{eqn}
although any linear combination of $\gamma_t, \gamma_x, \gamma_y$ would be equivalent since we are in EAdS. The boundary condition for $\bar \psi_{AdS}$ is,
\begin{eqn}
\lim_{z \to 0}\bar\psi_{AdS} P_- = \lim_{z\to 0 } z^{3/2} \bar \chi_{AdS}~,
\end{eqn}
where $\bar\chi_{AdS} = \chi_{AdS}^\dagg \gamma^t$ and it satisfies $\bar \chi_{AdS} = \bar\chi_{AdS} \gamma^z$. It can be checked that \eqref{barpsi} is consistent with the equation of motion for $\bar \psi_{AdS}$, 
\begin{eqn}
\bar \psi_{AdS} \overset{\leftarrow}{\slashed{D}} = 0~,
\end{eqn}
where $\overset{\leftarrow}{\slashed{D}}  = \gamma^z \overset{\leftarrow}{\partial}_z + \gamma_i \overset{\leftarrow}{\partial}_i - \frac32 \gamma^z $, which after acting on the spinor effectively becomes $- \gamma^z k - i \gamma^i  k_i  - \frac32 \gamma^z$. Hence the bulk-boundary propagator $\bar\psi_{AdS}$ is 
\begin{eqn}
\bar\psi_{AdS}(z, \vec x) = \int \frac{d^3 k }{(2\pi)^3} z^{3/2} e^{- k z} e^{- i \vec k \cdot \vec x}  \bar \chi_{AdS} \left( 1 + i \gammavec \right) ~.
\end{eqn}
The bulk-bulk propagator solves the following equation 
\begin{eqn}
\slashed D S_{AdS}(z, z', \vec k) = (z \gamma^z \p_z + i z \gamma^i k_i - \frac32 \gamma^z) S_{AdS}(z, z', \vec k) =   \delta(z, z')~,
\end{eqn}
which can be solved in the same way as the scalar case. The boundary condition that fixes the propagator is obtained from \eqref{fermion-bc1}
\begin{eqn}
\lim_{z \to0} P_- S_{AdS}(z, z', \vec k) = z^{3/2} (\cdots)~,
\end{eqn}
where $\cdots$ denote the terms regular in $z$. The final solution is given as 
\begin{align}
&S_{AdS}(z, z', \vec k) \nno \\
&= \frac{(zz')^{3/2}}{2} \Bigg[ \Theta(z - z') \left( \gamma^z - i \gammavec \right) e^{- k (z - z')} - \Theta(z' - z) \left( \gamma^z + i \gammavec \right) e^{- k (z' - z)} \nno \\
&\qquad\qquad\qquad - \left( 1 + i \gamma^z \gammavec \right) e^{- k (z+z')} \Bigg] ~.
\end{align}
As a consequence we also have $  \lim_{z' \to0}  S_{AdS}(z, z', \vec k) P_+ = 0$.  In section \ref{sec:massive} we also provide the explicit expression for the propagators of massive fermions in AdS$_4$.

%%%%%%%%%%%%%%%%%%%%%%%%%%%
%%%%%%%%%%%%%%%%%%%%%%%%%%%
\subsection{Spectral Representation}\label{sec:anreg}
It is often convenient to use an alternate representation of the bulk-bulk propagator for computing integrals for diverse masses or regulating divergences. One such representation is known as the {\it spectral representation}. This can be derived by expressing the fermionic bulk-bulk propagator as a derivative of the scalar propagator as given in equation  \eqref{fermionDscalar},
\begin{eqn*}
S(z_1, z_2, \vec k) = - (\gamma^z \p_{z_1} + i \vec\gamma\cdot \vec k) \mathcal G(z_1, z_2, \vec k)
\end{eqn*}
where $\mathcal G(z_1, z_2,\vec k) = P_+ G_D(z_1, z_2, \vec k) +  P_- G_N(z_1, z_2, \vec k)$ with their spectral representations given as 
\begin{eqn}
G_D(z_1, z_2, \vec k) &= \frac1\pi \intinf \frac{d\omega}{\omega^2 + k^2} \sin(\omega z_1) \sin(\omega z_2), \\
G_N(z_1, z_2, \vec  k) &=\frac1\pi \intinf \frac{d\omega}{\omega^2 + k^2 } \cos(\omega z_1) \cos(\omega z_2).
\end{eqn}
Using these, we have the following spectral representation for the fermionic bulk-bulk propagator in flat space,
\begin{eqn}\label{spectral}
S(z_1, z_2, \vec k) &= - \frac1\pi\intinf \frac{d\omega}{\omega^2 + k^2} \Bigg[ \omega \Big\{ \sin\big(\omega(z_1 + z_2)\big) - \gamma^z \sin\big(\omega(z_1 - z_2)\big)  \Big\} \\
&\qquad\qquad\qquad + i \vec \gamma \cdot \vec k \Big\{ \cos\big( \omega(z_1 - z_2) \big) - \gamma^z \cos\big( \omega(z_1 + z_2) \big) \Big\} \Bigg]
\end{eqn}
This representation naturally generalizes to a form that regularizes both IR and UV divergences in AdS, 
\begin{eqn}
&S^{AdS}_{reg}(z_1, z_2, \vec k) \\
&=  - \frac{(z_1z_2)^{3/2 + \kappa}}{\pi} \intinf \frac{d\omega}{(\omega^2 + k^2)^{1 + \kappa}} \Bigg[ \omega \Big\{ \sin\big(\omega(z_1 + z_2)\big) - \gamma^z \sin\big(\omega(z_1 - z_2)\big)  \Big\} \\
&\hspace{5cm}+ i \vec \gamma \cdot \vec k \Big\{ \cos\big( \omega(z_1 - z_2) \big) - \gamma^z \cos\big( \omega(z_1 + z_2) \big) \Big\} \Bigg]
\end{eqn}
where $\kappa$ is the regulator. This regularization technique is known as {\it analytic regularization} and was developed in \cite{Chowdhury:2023arc} for scalars in momentum space\footnote{
The spectral representation also allows one to derive the ``cut propagator' which enables the study of discontinuities of correlators w.r.t external momenta. These are analogous to Cutting rules \cite{Cutkosky:1960sp} for the flat space S-matrix where the propagator goes on-shell. For tree-level correlators, it is shown \cite{Meltzer:2020qbr, Baumann:2020dch} how the discontinuities of AdS correlators and dS wave functions are constrained via the spectral propagator going on-shell. It is also possible to extend this for fermionic correlators via picking the pole at $\omega = i k$ in equation \eqref{spectral} to obtain the fermionic ``cut propagator''
\begin{eqn}
S_{cut}(z_1, z_2, \vec k) = \frac12 &\Bigg[ k \Big\{ \sin\big(k(z_1 + z_2)\big) - \gamma^z \sin\big(k(z_1 - z_2)\big)  \Big\} \\
&\qquad + i \vec \gamma \cdot \vec k \Big\{ \cos\big( k(z_1 - z_2) \big) - \gamma^z \cos\big( k(z_1 + z_2) \big) \Big\} \Bigg]
\end{eqn}
It would be interesting to use the cut propagators to constrain the behavior of loop-level correlators which we leave for future work. 
}. This representation enables one to perform integrals for the cosmological correlator due to a hidden structure of rewriting them as flat space integrals \cite{Chowdhury:2023arc}. However, is not trivial to implement the same algorithm AdS correlators and we hope to explore these integrals in future work. 

%%%%%%%%%%%%%%%%%%%%%%%%%%%
\section{Soft Limit of Photons attached to Internal Legs} \label{app:class2}
%%%%%%%%%%%%%%%%%%%%%%%%%%%
In this appendix we consider the class of diagrams where the photon is attached to an internal leg in the Witten diagram as shown below. Together with the analysis in section \ref{sec:soft}, this completes the analysis of soft limits for fermionic QED.
\begin{eqn}
Y &\equiv \begin{tikzpicture}[baseline]
\draw[very thick] (0, 0) circle (2);
\node at (-1.75, 0) {\vdots};
\node at (1.75, 0) {\vdots};
\draw[fill = lightgray] (-1, 0) circle (0.5);
\draw[fill = lightgray] (1, 0) circle (0.5);

\draw[fermion] ({2*cos(150)},{(2*sin(150)}) -- (-1.3, 0.4);
\draw[fermionbar] ({2*cos(210)},{(2*sin(210)}) -- (-1.3, -0.4);

\draw[fermion] ({2*cos(30)},{(2*sin(30)}) -- (1.3, 0.4);
\draw[fermionbar] ({2*cos(-30)},{(2*sin(-30)}) -- (1.3, -0.4);

\draw[fermion] (-.5, 0) -- (0.5, 0);

\draw[photon] ({2*cos(90)},{(2*sin(90)}) -- (0, 0);

\node at ({2.3*cos(90)},{(2.3*sin(90)}) {$\vec k_s$};

\node at (0, -0.3) {$\vec k$};
\node at (-1, 0) {$F_L$};
\node at (1, 0) {$F_R$};
\end{tikzpicture}\\
&= \intsinf dz_1 dz_2 dz_3 e^{- k_s z_2} F_L(z_1) S(z_1, z_2, \vec k) \gamma_i \e^i_s S(z_2, z_3, \vec k+ \vec k_s) F_R(z_3)
\end{eqn}
where the gray blobs are denoted by $F_L(z_1)$ and $F_R(z_3)$ and $\e_s^i \equiv \e^i(\vec k_s)$. Since $F_L(z_1)$ and $F_R(z_3)$ are arbitrary functions of momenta, the soft limit derived for this diagram is equally applicable for loop diagrams where the soft leg is attached to an edge which carries loop momenta. In the soft limit, by momentum conservation at the vertex in the middle, we have 
\begin{eqn}\label{class2soft1}
\lim_{\vec k_s \to 0} Y = \intsinf dz_1 dz_2 dz_3 F_L(z_1) S(z_1, z_2, \vec k) \gamma_i \e^i_s  S(z_2, z_3, \vec k) F_R(z_3)~.
\end{eqn}
After some heavy algebra, we obtain a simple result for the $z_2$-integral,
\begin{eqn}
&\intsinf dz_2  S(z_1, z_2, \vec k) \gamma_i S(z_2, z_3, \vec k) \\
&= 2 i \frac{k^j}{k} \Bigg\{ 2 \delta_{ij}\pd{}{k} S(z_1, z_3, \vec k)+\frac{\Sigma_{ij} }{k}  \big[S(z_1, z_3, \vec k) - S(z_1, z_3, - \vec k) \big] \Bigg\}~.
\end{eqn}
Which gives the following expression for \eqref{class2soft1} 
\begin{eqn}
\lim_{\vec k_s \to 0} Y &= 4 i \frac{k^j\e_s^i}{k}  \delta_{ij}\pd{}{k}  \intsinf dz_1  dz_3 F_L(z_1) S(z_1, z_3, \vec k) F_R(z_3) \\
&\qquad +2 i  \frac{k^j\e_s^i}{k^2}  \intsinf dz_1  dz_3 F_L(z_1)\Sigma^{ij}  \big[S(z_1, z_3, \vec k) - S(z_1, z_3, - \vec k) \big] F_R(z_3)~.
\end{eqn}
By using 
\begin{eqn}
S(z_1, z_3, - \vec k) = \gamma^z S(z_1, z_3, \vec k) \gamma^z,
\end{eqn}
the term proportional to $\Sigma^{ij}$ can also be written as 
\begin{eqn}
& 2i\frac{k_j}{k^2} \Sigma^{ij} \Big[S(z_1, z_3, \vec k) - \gamma^z S(z_1, z_3, \vec k) \gamma^z \Big] = 2i\frac{k_j}{k^2} \Sigma^{ij} \Big[ P_- S(z_1, z_3, \vec k) P_+ 
+ P_+  S(z_1, z_3, \vec k)  P_- \Big]~.
\end{eqn}
Hence we obtain the soft limit for the class of diagrams where the photon is attached to an internal leg,
\begin{eqn}\label{class2soft}
\lim_{\vec k_s \to 0} Y &= 4 i \frac{k^j\e_s^i}{k}  \delta_{ij}\pd{}{k}  \intsinf dz_1  dz_3 F_L(z_1) S(z_1, z_3, \vec k) F_R(z_3) \\
&\quad + 2 i  \frac{k^j\e_s^i}{k^2}  \intsinf dz_1  dz_3 F_L(z_1) \Sigma^{ij} P_- S(z_1, z_3, \vec k) P_+ F_R(z_3) \\
&\quad + 2 i  \frac{k^j\e_s^i}{k^2}  \intsinf dz_1  dz_3 F_L(z_1) \Sigma^{ij} P_+ S(z_1, z_3, \vec k) P_- F_R(z_3) ~.
\end{eqn}
The terms proportional to $\Sigma^{ij}$ would be absent for scalars coupled to photons and we would only have the term with the derivative $\pd{}{k}$. We summarize the relation diagrammatically below 
\begin{eqn}
\lim_{\vec k_s \to 0}
 \begin{tikzpicture}[baseline]
\draw[very thick] (0, 0) circle (2);
\node at (-1.75, 0) {\vdots};
\node at (1.75, 0) {\vdots};
\draw[fill = lightgray] (-1, 0) circle (0.5);
\draw[fill = lightgray] (1, 0) circle (0.5);

\draw[fermion] ({2*cos(150)},{(2*sin(150)}) -- (-1.3, 0.4);
\draw[fermionbar] ({2*cos(210)},{(2*sin(210)}) -- (-1.3, -0.4);

\draw[fermion] ({2*cos(30)},{(2*sin(30)}) -- (1.3, 0.4);
\draw[fermionbar] ({2*cos(-30)},{(2*sin(-30)}) -- (1.3, -0.4);

\draw[fermion] (-.5, 0) -- (0.5, 0);

\draw[photon] ({2*cos(90)},{(2*sin(90)}) -- (0, 0);

\node at ({2.3*cos(90)},{(2.3*sin(90)}) {$\vec k_s$};

\node at (0, -0.3) {$\vec k$};
\node at (-1, 0) {$F_L$};
\node at (1, 0) {$F_R$};
\end{tikzpicture}
= 
\hat T_{r_1 r_2}
 \begin{tikzpicture}[baseline]
\draw[very thick] (0, 0) circle (2);
\node at (-1.75, 0) {\vdots};
\node at (1.75, 0) {\vdots};
\draw[fill = lightgray] (-1, 0) circle (0.5);
\draw[fill = lightgray] (1, 0) circle (0.5);

\draw[fermion] ({2*cos(150)},{(2*sin(150)}) -- (-1.3, 0.4);
\draw[fermionbar] ({2*cos(210)},{(2*sin(210)}) -- (-1.3, -0.4);

\draw[fermion] ({2*cos(30)},{(2*sin(30)}) -- (1.3, 0.4);
\draw[fermionbar] ({2*cos(-30)},{(2*sin(-30)}) -- (1.3, -0.4);

\draw[fermion] (-.5, 0) -- (0.5, 0);

%\draw[photon] ({2*cos(90)},{(2*sin(90)}) -- (0, 0);

%\node at ({2.3*cos(90)},{(2.3*sin(90)}) {$\vec k_s$};

\node at (0, -0.3) {$\vec k$};
\node at (-1, 0) {$F^{r_1}_L$};
\node at (1, 0) {$F^{r_2}_R$};
\end{tikzpicture}
\end{eqn}
where 
\begin{eqn}
\hat T_{r_1 r_2} = 2i k^j \e_s^i \left[  2 \frac{\delta_{ij}}{k}  \pd{}{k} + \frac{1}{k^2} \big\{ (\Sigma_{ij} P_-)_{r_1 r_2} + (P_- \Sigma_{ij})_{r_2 r_1} \big\} \right] ~.
\end{eqn}
 Hence we see that the soft limit of this class of diagrams does not factorize into the original lower point correlators but does so into diagrams with a modified bulk-bulk propagator. Unlike the case of the soft limit for the diagrams discussed in section \ref{sec:soft}, the individual terms in \eqref{class2soft} are not trivially discernible from each other as they have the same pole structure. However, for this specific theory, they can still be discerned in a unique way from the rest of the terms due to the order of poles in the exchange momenta and therefore provide useful consistency checks.

%%%%%%%%%%%%%%%%%%%%%%%%%%%
\section{Massive Loop Integrals} \label{app:evaluatingloop}
%%%%%%%%%%%%%%%%%%%%%%%%%%%
In this appendix, we demonstrate how the recursion relations described for the massive fields in section \ref{sec:massive} work at 1-loop. For simplicity, we consider the examples of scalar fields in the internal legs, but a similar strategy also works for fermions. Consider the two-site bubble graph with a scalar of mass $m = 0$ in one internal leg of the loop and a scalar of mass $m^2 = - 2$ in another internal leg of the loop. To illustrate the structure of the integral we consider an interaction term $z \phi^3 \zeta$ where $\phi$ is conformally coupled and $\zeta$ is massless. The bubble diagram which appears in the 4-pt function $\braket{\phi(\vec k_1)\phi(\vec k_2)\phi(\vec k_3) \phi(\vec k_4)}$ is   
\begin{eqn}\label{massiveloop1}
\begin{tikzpicture}[baseline]

%\draw (0,0) circle (1);
%\node at (0, -1.35) {$\chi, \vec l_2$};
%\node at (0, 1.35) {$\phi, \vec l_1$};

\draw[black] (1,0) arc (0:180:1);
\draw[red] (1,0) arc (0:-180:1);

\node at (-1.3, 0) {$x_1$};
\node at (1.3, 0) {$x_2$};

\node at (-1, 0) {\textbullet};
\node at (1, 0) {\textbullet};
\end{tikzpicture}
= \intsinf dz_1 dz_2 e^{- x_1 z_1} e^{- x_2 z_2}  G_{\phi}(z_1, z_2, \vec y_1)   G_{\zeta}(z_2, z_1, \vec y_2) 
\end{eqn}
where $x_1 = k_{12}, \ x_2 = k_{34}$, $y_1 = |\vec l|, y_2 = |\vec l + \vec k_1 + \vec k_2|$. The field $\zeta$ is represented by red lines and the field $\phi$, by black lines. The bulk-bulk propagators for $\phi$ and $\zeta$ are
\begin{eqn}
G_{\phi}(z_1, z_2, \vec y_1) &= (z_1 z_2)^{1/2} \intinf \frac{d\omega}{\omega^2 + y_1^2} J_{1/2}(\omega z_1) J_{1/2}(\omega z_2)
 \\
G_{\zeta}(z_1, z_2, \vec y_2) &= (z_1 z_2)^{3/2}  \intinf \frac{d\omega}{\omega^2 + y_2^2} J_{3/2}(\omega z_1) J_{3/2}(\omega z_2)~.
\end{eqn}
By using the IBP to evaluate the $z$-integrals we obtain
\begin{align}\label{massivebubble1}
\begin{tikzpicture}[baseline]
%\draw (0,0) circle (1);
%\node at (0, -1.35) {$\chi, \vec l_2$};
%\node at (0, 1.35) {$\phi, \vec l_1$};
\draw[black] (1,0) arc (0:180:1);
\draw[red] (1,0) arc (0:-180:1);
\node at (-1.3, 0) {$x_1$};
\node at (1.3, 0) {$x_2$};
\node at (-1, 0) {\textbullet};
\node at (1, 0) {\textbullet};
\end{tikzpicture}= \frac{1}{x_1 + x_2} &\Bigg\{ 
- \p_{x_1} \p_{x_2} 
\begin{tikzpicture}[baseline]
\draw (-1, 0) -- (1, 0);
\node at (-1, 0) {\textbullet};
\node at (1, 0) {\textbullet};
\node at (-1, -0.25) {$x_1 + y_2$};
\node at (1, -0.25) {$x_2 + y_2$};
\node at (0, 0.25) {$ y_1$};
\end{tikzpicture}
+ 
\begin{tikzpicture}[baseline]
\draw[red] (-1, 0) -- (1, 0);
\node at (-1, 0) {\textbullet};
\node at (1, 0) {\textbullet};
\node at (-1, -0.25) {$x_1 + y_1$};
\node at (1, -0.25) {$x_2 + y_1$};
\node at (0, +0.25) {$y_2$};
\end{tikzpicture} \nno \\
&\quad+ \Big( \p_{x_1} + \p_{x_2} \Big) 
\begin{tikzpicture}[baseline]
\node at (-1, 0) {\textbullet};
\node at (1, 0) {\textbullet};
\draw (0,0) circle (1);
%\node at (0, -1.25) {$\frac12, y_2$};
%\node at (0, 1.25) {$\frac12, y_1$};
\node at (-1.3, 0) {$x_1$};
\node at (1.3, 0) {$x_2$};
\end{tikzpicture}
\Bigg\}~.
\end{align}
This shows how the loop diagram with a massless leg can be expressed in terms of differential operators acting on tree level conformally coupled graph combined with a loop level conformally coupled bubble. However, it is not obvious on how to express this as a differential operator solely acting on a conformally coupled bubble. The second line in \eqref{massivebubble1} is obtained by differentiating the result of the bubble diagram for the conformally coupled scalar (given in equation (3.25) of \cite{Chowdhury:2023khl}). We evaluate the other terms below. The first term of \eqref{massivebubble1} is 
\begin{eqn}
&- \frac{1}{x_1 + x_2}\p_{x_1} \p_{x_2} \int d^3 l \begin{tikzpicture}[baseline]
\draw (-1, 0) -- (1, 0);
\node at (-1, 0) {\textbullet};
\node at (1, 0) {\textbullet};

\node at (-1, -0.25) {$x_1 + y_2$};
\node at (1, -0.25) {$x_2 + y_2$};

\node at (0, 0.25) {$ y_1$};
\end{tikzpicture} \\
&= - \frac{1}{x_1 + x_2}\p_{x_1} \p_{x_2} \int d^3 l \frac{1}{(x_1 + x_2 + 2 y_2) (x_1 + y_1 + y_2) (x_2 + y_1 + y_2)}
\end{eqn}
This term already appears in the loop integrand for the bubble (see equation (3.23) of \cite{Chowdhury:2023khl})  and upon evaluating the loop integral gives a $\mbox{Li}_2$ in the external momenta. The explicit integral is similar to the one evaluated in equation \eqref{bubble-x1x2-ans} and we record it in the Mathematica notebook found in \cite{github}.  The remaining term in \eqref{massivebubble1} is given as 
\begin{eqn}\label{massivebubble1-term3}
&\frac{1}{x_1 + x_2} \int d^3l 
\begin{tikzpicture}[baseline]
\draw[red] (-0.5, 0) -- (0.5, 0);
\node at (-0.5, 0) {\textbullet};
\node at (0.5, 0) {\textbullet};

\node at (-0.5, -0.25) {$x_1'$};
\node at (0.5, -0.25) {$x_2'$};

\node at (0, +0.25) {$y_2$};
\end{tikzpicture}\\
&= - \frac{1}{(x_1 + x_2)}  \int \frac{d^3 l}{x_1' + x_2'} \Big[ ( \p_{x_1}  + \p_{x_2} )
\begin{tikzpicture}[baseline]
\draw (-0.5, 0) -- (0.5, 0);
\node at (-0.5, 0) {\textbullet};
\node at (0.5, 0) {\textbullet};

\node at (-0.5, -0.25) {$x_1'$};
\node at (0.5, -0.25) {$x_2'$};

\node at (0, +0.25) {$y_2$};
\end{tikzpicture}
+ \p_{x_1} \p_{x_2}
\begin{tikzpicture}[baseline]
%\draw (-0.5, 0) -- (0.5, 0);
\node at (-0.5, 0) {\textbullet};
\node at (0.5, 0) {\textbullet};

\node at (-0.5, -0.25) {$x_1'$};
\node at (0.5, -0.25) {$x_2'$};

%\node at (0, +0.25) {$y_2$};
\end{tikzpicture}
\Big] \end{eqn}
where $x_1' = x_1 + y_1, \ x_2' = x_2 + y_1$ and we have used equation \eqref{massiverecursion2} to obain the RHS\footnote{We also note the following useful relation that simplifies many computations
\begin{eqn}\label{recsim1}
\p_{x_1} \p_{x_2} \Big[ \begin{tikzpicture}[baseline]
\node at (-0.5, 0) {\textbullet};
\node at (0.5, 0) {\textbullet};
%\draw (-0.5,0) -- (0.5,0);
\end{tikzpicture} 
 \Big]
 = \Big\{ (x_1 + x_2) \p_{x_1} \p_{x_2} + (\p_{x_1} + \p_{x_2}) \Big\}\begin{tikzpicture}[baseline]
\node at (-0.5, 0) {\textbullet};
\node at (0.5, 0) {\textbullet};
\draw (-0.5,0) -- (0.5,0);
\end{tikzpicture} 
\end{eqn}}. This results in the following loop integral for \eqref{massivebubble1-term3}
\begin{align}
&- \frac{1}{(x_1 + x_2)} 
\Bigg[ \p_{x_1} \p_{x_2} \int d^3 l \frac{1}{(x_1 + x_2 + 2 y_1)(x_1 + y_1 + y_2) (x_2 + y_1 + y_2)}  \\
&   + (\p_{x_1} + \p_{x_2}) \int d^3 l \frac{1}{(x_1 + x_2 + 2 y_1)(x_1  + y_1 + y_2)(x_2 + y_1 + y_2) (x_1 + x_2 + y_1 + y_2)} \Bigg]\nno
\end{align}
and hence these integral reduces to the kind that appear for the conformally coupled bubble diagram and are contained in the \reference{Mathematica} notebook found in \cite{github}. By evaluating all the integrals explicitly, we find that the final answer is a Li$_2$, which is same as that of the conformally coupled bubble. Therefore, in cases without any IR divergences, the bubble diagram with massive internal legs with specific values of the mass, has the same transcendentality as the conformally coupled bubble.

There is an alternate way to evaluate the diagram in \eqref{massiveloop1} which make use of the tree theorem developed in section \ref{sec:treethm}. Since the external fields are conformally coupled, by using the tree theorem in section \ref{sec:treethm} we obtain,
\begin{eqn}
\begin{tikzpicture}[baseline]

%\draw (0,0) circle (1);
%\node at (0, -1.35) {$\chi, \vec l_2$};
%\node at (0, 1.35) {$\phi, \vec l_1$};
\draw[black] (1,0) arc (0:180:1);
\draw[red] (1,0) arc (0:-180:1);
\node at (-1.3, 0) {$x_1$};
\node at (1.3, 0) {$x_2$};
\node at (-1, 0) {\textbullet};
\node at (1, 0) {\textbullet};
\end{tikzpicture}
&= \int_{-i\infty}^{i\infty} \frac{d\e}{2\pi i} 
\begin{tikzpicture}[baseline]

\node at (-1, +0.25) {$\frac{x_1}{2} - \e$};
\node at (0, +0.25) {$x_2$};
\node at (1, +0.25) {$\frac{x_1}{2} + \e$};

\draw (-1, 0) -- (0, 0);
\draw[red] (0, 0) -- (1, 0);
\node at (-1, 0) {\textbullet};
\node at (0, 0) {\textbullet};
\node at (1, 0) {\textbullet};
\end{tikzpicture}
\end{eqn}
where the diagram on the right can be written in terms of derivatives acting on conformally coupled graphs and is given as  ($x_1' = \frac{x_1}{2} - \e$, $x_3'= \frac{x_1}{2} + \e$, $x_3 = \frac{x_1}2$)
\begin{eqn*}
\begin{tikzpicture}[baseline]
\draw[red] (0, 0) -- (1, 0);
\node at (-1, 0) {\textbullet};
\node at (0, 0) {\textbullet};
\node at (1, 0) {\textbullet};
\node at (-1, +0.25) {$x_1 - \e$};
\node at (0, +0.25) {$x_2$};
\node at (1, +0.25) {$x_3 + \e$};
%\node at (0.5, -0.25) {$\frac32$};
%\node at (-0.5, -0.25) {$\frac12$};

\draw (-1, 0) -- (0, 0);

\end{tikzpicture}
= \frac{1}{x_2 + x_3} \Big[ - (\p_{x_2} + \p_{x_3}) 
\begin{tikzpicture}[baseline]
\node at (-1, 0) {\textbullet};
\node at (0, 0) {\textbullet};
\node at (1, 0) {\textbullet};
\node at (-1, +0.25) {$x_1'$};
\node at (0, +0.25) {$x_2$};
\node at (1, +0.25) {$x_3'$};
%\node at (0.5, -0.25) {$\frac12$};
%\node at (-0.5, -0.25) {$\frac12$};

\draw (-1, 0) -- (1, 0);
\end{tikzpicture}
+ \p_{x_2} \p_{x_3}
\begin{tikzpicture}[baseline]

\node at (-1, +0.25) {$x_1'$};
\node at (0, +0.25) {$x_2 + y_2$};
\node at (1, -0.25) {$x_3' + y_2$};

%\node at (-0.5, -0.25) {$\frac12$};

\draw (-1, 0) -- (0, 0);

\node at (-1, 0) {\textbullet};
\node at (0, 0) {\textbullet};
\node at (1, 0) {\textbullet};
\end{tikzpicture}
 \Big] 
\end{eqn*}
which results in simple graphs with conformally coupled exchanges. This shows that the intermediate steps indeed simplify by using the tree theorem and also shows how the tree theorem is also useful for diagrams with non-conformally coupled internal legs. By evaluating the integral in $\vec l$ we obtain a Li$_2$ in the external momenta which is similar to the conformally coupled case. It would be interesting to generalize this approach to studying the transcendentality of functions obtained for generic masses including cases when we have IR divergences.

%%%%%%%%%%%%%%%%%%%%%%%%%%%
\section{In-In Correlators with External Fermions}\label{app:ferm-inin}
%%%%%%%%%%%%%%%%%%%%%%%%%%%
In this appendix, we provide further details on computing the path integral for the cosmological correlator with external fermions. To be concrete, we evaluate the following correlator in Yukawa theory,
\begin{eqn}
\braket{\Psi_{dS}| \bar\psi_i(\vec k_1) \phi(\vec k_2) \psi_j(\vec k_3) \phi(\vec k_4) | \Psi_{dS}}
= \int D\phi D\psi D\bar\psi \bar\psi_i(\vec k_1) \phi(\vec k_2) \psi_j(\vec k_3) \phi(\vec k_4) |\Psi_{dS}|^2
\end{eqn}

\subsubsection*{Tree-Level: $\bm{O(\lambda^2)}$}
The 2-pt functions are normalized as, 
\begin{eqn}
\braket{\bar\psi_a(\vec q) \psi_b(\vec q')} \equiv \frac{1}{\re W^{ab}_2(q,  q')} \delta(\vec q - \vec q') &= \frac{(\gamma \cdot q)^{ab}}{q} \delta(\vec q - \vec q'), \\
\braket{\phi(\vec q) \phi(\vec q')} \equiv \frac{1}{\re W^{(\phi)}_2( q,   q')} \delta(\vec q - \vec q') &= q \delta(\vec q - \vec q').
\end{eqn}

From the Witten diagram computation of the 3-pt function we get,
\begin{eqn}
\re W^{ab}_3(q_1, q_2, q_3) = \frac{\bar u^a(q_1) u^b(q_2)}{q_1 + q_2 + q_3} \delta(\vec q_1 + \vec q_2 + \vec q_3)
\end{eqn}

The 3-pt function at tree level is given as,
\begin{align}
\braket{\bar\psi^i(p_1) \psi^j(p_2) \phi(p_3)} &= \int D\bar\psi D\psi D\phi
\exp\Big(- \re\int \bar\psi_a \psi_b W^{ab}_2[\bar\psi, \psi] -\re  \int \phi \phi W_2^{(\phi)}[\phi, \phi] \Big) \nno \\
&\quad \times \bar\psi^i(p_1) \psi^j(p_2) \phi(p_3) \Big( 1 - \re \int \bar\psi_a \psi_b \phi W_3^{ab}[\bar \psi, \psi, \phi]  \Big) 
\end{align}

Plugging everything explicitly, we obtain the following
\begin{align}
&\braket{\bar\psi^i(p_1) \psi^j(p_2) \phi(p_3)} \nno\\
 &= \int dq_1 dq_2 dq_3 \frac{ \delta(q_3 - p_3)}{\re W_2^\phi(q_3, p_3)} \re W^{ab}_3 (q_1, q_2, q_3) \frac{\delta(q_1 - p_2)}{q_1} (\gamma \cdot q_1)^{ja} \frac{\delta(q_2 - p_1)}{q_2} (\gamma \cdot q_2)^{ib} \nno\\
  &=  \frac{1}{\re W_2^\phi(p_3, p_3)} \re W^{ab}_3 (p_2, p_1, p_3) \frac{1}{p_1} (\gamma \cdot p_2)^{ja} \frac{1}{p_2} (\gamma \cdot p_1)^{ib} 
\end{align}

For the 4-pt function, we divide the contribution into two terms, one coming from the product of $W_3$'s and the other coming from $W_4$. 
\begin{eqn}
\braket{\bar\psi^i(p_1) \psi^j(p_2) \phi(p_3) \phi(p_4)} = \braket{\bar\psi^i(p_1) \bar\psi^j(p_2) \phi(p_3) \phi(p_4)}^{(1)}  + \braket{\bar\psi^i(p_1) \psi^j(p_2) \phi(p_3) \phi(p_4)}^{(2)} 
\end{eqn}
where 
\begin{eqn}
&\braket{\bar\psi^i(p_1) \bar\psi^j(p_2) \phi(p_3) \phi(p_4)}^{(1)}  = \int dq_1 \cdots dq_6  \re W^{ab}_3 (q_1, q_2, q_3) \re W^{cd}_3 (q_4, q_5, q_6) \\
&\qquad\times \int D\phi D\bar\psi D\psi \bar\psi_a(q_1)  \psi_b(q_2)  \phi(q_3)  \bar\psi_c(q_4)  \psi_d(q_5)  \phi(q_6)  \times 
\bar\psi^i(p_1) \psi^j(p_2) \phi(p_3) \phi(p_4)  e^{- S_{quad}}
\end{eqn}
and
\begin{eqn}
&\braket{\bar\psi^i(p_1) \bar\psi^j(p_2) \phi(p_3) \phi(p_4)}^{(2)}  = \int dq_1 \cdots dq_4  \re W^{ab}_4 (q_1, q_2, q_3, q_4)\\
&\qquad\times \int D\phi D\bar\psi D\psi \bar\psi_a(q_1)  \psi_b(q_2)  \phi(q_3)  \phi(q_4) \times 
\bar\psi^i(p_1) \psi^j(p_2) \phi(p_3) \phi(p_4)  e^{- S_{quad}}
\end{eqn}

The correlators also receive contributions from disconnected diagrams but we only list the connected contributions below. The first term is given as
\begin{align}
&\braket{\bar\psi^i(p_1) \bar\psi^j(p_2) \phi(p_3) \phi(p_4)}^{(1)}  = \int dq_1 \cdots dq_6  \re W^{ab}_3 (q_1, q_2, q_3) \re W^{cd}_3 (q_4, q_5, q_6) \nno\\
&\qquad\times \int D\phi D\bar\psi D\psi \bar\psi_a(q_1)  \psi_b(q_2)  \phi(q_3)  \bar\psi_c(q_4)  \psi_d(q_5)  \phi(q_6)  \times 
\bar\psi^i(p_1) \psi^j(p_2) \phi(p_3) \phi(p_4)  e^{- S_{quad}} \nno\\
& = \int dq_1 \cdots dq_6  \re W^{ab}_3 (q_1, q_2, q_3) \re W^{cd}_3 (q_4, q_5, q_6) \\
&\times
\frac{\delta( \vec q_1 -  \vec p_2)}{ \re W_2^{aj}(q_1, p_2)}  \frac{\delta( \vec q_2 -  \vec q_4)}{ \re W_2^{bc}(q_2, q_4)} \frac{\delta( \vec q_5 -  \vec p_1)}{ \re W_2^{id}(q_5, p_1)} \times  \frac{\delta(\vec q_3 -\vec p_3)}{W_2^{\phi} (q_3, p_3)} \frac{\delta(\vec q_6 -\vec p_4)}{W_2^{\phi} (q_6, p_4)} \nno \\ 
&= \int d q_2  \frac{\re W^{ab}_3 (p_2, q_2, p_3) \re W^{cd}_3 (q_2, p_1, p_4)}{\re W_2^{aj}(p_2, p_2) \re W_2^{bc}(q_2, q_2)\re W_2^{di}(p_1, p_1) W_2^{\phi} (p_4, p_4) W_2^{\phi} (p_3, p_3)}\nno
\end{align}
Note that the momentum conservations in $W_3$ would give $\vec p_2 + \vec p_3 + \vec q_2 = \vec p_1 + \vec p_4 + \vec q_2 = 0 $ which would eliminate the $q_2$ integral. The second term's contribution to the correlator is given as 
\begin{eqn}
&\braket{\bar\psi^i(p_1) \bar\psi^j(p_2) \phi(p_3) \phi(p_4)}^{(2)}  = \int dq_1 \cdots dq_4  \re W^{ab}_4 (q_1, q_2, q_3, q_4)\\
&\qquad\times \int D\phi D\bar\psi D\psi \bar\psi_a(q_1)  \psi_b(q_2)  \phi(q_3)  \phi(q_4) \times 
\bar\psi^i(p_1) \psi^j(p_2) \phi(p_3) \phi(p_4)  e^{- S_{quad}} \\
&= \frac{\re W^{ac}_4 (p_2, p_1, p_3, p_4)}{\re W_2^{aj}(p_1, p_1) \re W_2^{ci}(p_2, p_2) W_2^{\phi} (p_3, p_3) W_2^{\phi} (p_4, p_4)}
\end{eqn}

By adding the two we obtain the full correlator, 
\begin{align}
\braket{\bar\psi^i(p_1) \psi^j(p_2) \phi(p_3) \phi(p_4)}&= \frac{1}{\re W_2^{aj}(p_1, p_1) \re W_2^{di}(p_2, p_2) W_2^{\phi} (p_3, p_3) W_2^{\phi} (p_4, p_4)} \\
 &\times\left\{ \frac{\re W^{ab}_3 (p_2, q_2, p_3) \re W^{cd}_3 (q_2, p_1, p_4)}{W_2^{bc}(q_2, q_2)}
 + \re W^{ac}_4 (p_2, p_1, p_3, p_4) \right\}\nno
\end{align}

%%%%%%%%%%%%%%%%%%%%%%%%%%%
\subsubsection*{One-Loop: $\bm{O(\lambda^4)}$}
We can similarly compute the contribution at the next order. It is seen that $O(\lambda^3)$ the contribution is zero. However at $O(\lambda^4)$ receives contributions from multiple Witten diagrams, including 
\begin{eqn}\label{wavefunctioncoeff-loop}
W_{3}^{(1)ab} &=  \scalebox{0.75}{\begin{tikzpicture}[baseline]
\draw[very thick] (0, 0) circle (2);
\draw[fermion] ({2*cos(-30)},{(2*sin(-30)}) -- (0, 0);
\draw[fermionbar] ({2*cos(210)},{(2*sin(210)}) -- (0, 0);
\draw ({2*cos(90)},{(2*sin(90)}) -- (0, 0);

\end{tikzpicture}}, \quad 
W_{4}^{(1) ab} =    \scalebox{0.75}{\begin{tikzpicture}[baseline]
\draw[very thick] (0, 0) circle (2);

\draw[fermionbar] ({2*cos(210)},{(2*sin(210)}) -- (-1, 0);
\draw ({2*cos(150)},{(2*sin(150)}) -- (-1, 0);
\draw[fermionbar] (-1, 0) -- (1,0); 
\draw[fermion] ({2*cos(-30)},{(2*sin(-30)}) -- (1, 0);
\draw ({2*cos(30)},{(2*sin(30)}) -- (1, 0);

\end{tikzpicture}}, \quad 
W_{3}^{(3) ab} =   \scalebox{0.75}{\begin{tikzpicture}[baseline]
\draw[very thick] (0, 0) circle (2);
\draw[fermion] ({2*cos(-30)},{(2*sin(-30)}) -- (0.75, -0.75);
\draw[fermionbar] ({2*cos(210)},{(2*sin(210)}) -- (-0.75, -0.75);
\draw (0,2) -- (0, 0.75);

\draw[fermion] (0.75, -0.75) -- (0, 0.75);
\draw[fermion] (0, 0.75) -- (-0.75, -0.75);
\draw (-0.75, -0.75) -- (0.75, -0.75);
\end{tikzpicture}}, \\
W_{4}^{(4) ab} &=  \scalebox{0.75}{  \begin{tikzpicture}[baseline]
\draw[very thick] (0, 0) circle (2);
\draw[fermion] ({2*cos(-30)},{(2*sin(-30)}) -- (0.75, -0.75);
\draw ({2*cos(30)},{(2*sin(30)}) -- (0.75, 0.75);
\draw ({2*cos(150)},{(2*sin(150)}) -- (-0.75, 0.75);
\draw[fermionbar] ({2*cos(210)},{(2*sin(210)}) -- (-0.75, -0.75);
\draw[fermion] (0.75, -0.75) -- (0.75, 0.75);
\draw[fermionbar] (-0.75, 0.75) -- (0.75, 0.75);
\draw[fermion] (-0.75, 0.75) -- (-0.75, -0.75);
\draw (-0.75, -0.75) -- (0.75, -0.75);

\end{tikzpicture}},\quad 
W_{5}^{ab} =  \scalebox{0.75}{\begin{tikzpicture}[baseline]
\draw[very thick] (0, 0) circle (2);

\draw[fermionbar] ({2*cos(210)},{(2*sin(210)}) -- (-1, 0);
\draw ({2*cos(150)},{(2*sin(150)}) -- (-1, 0);
\draw[fermionbar] (-1, 0) -- (0,0); 
\draw (0, 0) -- (0,2); 
\draw[fermionbar] (0, 0) -- (1,0); 
\draw[fermion] ({2*cos(-30)},{(2*sin(-30)}) -- (1, 0);
\draw ({2*cos(30)},{(2*sin(30)}) -- (1, 0);

\end{tikzpicture}}, \quad
W_{6}^{ab} =  \scalebox{0.75}{\begin{tikzpicture}[baseline]
\draw[very thick] (0, 0) circle (2);

\draw[fermionbar] ({2*cos(210)},{(2*sin(210)}) -- (-1.5, 0);
\draw ({2*cos(150)},{(2*sin(150)}) -- (-1.5, 0);

\draw (0.75, 0) -- (0.75,1.85); 
\draw (-0.75, 0) -- (-0.75,1.85); 

\draw[fermionbar] (-1.5, 0) -- (-0.75,0); 
\draw[fermionbar] (-0.75, 0) -- (0.75,0); 
\draw[fermionbar] (0.75, 0) -- (1.5,0); 

\draw[fermion] ({2*cos(-30)},{(2*sin(-30)}) -- (1.5, 0);
\draw ({2*cos(30)},{(2*sin(30)}) -- (1.5, 0);

\end{tikzpicture}}
\end{eqn}
The 4-pt function at $O(\lambda^4)$ is given as 
\begin{eqn}
&\braket{\Psi| \bar\psi^i(p_1) \psi^j(p_2) \phi(p_3) \phi(p_4) |\Psi}= -\int D\psi D\bar\psi D\phi \bar\psi^i(p_1) \psi^j(p_2) \phi(p_3) \phi(p_4) \\
&\times \Bigg\{ \int W_6^{(1)ab}(q_1, \cdots, q_6) \bar \psi_a(q_1) \psi_b(q_2) \phi(q_3) \cdots \phi(q_6) 
+ \int W_4^{(4)ab}(q_1, \cdots, q_4) \bar \psi_a(q_1) \psi_b(q_2) \phi(q_3) \phi(q_4) \\
&\quad + \int W_{3}^{(1)ab}(q_1, q_2, q_3) \bar\psi_a(q_1) \psi_b(q_2) \phi(q_3) 
\int W_3^{(3)cd}(q_1, q_2, q_3) \psi_c(q_1) \psi_d(q_2) \phi(q_3) \\
&\quad + \int W_{3}^{(1)ab}(q_1, q_2, q_3) \bar\psi_a(q_1) \psi_b(q_2) \phi(q_3)  \int W_{5}^{(1)cd}(q_1, \cdots, q_5) \bar\psi_c(q_1) \psi_d(q_2) \phi(q_3) \phi(q_4) \phi(q_5), \\ 
&\quad + \int W_{4}^{(1)ab}(q_1, \cdots, q_4) \bar\psi_a(q_1) \psi_b(q_2) \phi(q_3) \phi(q_4)  
\int W_{4}^{(1)cd}(q_1,\cdots, q_4) \bar\psi_c(q_1) \psi_d(q_2) \phi(q_3) \phi(q_4)  
\Bigg\}
\end{eqn}
 Using the recursion relations it is easily possible to compute the loop integrands in \eqref{wavefunctioncoeff-loop}. For example, $W_4^{(4)ab}$ is given as 
 \begin{eqn}
 W_4^{(4)ab} &= \bar u_1\Big( 
\begin{tikzpicture}[baseline]
   \draw[fermionbar] (0, 0) -- (1, 0);
   \draw[fermionbar] (-1, 0) -- (0, 0);
 \node at (-1, 0) {\textbullet};
  \node at (0, 0) {\textbullet};
   \node at (1, 0) {\textbullet};
 \end{tikzpicture} 
 + 
 \begin{tikzpicture}[baseline]
   \draw[fermionbar] (0, 0) -- (1, 0);
   \draw (-1, 0) -- (0, 0);
 \node at (-1, 0) {\textbullet};
  \node at (0, 0) {\textbullet};
   \node at (1, 0) {\textbullet};
 \end{tikzpicture}  
 \Pi_1
  \begin{tikzpicture}[baseline]
  \node at (0, 0) {\textbullet};
 \end{tikzpicture} 
  + \mbox{perms.}
\Big)  u_4\\
&+ \bar u_1\Big(
  \begin{tikzpicture}[baseline]
   \draw[fermionbar] (0, 0) -- (1, 0);
  \node at (0, 0) {\textbullet};
   \node at (1, 0) {\textbullet};
 \end{tikzpicture}  P_+ \Pi_2
+\quad 
\Pi_2 P_-   \begin{tikzpicture}[baseline]
   \draw[fermionbar] (0, 0) -- (1, 0);
  \node at (0, 0) {\textbullet};
   \node at (1, 0) {\textbullet};
 \end{tikzpicture}   
 \Big) u_4~,
 \end{eqn}
 however performing the integrals requires some new ideas beyond the scope of the paper. These examples show how we can use the tools developed in the main sections to also compute cosmological correlators in perturbation theory. However, given the complexity of the answer above, it is clear that this approach is not suitable for computing loop diagrams beyond the first few orders. It would be interesting to develop alternate tools to compute such objects. For example, using the shadow formalism for spinors \cite{Schaub:2023scu} or the in-out formalism for in-in correlators \cite{Donath:2024utn} seem like interesting avenues to explore in the future.

\end{appendix}
%\printbibliography 
%\newpage
\bibliographystyle{JHEP}
\bibliography{references}
\end{document}

%% file: chandra-macros.tex
\usepackage{bm, bbm, bbold}
\usepackage{latexsym}
\usepackage{dcolumn}
\usepackage{amsmath,amsfonts,amssymb, mathrsfs, amsthm}
\usepackage[T1]{fontenc}

\usepackage{graphicx,epsfig}
\usepackage{environ}
\usepackage{mathtools}
\usepackage{braket}
\usepackage{fancyhdr}
\usepackage{hyperref}
\usepackage{graphicx,epstopdf}
\usepackage{tikz}
\usepackage{float}
\usepackage{framed}
\usepackage{soul}
\usetikzlibrary{positioning,decorations.markings, decorations.pathmorphing, calc}
\tikzset{snake it/.style={decorate, decoration=snake}}
\usetikzlibrary{angles,quotes}

\usepackage{stmaryrd}
\usepackage{slashed}
\usepackage{array,multirow}

\usepackage[framemethod=default]{mdframed}
\newmdenv[skipabove=7pt,
skipbelow=7pt,
rightline=false,
leftline=false,
topline=false,
bottomline=false,
backgroundcolor=gray!10,
linecolor=gray,
innerleftmargin=5pt,
innerrightmargin=5pt,
innertopmargin=5pt,
innerbottommargin=5pt,
leftmargin=0cm,
rightmargin=0cm,
linewidth=4pt]{eBox}

\newcommand{\dagg}{\dagger}

\def\a{\alpha}
\def\b{\beta}

\def\e{\epsilon}

\def\p{\partial}

\newcommand{\re}{\mbox{Re}}

\newcommand{\be}{\begin{equation}}
\newcommand{\ee}{\end{equation}}
\newcommand{\bes}{\begin{equation*}}
\newcommand{\ees}{\end{equation*}}

\NewEnviron{derivation}{
\begin{framed}
\begin{center}
{\bf-: Derivation :-}\\
\end{center}
  \BODY

\end{framed}
}

\NewEnviron{eqn}{
\begin{align}
\begin{split}
  \BODY
\end{split}
\end{align}
}

\NewEnviron{eqn*}{
\begin{align*}
\begin{split}
  \BODY
\end{split}
\end{align*}
}

\newcommand{\nno}{\nonumber}
\newcommand{\intinf}{\int_{-\infty}^{\infty}} 
\newcommand{\intsinf}{\int_{0}^{\infty}} 
\newcommand{\pd}[2]{\frac{\partial #1}{\partial #2}} 

\newcommand{\Res}[1]{\underset{#1}{\mbox{Res}}}

\newcommand{\reference}[1]{{\fontfamily{lmtt}\selectfont #1}}
\newcommand{\tr}{\mbox{tr}}

\usepackage{cancel}
\usepackage{tikz, pgf}
\usetikzlibrary{shapes.misc}
\usetikzlibrary{shapes}
\usetikzlibrary{fit}
\usetikzlibrary{arrows}

\usetikzlibrary{decorations.pathmorphing}	% For Feynman Diagrams
\usetikzlibrary{decorations.markings}
\tikzset{
    sugra/.style={decorate, decoration={snake}, draw=black},
    scalarphi/.style={dashed,draw=black, postaction={decorate},
        },
    hwbou/.style={draw=blue, postaction={decorate}, ultra thick
        },
    vector/.style={draw=blue,decorate, decoration={snake}, draw},
	provector/.style={decorate, decoration={snake,amplitude=2.5pt}, draw},
	antivector/.style={decorate, decoration={snake,amplitude=-2.5pt}, draw},
   	 fermion/.style={draw=cyan, postaction={decorate},
        decoration={markings,mark=at position .55 with {\arrow[draw=black]{>}}}},
    fermionbar/.style={draw=cyan, postaction={decorate},
        decoration={markings,mark=at position .55 with {\arrow[draw=black]{<}}}},
    fermionnoarrow/.style={draw=black},
    gluon/.style={decorate, draw=red,
        decoration={coil,amplitude=4pt, segment length=5pt}},
    scalar/.style={dashed,draw=black, postaction={decorate},
        decoration={markings,mark=at position .55 with {\arrow[draw=black]{>}}}},
    scalarbar/.style={dashed,draw=black, postaction={decorate},
        decoration={markings,mark=at position .55 with {\arrow[draw=black]{<}}}},
    electron/.style={draw=black, postaction={decorate},
        decoration={markings,mark=at position .55 with {\arrow[draw=black]{>}}}},
    scalarnoarrow/.style={dashed, draw=black},
    electron/.style={draw=black, postaction={decorate},
        decoration={markings, mark=at position .55 with {\arrow[draw=black]{>}}}},
	bigvector/.style={decorate, decoration={snake, amplitude=4pt}, draw},
    photon/.style={draw=red, decorate, decoration={zigzag}, draw},
    higgs/.style={dashed, draw=black, postaction={decorate},
        },	
        goldstone/.style={draw=brown, postaction={decorate},
        },    
          ghost/.style={dashed, draw=magenta, postaction={decorate},
        decoration={markings, mark=at position .55 with {\arrow[draw=black]{>}}}
        },  
          antighost/.style={dashed, draw=magenta, postaction={decorate},
        decoration={markings, mark=at position .55 with {\arrow[draw=black]{<}}}
        },  
          mphoton/.style={decorate, decoration={snake}, draw=violet},
            realscalar/.style={draw=black}, 
           mgluon/.style={decorate, draw=blue,
        decoration={coil,amplitude=4pt, segment length=5pt}},
         weylfermion/.style={draw=orange, postaction={decorate},
        decoration={markings,mark=at position .55 with {\arrow[draw=black]{>}}}},
         weylfermionbar/.style={draw=orange, postaction={decorate},
        decoration={markings,mark=at position .55 with {\arrow[draw=black]{<}}}}, 
   	wboson/.style={draw=blue,decorate, decoration={snake,amplitude=4pt}, draw},  
    zboson/.style={draw=violet, decorate, decoration={snake}, draw},   
    lepton/.style={draw=black, postaction={decorate},
        decoration={markings,mark=at position .55 with {\arrow[draw=black]{>}}}},
    leptonbar/.style={draw=black, postaction={decorate},
        decoration={markings,mark=at position .55 with {\arrow[draw=black]{<}}}}, 
        graviton/.style={draw=blue,decorate, decoration={snake,amplitude=4pt}, draw},  
        gravitino/.style={draw=red, postaction={decorate},
        decoration={snake, markings, mark=at position .55 with {\arrow[draw=black]{>}}}},
    gravitinobar/.style={draw=red, postaction={decorate},
        decoration={snake, markings,mark=at position .55 with {\arrow[draw=black]{<}}} },    
}